\pdfoutput=1
\documentclass[12pt]{article}
\usepackage{caption}
\usepackage{amsmath,amsfonts,amsthm,amssymb}
\usepackage{subfigure}
\usepackage{setspace}
\usepackage{Tabbing}
\usepackage{lastpage}
\usepackage{extramarks}
\usepackage{chngpage}
\usepackage[usenames,dvipsnames]{color}
\usepackage{graphicx,float,wrapfig}
\usepackage{lineno}
\usepackage{float}
\usepackage{xcolor}
\usepackage[colorlinks=true, a4paper=true, pdfstartview=FitV,
linkcolor=blue, citecolor=blue, urlcolor=blue]{hyperref}

\def\rf#1{(\ref{eq:#1})}
\def\lab#1{\label{eq:#1}}

\def\br{\begin{eqnarray}}
\def\er{\end{eqnarray}}
\def\be{\begin{equation}}
\def\ee{\end{equation}}
\def\({\left(}
\def\){\right)}

\def\rlx{\relax\leavevmode}
\def\IR{\rlx\hbox{\rm I\kern-.18em R}}
\def\vp{\varphi}

\def\IZ{\rlx\hbox{\sf Z\kern-.4em Z}}
\def\IR{\rlx\hbox{\rm I\kern-.18em R}}
\def\IC{\rlx\hbox{\,$\inbar\kern-.3em{\rm C}$}}
\def\one{\hbox{{1}\kern-.25em\hbox{l}}}

\begin{document}

\begin{titlepage}
\vspace*{-1cm}


\vspace{.2in}
\begin{center}
{\large\bf Some Comments on BPS systems}
\end{center}

\vspace{.5cm}

\begin{center}
L.A. Ferreira~$^{\star}$,
P. Klimas~$^{\star\star}$
 A. Wereszczy\'nski~$^{\nabla}$ and W.J. Zakrzewski~$^{\dagger}$

\small
\par \vskip .2in \noindent
$^{(\star)}$Instituto de F\'\i sica de S\~ao Carlos; IFSC/USP;\\
Universidade de S\~ao Paulo  \\ 
Caixa Postal 369, CEP 13560-970, S\~ao Carlos-SP, Brazil\\
email: laf@ifsc.usp.br
\small

 \par \vskip .2in \noindent
$^{(\star\star)}$Departamento de F\'isica,\\Universidade Federal de Santa Catarina,\\
 Trindade, CEP 88040-900, Florian\'opolis-SC, Brazil\\
email: pawel.klimas@ufsc.br

\small

\par \vskip .2in \noindent
$^{(\nabla)}$Institute of Physics\\
Jagiellonian University \\ 
\L{}ojasiewicza 11, Krak\'ow, Poland\\
email: andwereszczynski@gmail.com

\small 
\par \vskip .2in \noindent
$^{(\dagger)}$~Department of Mathematical Sciences,\\
 University of Durham, Durham DH1 3LE, U.K.\\
email: W.J.Zakrzewski@durham.ac.uk

\normalsize
\end{center}


\begin{abstract}

We look at simple BPS systems involving more than one field. We discuss the conditions that have to be imposed
on various terms in Lagrangians involving many fields to produce BPS systems and then look in more detail
at the simplest of such cases.
We analyse in detail BPS systems involving 2 interacting Sine-Gordon like fields, both when one of them has a kink solution and the second one either a kink or an antikink solution. We take their solitonic static solutions and use them as initial conditions for their evolution in Lorentz covariant versions of such models.

We send these structures towards themselves
and find that when they interact weakly they can pass through each other with a phase shift which is related
to the strength of their interaction. When they interact strongly they repel and reflect on each other. We use the method  of a modified gradient flow in order to visualize the  solutions in the space of fields.

\end{abstract} 
\end{titlepage}


\section{Introduction}
\label{sec:intro}
\setcounter{equation}{0}

The BPS property is a very powerful tool in solitonic models, in especially higher than one spatial dimensions, often allowing for an analytical insight into nonlinear mathematical features of topological solitons. Here, being a BPS theory means that {\it (i)} there exists a lower energy bound which is {\it (ii)} {\it saturated} by solutions of pertinent differential equations (called Bogomolny or BPS equations). The order of the Bogomolny equation is lower than the order of Euler-Lagrange equations which may lead to solvability of the BPS sector. Furthermore, the energy of a BPS solution depends entirely in its asymptotical behaviour related to the carried topological charge and not on particularities of its spatial dependence. In addition, BPS solitons can form multi particle states where constituents do not interact forming a large moduli space. 

In contra-distinction to higher dimensions, the BPS property seems to be a generic feature of models in $(1+1)$ dimensions. This property is not only shared by the more conventional scalar field models with arbitrary potentials (which possess at least two vacua) \cite{MS} but also by models involving a set of scalar fields \cite{Adam}-\cite{AFHWZ} (spanning an arbitrary target space) with non-canonical derivative parts (any function of first derivatives \cite{Adam}, \cite{k-models}  as well as of their higher derivative generalisations \cite{higher-der1}, \cite{higher-der2}). This differs significantly from the integrability which restricts the form of the models also in $(1+1)$ dimensions. 

One important quantity related to the BPS property is the existence of the so-called pre-potential (see below) which is related to the potential of a particular model by a differential equation on the target space. In particular, BPS solutions follow a sort of a gradient flow equation of the pre-potential \cite{Adam}, \cite{e}. However, the relation between the pre-potential and the BPS solutions as well as the role of the pre-potential in the time dependent processes (relaxation and scattering) has not been systematically investigated. One of the aims of the present paper to analyse this issue in some detail. 

To do this we have chosen a set of two coupled sine-Gordon models. This is the simplest model which possesses a non-trivial pre-potential.  This model, being a generalisation of the simple Sine-Gordon involves two real fields, has also BPS property and its BPS conditions give us lower bounds on the energy and so guarantee the stability of its BPS solutions. 
 It is very well known that the one field Sine-Gordon model in (1+1) dimensions possesses many interesting and useful properties. The model is integrable; hence all its solutions can be constructed explicitly in an analytical form. 

This does not seem to be the case for its generalisation to two (or more) fields. However, the model we are studying here
has, like the one field Sine-Gordon model, the BPS properties
and so its one soliton (per field) solutions are stable and can be constructed by solving two first order equations.

Various (1+1) dimensional models involving two or more scalar fields have been already comprehensively studied - see for example \cite{bazeia}, \cite{shnir}-\cite{gani}, however, not from the pre-potential point of view. Recently, some of us have discussed this problem, in a more general setting, in \cite{e}. That paper, presented a discussion of the construction of new BPS models with various symmetries of the fields. It also introduced a modified gradient flow and showed 
how this flow can be used to understand  better various properties of the derived solutions. The approach used in \cite{e} exploited the ideas of another relatively recent paper \cite{AFHWZ}, which considered BPS conditions in several spatial dimensions. Both papers were quite
general and did not look in much detail at the properties of the solutions of such equations. However, such properties
are very interesting and in this paper we look in detail at them for the simplest important cases, namely, the system
of two Sine-Gordon fields in (1+1) dimensional theory. To make the paper more self-contained we recall here the relevant parts of \cite{e} and \cite{AFHWZ}.

Hence are considering here various properties of solutions of a scalar (1+1)
 dimensional field theory defined by the Lagrangian ($\eta_{ab}$ is the target space metric)
\be
{\cal L}= \frac{1}{2}\,\eta_{ab}\,\partial_{\mu}\vp_a\,\partial^{\mu}\vp_b - V\(\vec\vp\),
\lab{action}
\ee
where $\vec\vp$ stands for a set of scalar fields $\vp_1,\vp_2,\ldots$ Its static energy is given by
\be
E=\int_{-\infty}^{\infty}dx\,\left[\frac{1}{2}\,\eta_{ab}\,\partial_x\vp_a\,\partial_x\vp_b + V\(\vec\vp\) \right]
\lab{energy}
\ee
and its Euler-Lagrange equations are given by
\be
\eta_{ab}\,\partial^2\vp_b+\frac{\partial \, V}{\partial\,\vp_a}=0,
\lab{eleq}
\ee
where $\partial^2=\partial_t^2-\partial_x^2$.

Following \cite{AFHWZ}, in order to get BPS equations for such a theory, we construct a topological charge $Q$ from a pre-potential $U$, that is a functional of the fields but not of their derivatives, and then we split its density into the sum of products of two terms, $ A_a$ and ${\tilde A}_a$, as follows: 
\be
Q=\int_{-\infty}^{\infty}dx\, \frac{\partial\,U}{\partial\, x}=\int_{-\infty}^{\infty}dx\, \frac{\partial\,U}{\partial\, \vp_a}\,\partial_x \vp_a=\int_{-\infty}^{\infty}dx\, A_a\,{\tilde A}_a=U\(\infty\)-U\(-\infty\).
\lab{topcharge}
\ee

 Clearly, the topological charge depends only on asymptotic values that the pre-potential tales at spatial infinity.
Of course, there is some freedom in the choice of $ A_a$ and ${\tilde A}_a$, and we shall take them as
\be
A_a=\Lambda_{ab}\,\partial_x \vp_b, \qquad\qquad \qquad 
{\tilde A}_a=\frac{\partial\,U}{\partial\,\vp_b}\,\Lambda^{-1}_{ba},
\lab{aatildadef}
\ee
where we have introduced an invertible matrix $\Lambda_{ab}$. As $Q$ is a topological quantity, this implies that it is invariant under any smooth variations of the fields. The equation $\delta Q=0$, for any smooth  variation $\delta \vp_a$, leads to identities for its density, which are bilinear in $ A_a$ and ${\tilde A}_a$, and have a structure similar to the Euler-Lagrange equations, {\it i.e.} they are linear in functional derivatives of $ A_a$ and ${\tilde A}_a$ {\it w.r.t.} to the fields and their first derivatives. By imposing the first order differential BPS equations
\be
A_a=\pm {\tilde A}_a
\lab{bpsequations}
\ee
it was shown in \cite{AFHWZ} that the identity $\delta Q=0$, together with \rf{bpsequations}, imply the Euler-Lagrange equations for the functional 
\be
E=\int_{-\infty}^{\infty}dx\,\frac{1}{2}\,\left[ A_a^2+{\tilde A}_a^2\right].
 \lab{charges}
\ee
which becomes \rf{energy} if we take the matrix $\Lambda_{ab}$ such that
\be
\Lambda^T\,\Lambda =  \eta
\lab{cond1}
\ee
and if the potential $V$ is related to the pre-potentail $U$ by  
\be
V= \frac{1}{2}\,\frac{\partial\,U}{\partial\,\vp_a}\,\eta^{-1}_{ab}\,\frac{\partial\,U}{\partial\,\vp_b}.
\lab{cond2}
\ee

In consequence, the solutions of the BPS equations \rf{bpsequations}, which can now be written as
\be
\partial_x\vp_a= 
\pm  \,\eta^{-1}_{ab}\, \frac{\partial\,U}{\partial\,\vp_b},
\label{bpseqs}
\ee
are static solutions of the theory \rf{action}.  If the energy functional is indeed positive definite, one gets, as a byproduct of this construction,  a bound on the energy functional. Indeed, one can write 
\be
E=\int_{-\infty}^{\infty}dx\,\frac{1}{2}\,\left[ A_a^2+{\tilde A}_a^2\right]=\int_{-\infty}^{\infty}dx\,\frac{1}{2}\,\left[ A_a \pm {\tilde A}_a\right]^2 \mp  \int_{-\infty}^{\infty}dx\,A_a\, {\tilde A}_a\geq \mid Q\mid
\ee
and so we see that when the BPS equations are satisfied we have
 \be
 E=\mid Q\mid.
 \ee

Note that, since the potential $V$ in \rf{cond2} is quadratic in derivatives of the pre-potential $U$, one can think of the energy functional as a ``gauged'' version of the free theory. Indeed, we can write $E$ as
\be
E=\int_{-\infty}^{\infty}dx\,\left({\textstyle \frac{1}{2}}\left[\eta_{ab}\,\(\partial_x\vp_a-F_a\)\,\(\partial_x\vp_b-F_b\)\right] +\partial_x \Omega\right),
\lab{energygauged}
\ee
where we have added a total derivative that does not interfere with the Euler-Lagrange equations but that may change the actual value of $E$ if $\Omega$ is topologically non-trivial. In such a case the potential has to be identified with
\be
V= \frac{1}{2}\,\eta_{ab}\,F_a\,F_b
\lab{vintermsf}
\ee
and so, from \rf{cond2} the $F_a$'s should be related to the derivatives of the pre-potential as
\be
F_a=\pm\,\eta^{-1}_{ab}\,\frac{\partial\,U}{\partial\,\vp_b}.
\lab{furelation}
\ee

Thus, in terms of $F_a$'s, the self-duality equations \eqref{bpseqs} become 
\be
\partial_x\vp_a=F_a
\label{BPSeq}
\ee
The crossed term in \rf{energygauged}  does not interfere with the Euler-Lagrange equations either, since it is the topological charge, {\it i.e.}
\be
\int_{-\infty}^{\infty}dx\,\eta_{ab}\,\partial_x\vp_a\,F_b=\pm\, \int_{-\infty}^{\infty}dx\,\partial_x\vp_a\,\frac{\partial\,U}{\partial\,\vp_a}= \pm \, Q,
\ee
where we have used \rf{furelation} and \rf{topcharge}. Thus, when the BPS equations \eqref{bpseqs}, or equivalently \eqref{BPSeq}, are
satisfied, the energy $E$ given in  \rf{energygauged}, becomes $E=\Omega\(\infty\)-\Omega\(-\infty\)$, and so from \rf{topcharge}, one observes that $\Omega$ should be identified with the pre-potential $U$, {\it i.e.} $\Omega=\pm U$.

In this paper we  are interested only in theories with two scalar fields, $\vp_1$ and $\vp_2$, and we so we take the matrix $\eta$, introduced in \rf{action} (see \rf{cond1}), to be of the form
\be
\eta_{ab}\,=\, \left(\begin{array}{cc}
1& -\frac{\lambda}{2}\\
-\frac{\lambda}{2}& 1 \end{array}\right).
\lab{eta}
\ee
Then, choosing the $(+)$ sign in \rf{furelation} we find that 
\be
F_1=\frac{4}{4-\lambda^2} \left( \frac{\partial U}{\partial \varphi_1}+\frac{\lambda}{2} \frac{\partial U}{\partial \varphi_2} \right),\label{F1}
\ee
\be
F_2=\frac{4}{4-\lambda^2} \left( \frac{\lambda}{2} \frac{\partial U}{\partial \varphi_1} + \frac{\partial U}{\partial \varphi_2}\right).\label{F2}
\ee
The simplest case corresponds to the choice 
\be
\frac{\partial U}{\partial \varphi_1}\,=\, M(\varphi_1)\qquad \hbox{and}\qquad \frac{\partial U}{\partial \varphi_2}\,=\,N(\varphi_2) 
\ee
and this is the case we have studied in detail and we will discuss in this paper.

As we mentioned this in \cite{e} the equation \eqref{BPSeq} suggests a nice geometric interpretation of the BPS solutions; namely, in the space of fields $\vp_1,\vp_2,\ldots$ they are represented by curves that must follow the modified gradient flow of the pre-potential
$
\vec F\equiv\vec\nabla_\eta U:=\eta^{-1}\cdot\vec\nabla U.
$

This observation allows us to view the BPS solutions in a new way. Even more so, it allows us to visualise the
time dependence of  non-BPS solutions  ( {\it i.e.} time dependent fields) by comparing them with the static BPS-curves and so think of them as a motion of curves.  We will exploit this technique in many examples presented in this paper. 

The paper is organized as follows. In Section 2 we introduce the model with two interacting Sine-Gordon fields, then in Section 3 we construct numerical BPS solutions of their (static) equations and discuss some of their properties. As the two  Sine-Gordon fields are very strongly localised we also consider the evolution of slightly deformed initial field configurations which are not exactly BPS solutions. This is discussed in the next two sections of the paper. Our deformations correspond 
to either taking the static fields corresponding to one value of the parameter of the interaction and then evolving these
initial fields with a different value of this parameter or by modifying the initial
conditions of the soliton kinks (or anti-kinks) to make them move towards each other. The last section contains some of our conclusions and plans for further studies.

\section{The model}\label{s3}
\label{sec:model}
\setcounter{equation}{0}

Our model is a $(1+1)$-dimensional Minkowski space-time theory involving two coupled real scalar fields $\varphi_a$, $a=1,2$, defined by the Lagrangian \rf{action} for which we chose the topological charges to be given 
by the generalisations of the Sine-Gordon terms. We take
\be
M(\varphi_1)\,=\,4\sin(\varphi_1),\qquad  N(\varphi_2)\,=\,4\sin(\varphi_2).
\label{topo2d}
\ee
Thus our $(1+1)$-dimensional Minkowski  two field Lagrangian density is given by
\be
{\cal L}= \frac{1}{2}\left[ (\partial_{\mu}\varphi_1)^2 + 
(\partial_{\mu}\varphi_2)^2 - \lambda \partial_{\mu}\varphi_1\,\partial^{\mu}\varphi_2\right] -V(\varphi_1,  \varphi_2),
\label{lagrangiannewtwo}
\ee
where $V(\varphi_1,\varphi_2)$ is then given by (see \rf{cond2} and \rf{vintermsf}) 
\be
 V=\frac{1}{2}\(F_1^2+F_2^2-\lambda F_1F_2\)\,= \frac{2}{4-\lambda^2}\left[M^2\(\vp_1\)+N^2\(\vp_2\)+\lambda M\(\vp_1\)\,N\(\vp_2\)\right].
\lab{poten}
\ee
The self-duality (BPS) equations are given by \eqref{BPSeq} and the corresponding Euler-Lagrange equations are given by (see \rf{eleq}):
\br
\partial^2\vp_1-\frac{\lambda}{2}\,\partial^2 \vp_2=-\frac{\delta\, V}{\delta\,\varphi_1}\;,\qquad\qquad\qquad 
\partial^2\vp_2-\frac{\lambda}{2}\,\partial^2 \vp_1=-  \frac{\delta\, V}{\delta\,\varphi_2}.
\label{generaleq}
\er

Note that the energy $E$ is given by
\be
E\,=\,\int_{-\infty}^{\infty}\,dx \Big(\partial_x\varphi_1\,M(\varphi_1)\,+\,\partial_x \varphi_2\,N(\varphi_2)\Big)\,=\,
-4\Big[\cos(\varphi_1)+\cos(\varphi_2)\Big]_{-\infty}^{\infty},
\label{ener}
\ee
which, if $\lim_{x\rightarrow-\infty}\varphi_a(x)=0$ and $\lim_{x\rightarrow \infty}\varphi_a(x)=\pi$, equals 16.

\section{Solutions of static equations}
\setcounter{equation}{0}

We have solved the BPS equations  \eqref{BPSeq} for the case \rf{poten} and some others too. This can be done in two
different ways. First of all one can solve \eqref{BPSeq} directly. In this case we have 2 first order equations
for $\varphi_1$ and $\varphi_2$ and so we need the initial values, for each field (say $\varphi_1(a)$ and $\varphi_2(a)$).
To get the fields for all values of $x$ we need to perform separate simulations for $x$ larger than $a$
and $x$ smaller than  $a$ and then combine them. The other approach would involve using \eqref{BPSeq} 
and then, differentiating one of these equations, obtain from them a second order equation for $\varphi_1$.
This equation then requires two conditions (say $\varphi_1(a)$ and $\partial_x\varphi_1(a)$) and then the solution
for $\varphi_2(x)$ is uniquely determined by $\varphi_1(x)$. We have used both methods of deriving
the static solutions $\varphi_1(x)$ and $\varphi_2(x)$ and they gave us the same fields (though in the second
method one has to be very careful with taking the proper signs of $\cos(x)$ and $\sin(x)$ which arise
in the intermediate steps). So in our calculations, in most cases, we used the first method.

Note that our solutions do depend on the value of parameter $\lambda$ in (\ref{lagrangiannewtwo}).
Clearly when $\lambda=0$ the equations separate and we have two independent Sine-Gordon equations,
but for other values of $\lambda$ they are coupled. Note that the potential \rf{poten} is singular
at $\lambda=\pm2$ so in this paper we restrict our attention to $\vert\lambda\vert<2$.

\begin{figure}[h!]
  \centering
  \subfigure[]{\includegraphics[width=0.4\textwidth,height=0.25\textwidth, angle =0]{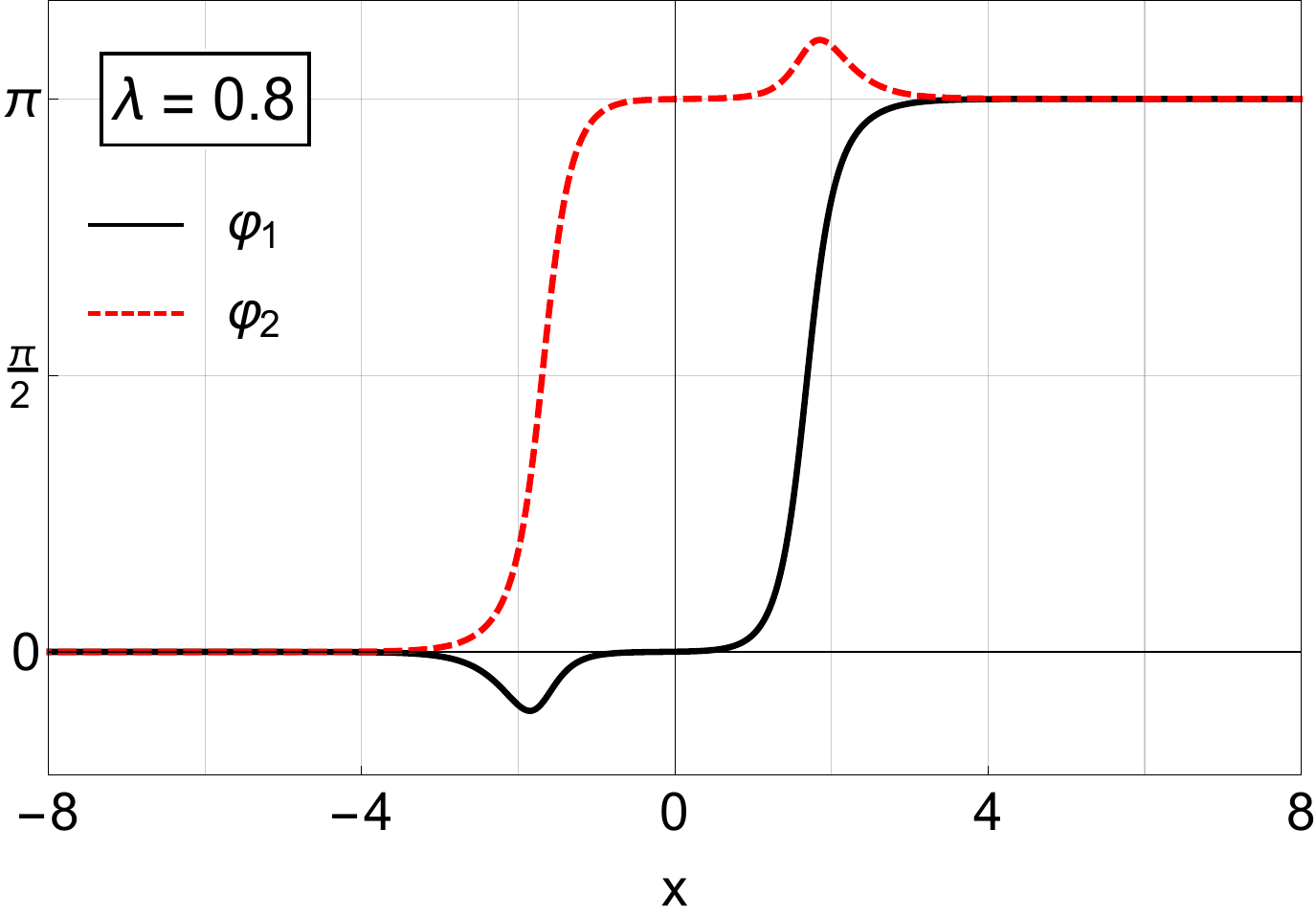}} 
  \hskip0.5cm
   \subfigure[]{\includegraphics[width=0.4\textwidth,height=0.25\textwidth, angle =0]{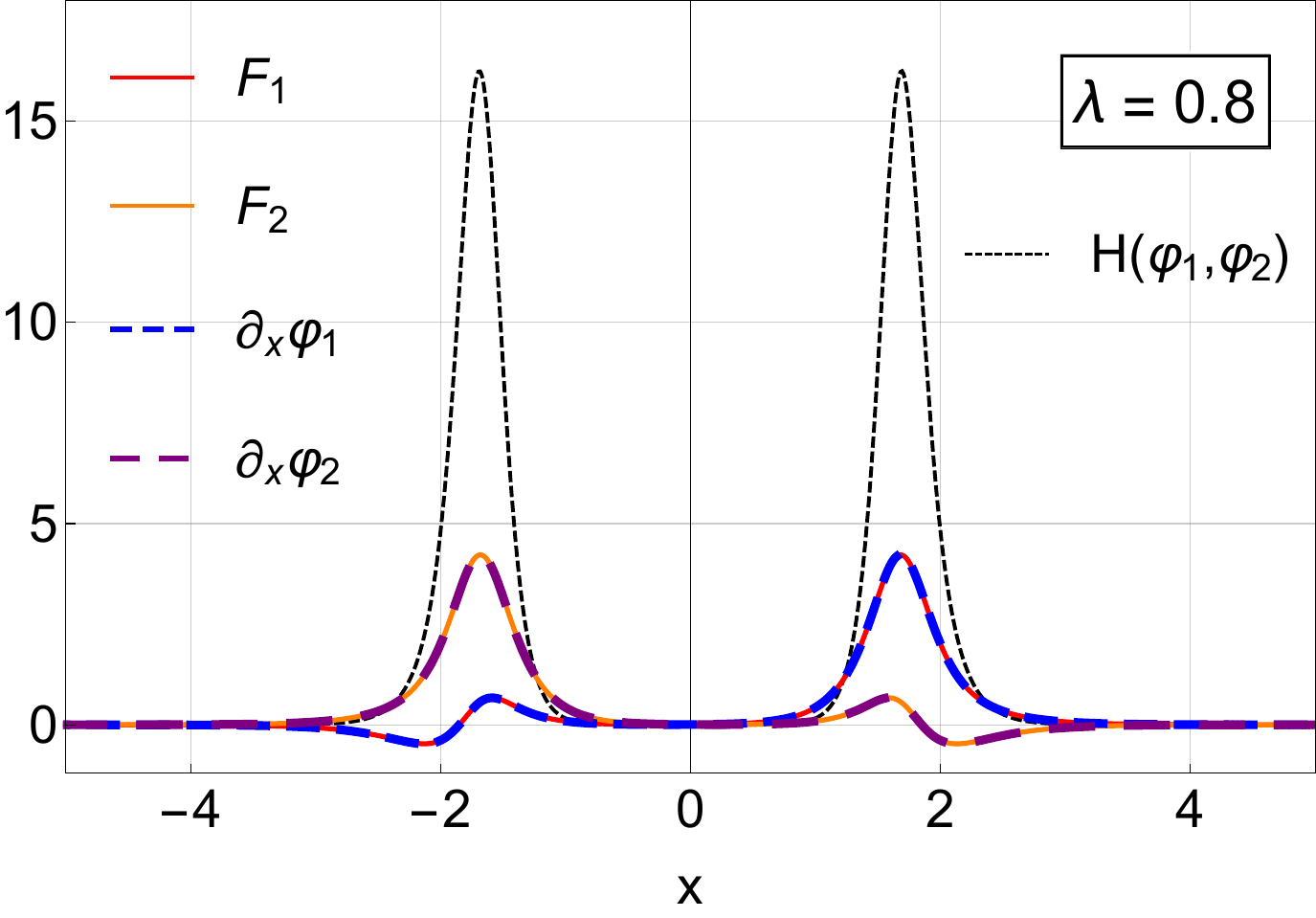}}   
   \subfigure[]{\includegraphics[width=0.4\textwidth,height=0.25\textwidth, angle =0]{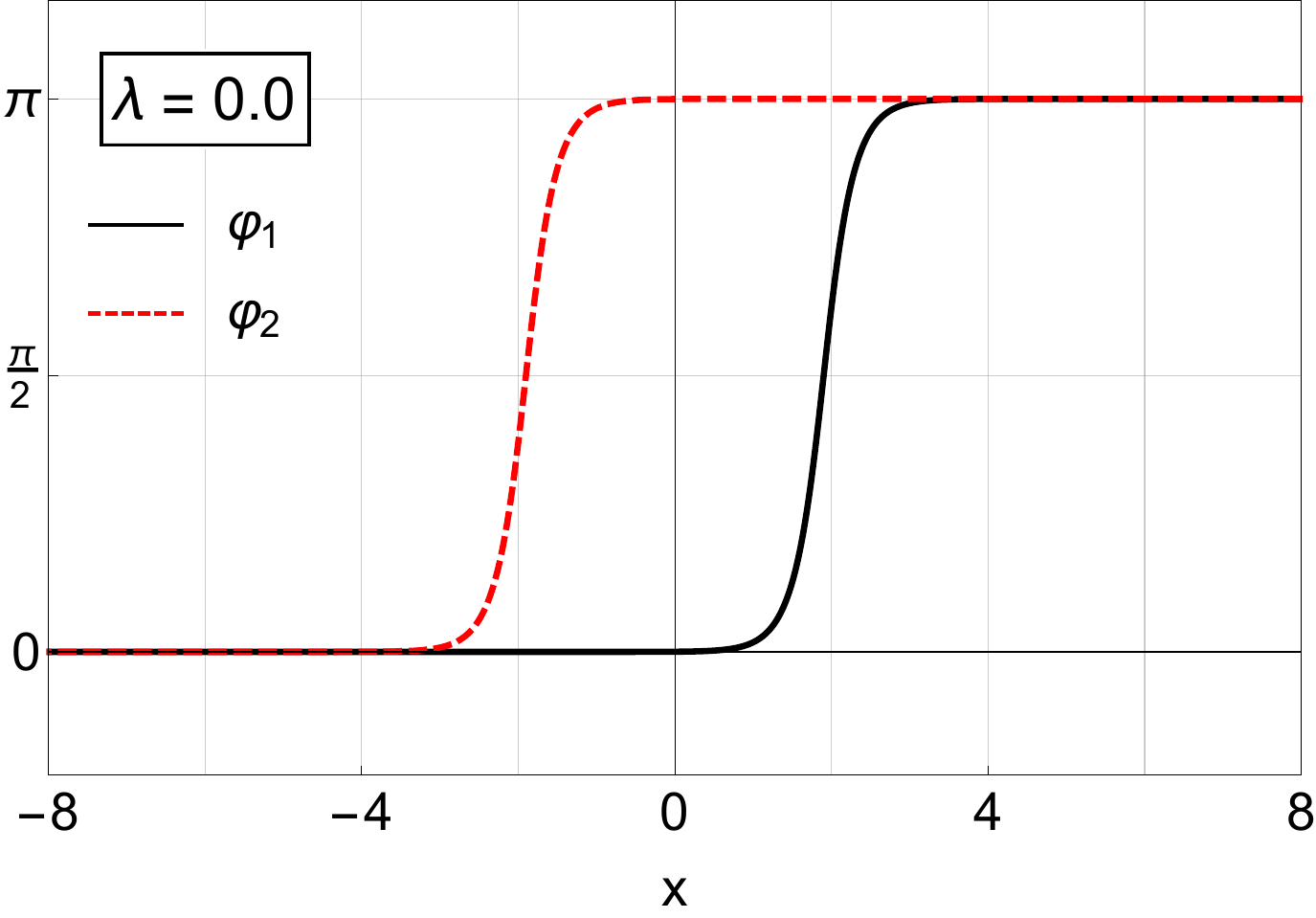}}
   \hskip0.5cm
    \subfigure[]{\includegraphics[width=0.4\textwidth,height=0.25\textwidth, angle =0]{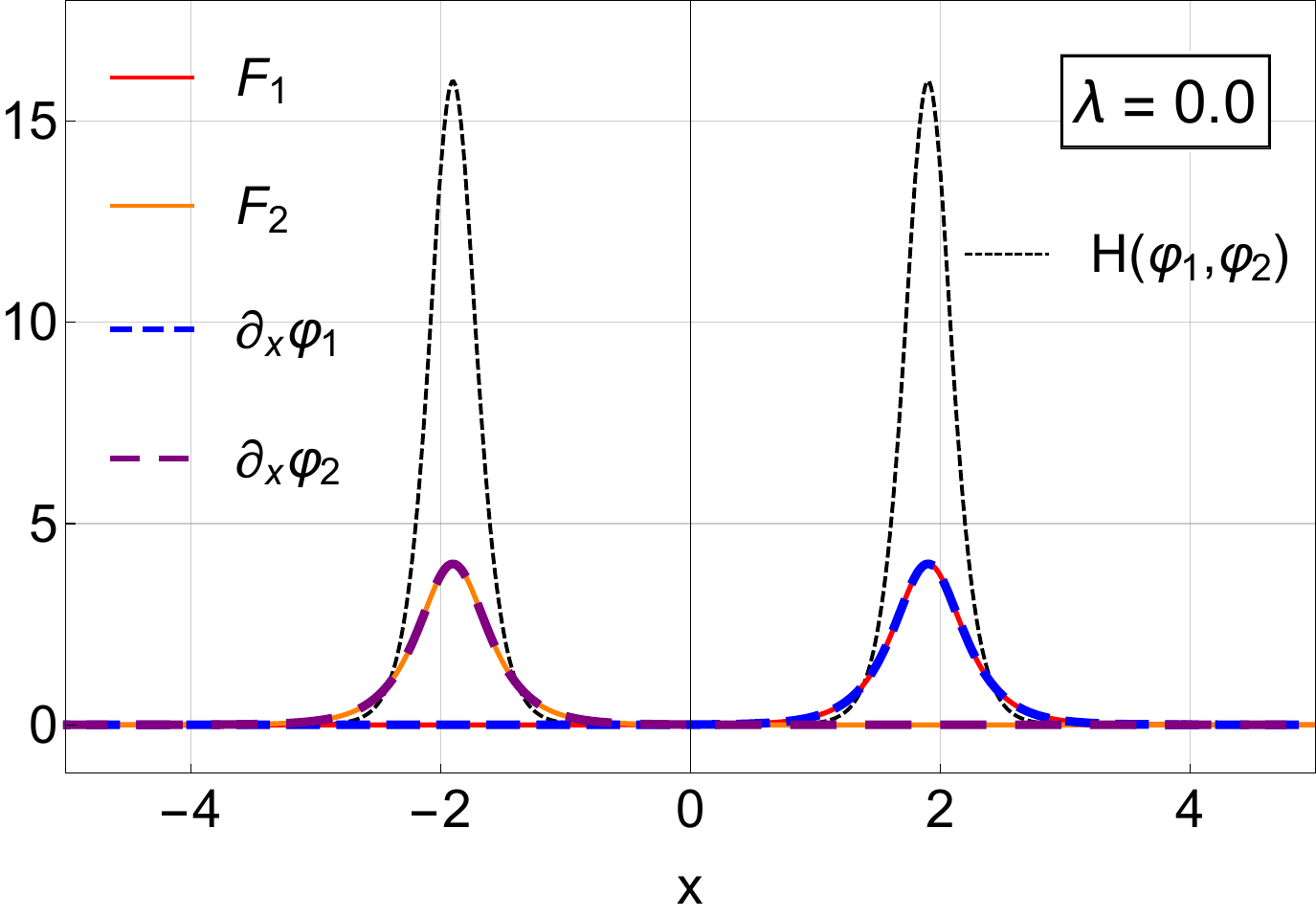}}            
   \subfigure[]{\includegraphics[width=0.4\textwidth,height=0.25\textwidth, angle =0]{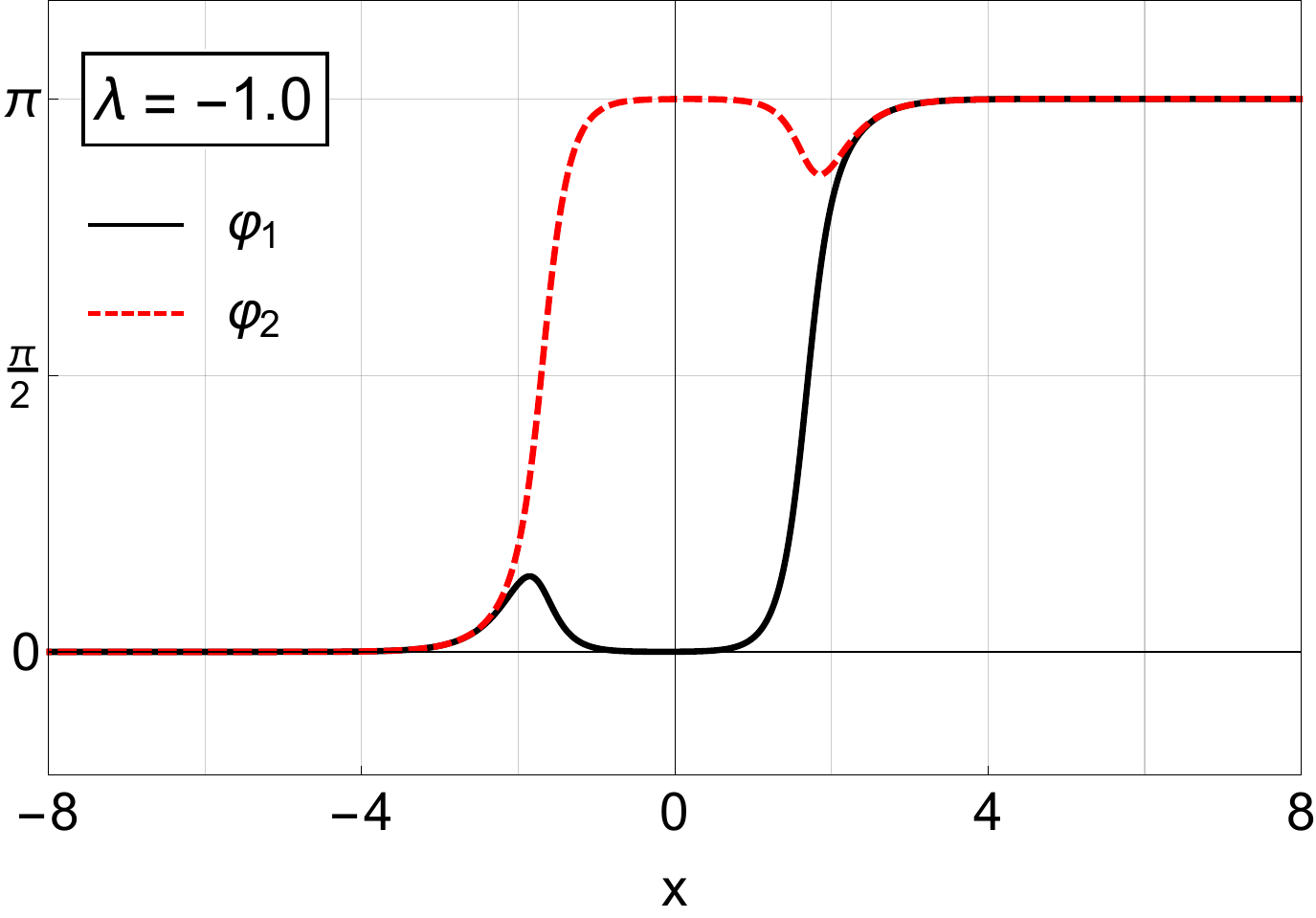}}
   \hskip0.5cm
    \subfigure[]{\includegraphics[width=0.4\textwidth,height=0.25\textwidth, angle =0]{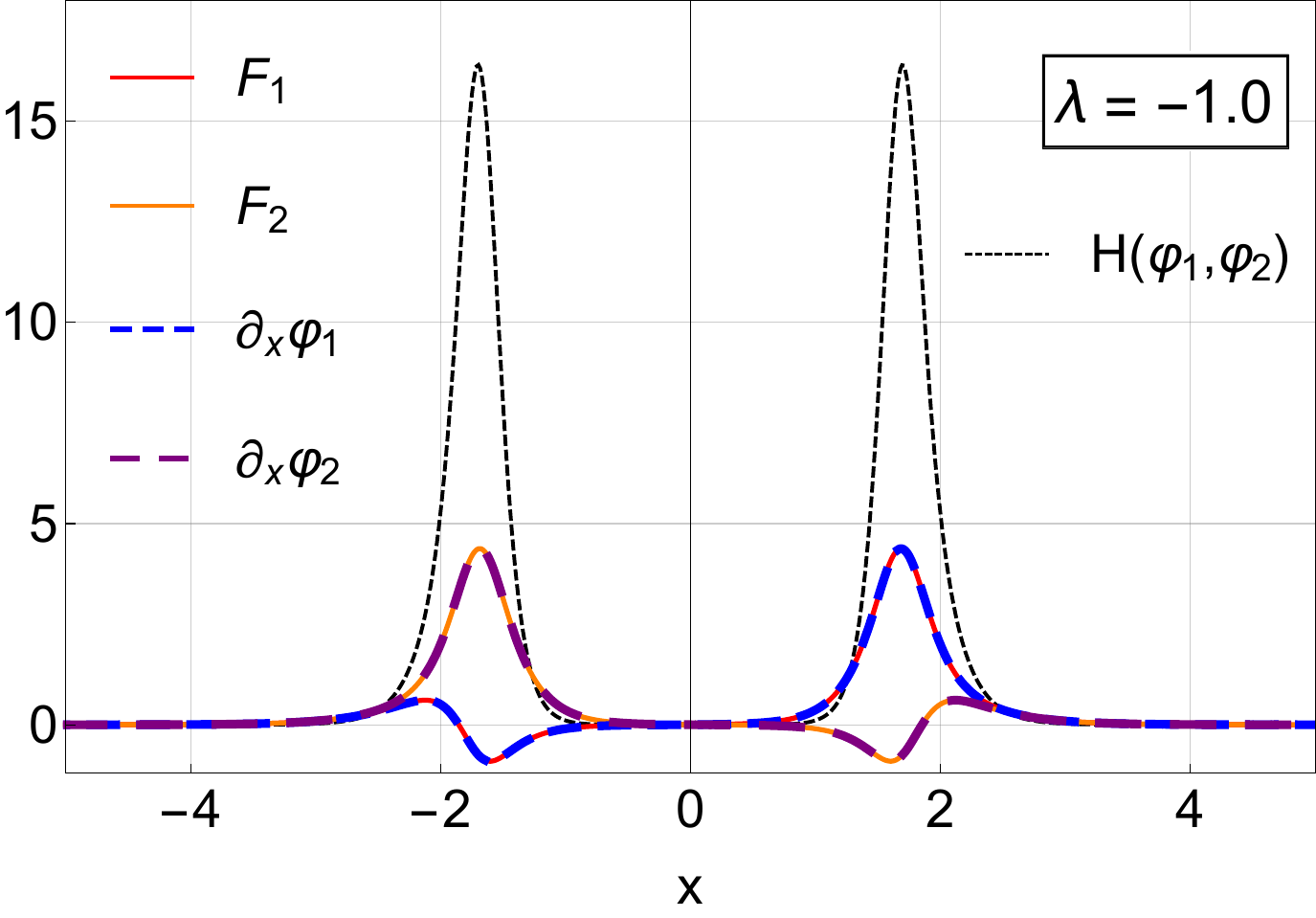}}
  \caption{Fields $\varphi_1$ and $\varphi_2$ for (a)  $\lambda=0.0$, (b) $\lambda=0.8$ and (c) $\lambda=-1.0$. Figures (b), (d) and (f) show the energy density {\color{green}}(H), spatial derivatives of fields and functions $F_1$ and $F_2$.}
  \label{fig:1sol}
\end{figure}

We have determined solutions of the BPS equations for many values of the parameter 
$\lambda$. Solutions for $\lambda>0$ are quite different in form from those for $\lambda<0$. In this section we present representative examples of BPS solutions that were obtained for three values of the coupling constant $\lambda$, namely $\lambda=0.8$, $\lambda=0$ and $\lambda=-1.0$. The case $\lambda=0$, of course, corresponds to two decoupled Sine-Gordon models. In this particular case we know exact solutions of  the model. They are given by
\be
\vp_1(x)=2\arctan\Big(e^{4(x-x_1)}\Big),\qquad\quad \vp_2(x)=2\arctan\Big(e^{4(x-x_2)}\Big),\label{sgkinks}
\ee
where $x_1$ and $x_2$ 
are two arbitrary constants that describe positions of individual kinks. In particular, one can rewrite these constants as $x_1=\frac{1}{2}(x^{(+)}+x^{(-)})$ and $x_2=\frac{1}{2}(x^{(+)}-x^{(-)})$. Then  $x^{(+)}$ describes an overall position of the system of kinks and $x^{(-)}$ the relative distance between the kinks. The solution $\vp_1(x)$ and $\vp_2(x)$ can be obtained from a particular solution $\vp^{(0)}_1(x)=\vp^{(0)}_2(x)=2\arctan(e^{4x-2x^{(+)}})$ by performing translations by $\mp \frac{x^{(-)}}{2}$ {\it i.e.}
\be
\tan\frac{\vp_1}{2}=\tan\frac{\vp^{(0)}_1}{2}e^{-2x^{(-)}},\qquad\quad
\tan\frac{\vp_2}{2}=\tan\frac{\vp^{(0)}_2}{2}e^{+2x^{(-)}}.\label{transf}
\ee
The constant $x^{(+)}$ can always be absorbed into definition of variable $x$ due to the symmetry of the system under spatial translations. Thus we can choose $x^{(+)}=0$. An example of such a solution for $\lambda=0$ is shown in Fig.\ref{fig:1sol}(c). It has  kinks at $x_1=-x_2=\frac{1}{2}x^{(-)}=1.9002$.

A typical solution for positive $\lambda$ is shown in Fig.\ref{fig:1sol}(a).
We note that the plots look very much like the plots of Sine-Gordon kinks going from 0 to $\pi$ except 
for an extra bump in the shape of one soliton function at the position of the other soliton. The size of the bumps depends on $\lambda$ and as $\lambda$ goes to 0 they vanish.
The direction (up or down) of the bumps depends on the sign of $\lambda$, as can be seen by looking at figures Fig.\ref{fig:1sol}(a) and Fig.\ref{fig:1sol}(e). In figures Fig.\ref{fig:1sol}(b), Fig.\ref{fig:1sol}(d) and Fig.\ref{fig:1sol}(f) we  plot the spatial derivatives of fields $\partial_x\vp_a$  as well as the functions $F_a(\vp_1,\vp_2)$, where $a=1,2$. A very good agreement between the curves representing spatial derivatives of fields and plots of functions $F_a$ is possible only for the BPS solutions. This confirms the BPS nature of our solutions. In the same three figures we have also plotted the energy density ($H(\vp_1,\vp_2)$) of each field configuration.
\begin{figure}[h!]
 \centering
 \subfigure[]{\includegraphics[width=0.3\textwidth,height=0.25\textwidth, angle =0]{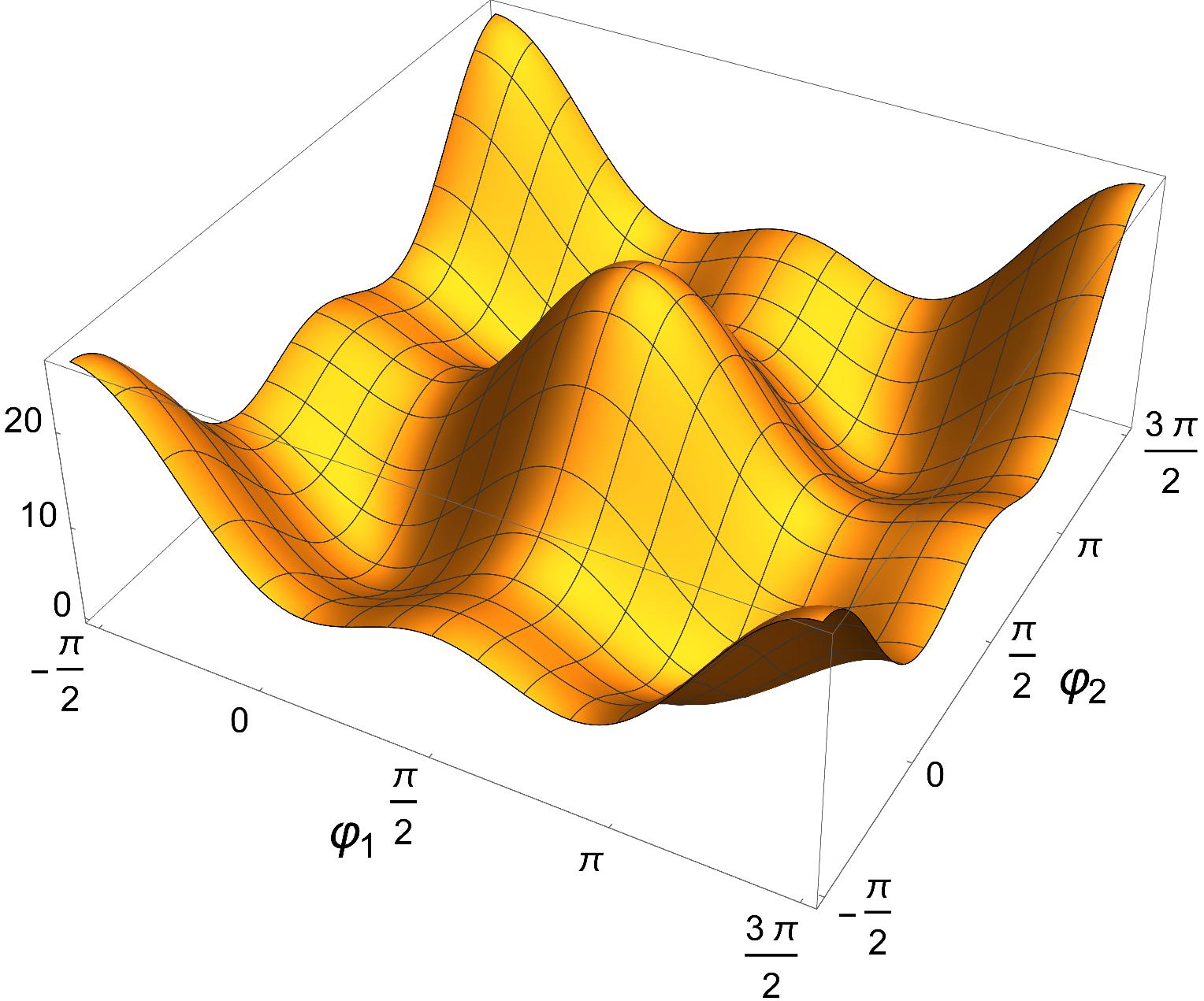}}                
 \subfigure[]{\includegraphics[width=0.3\textwidth,height=0.25\textwidth, angle =0]{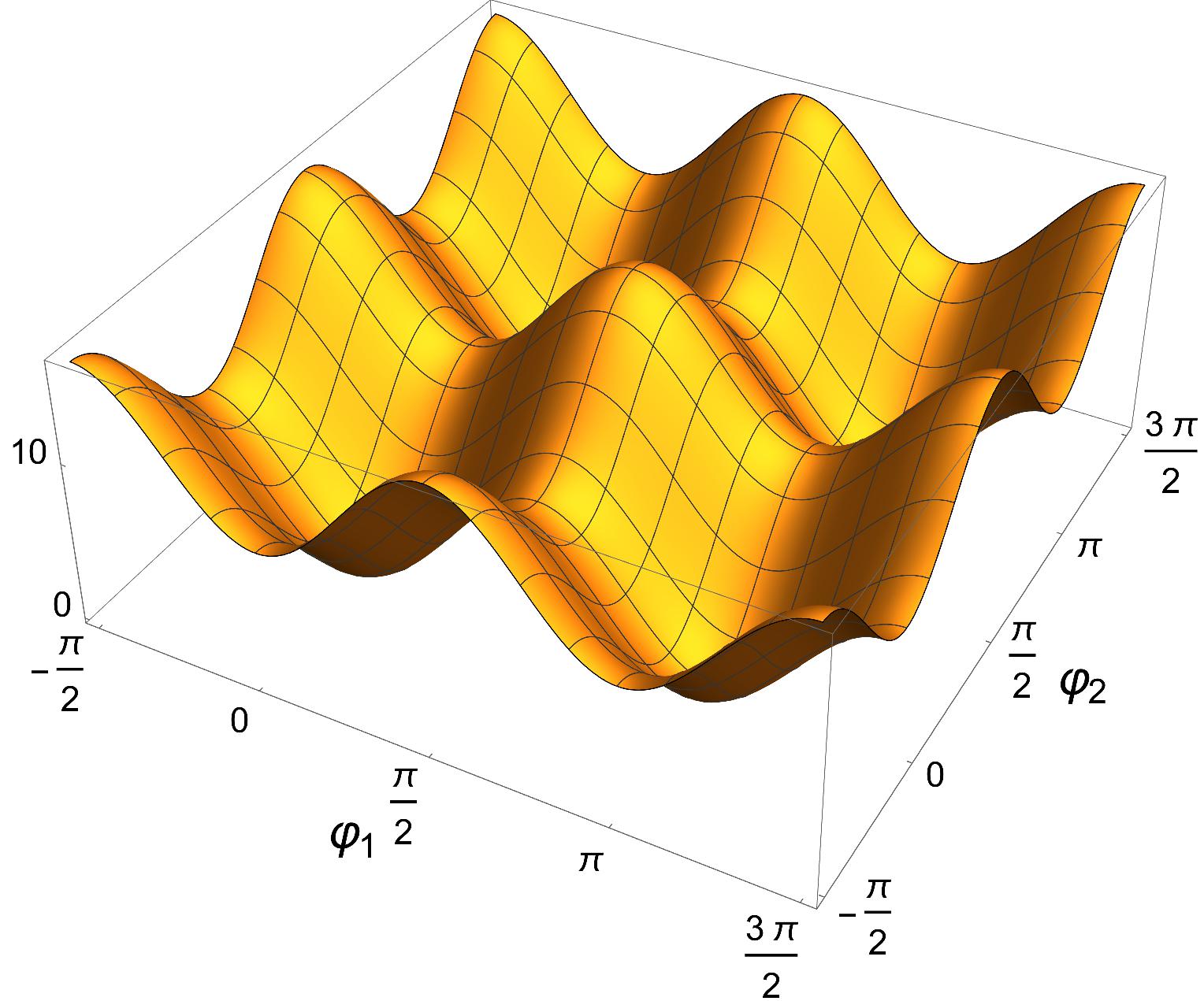}}
  \subfigure[]{\includegraphics[width=0.3\textwidth,height=0.25\textwidth, angle =0]{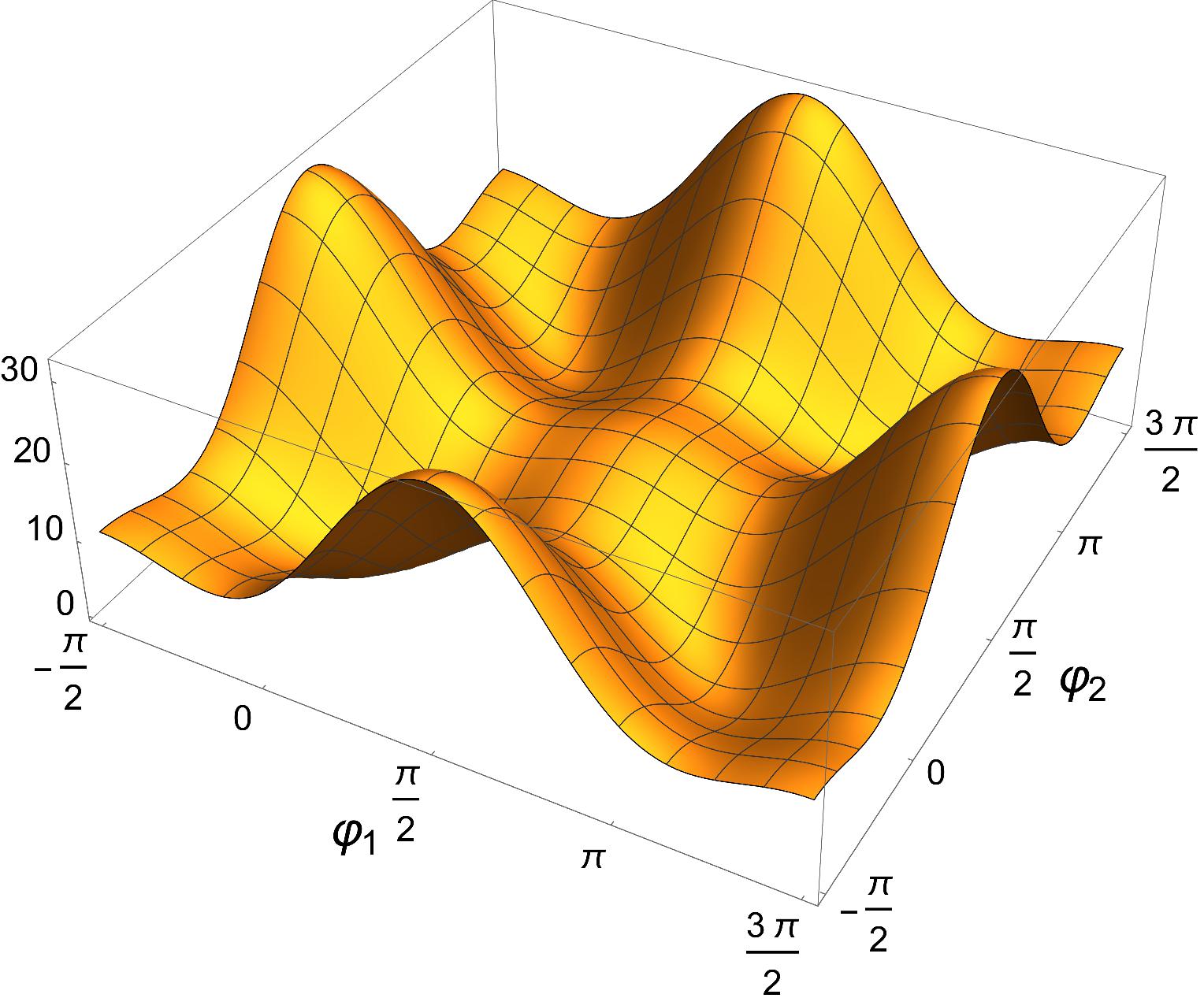}}
   \subfigure[]{\includegraphics[width=0.3\textwidth, angle =0]{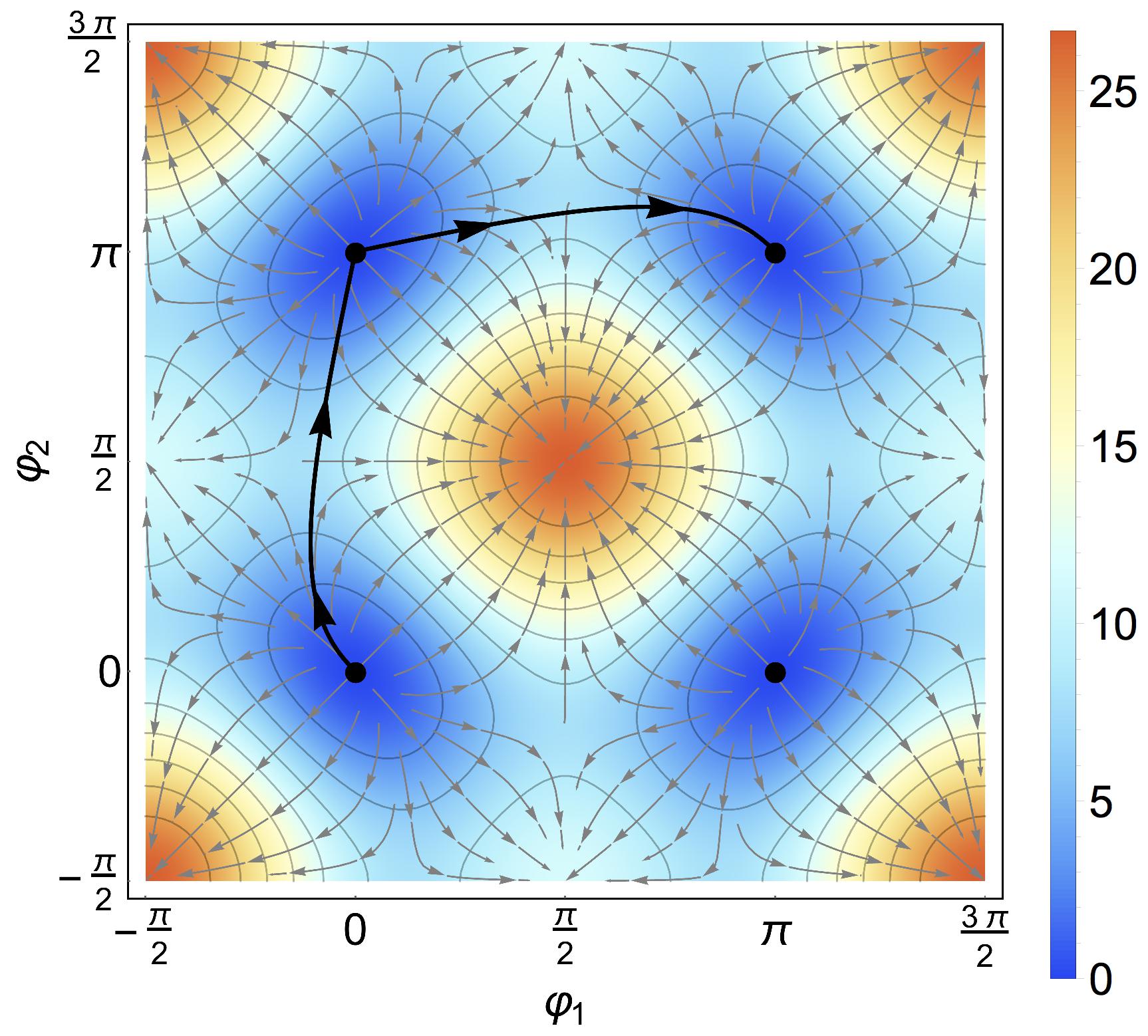}}                
  \subfigure[]{\includegraphics[width=0.3\textwidth, angle =0]{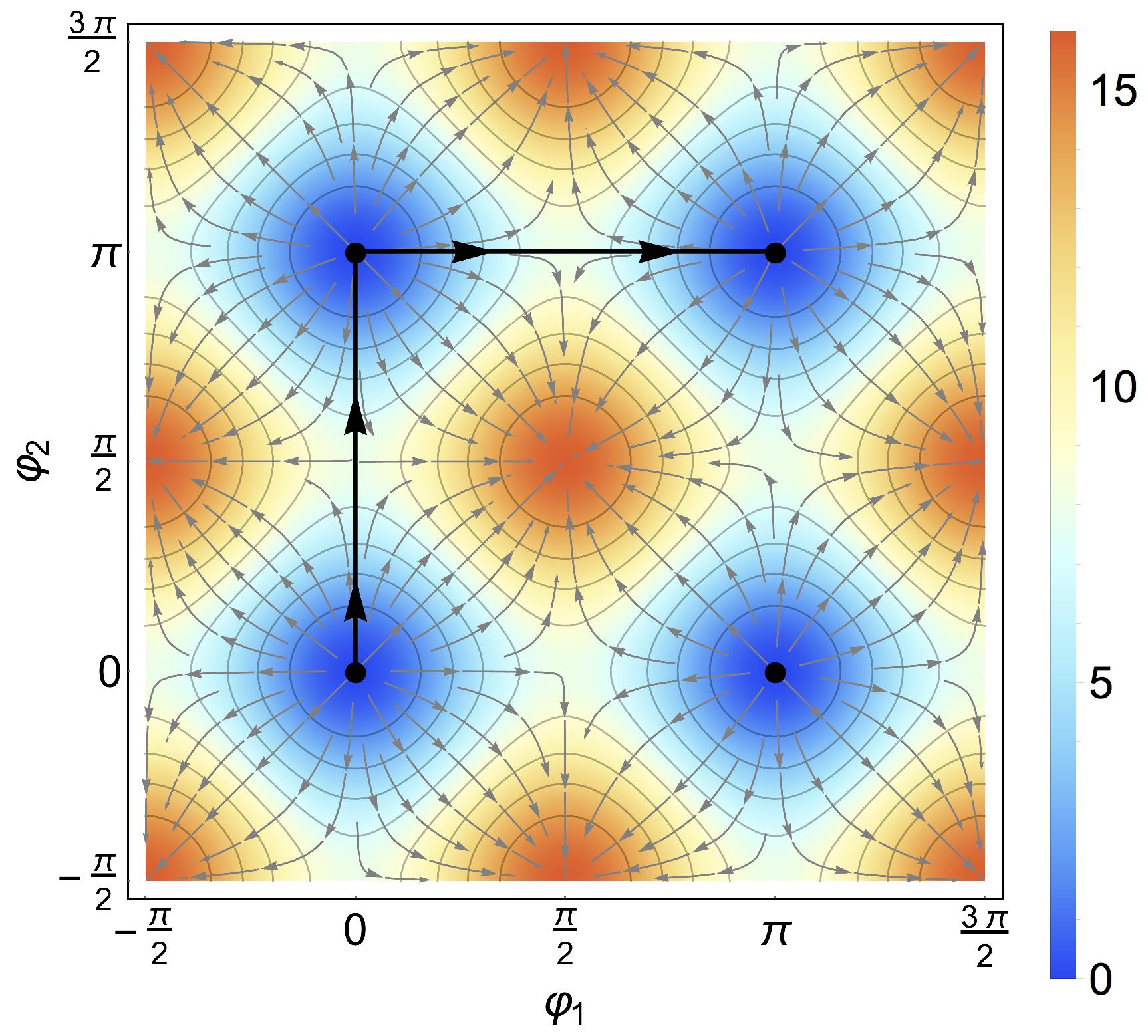}}
  \subfigure[]{\includegraphics[width=0.3\textwidth, angle =0]{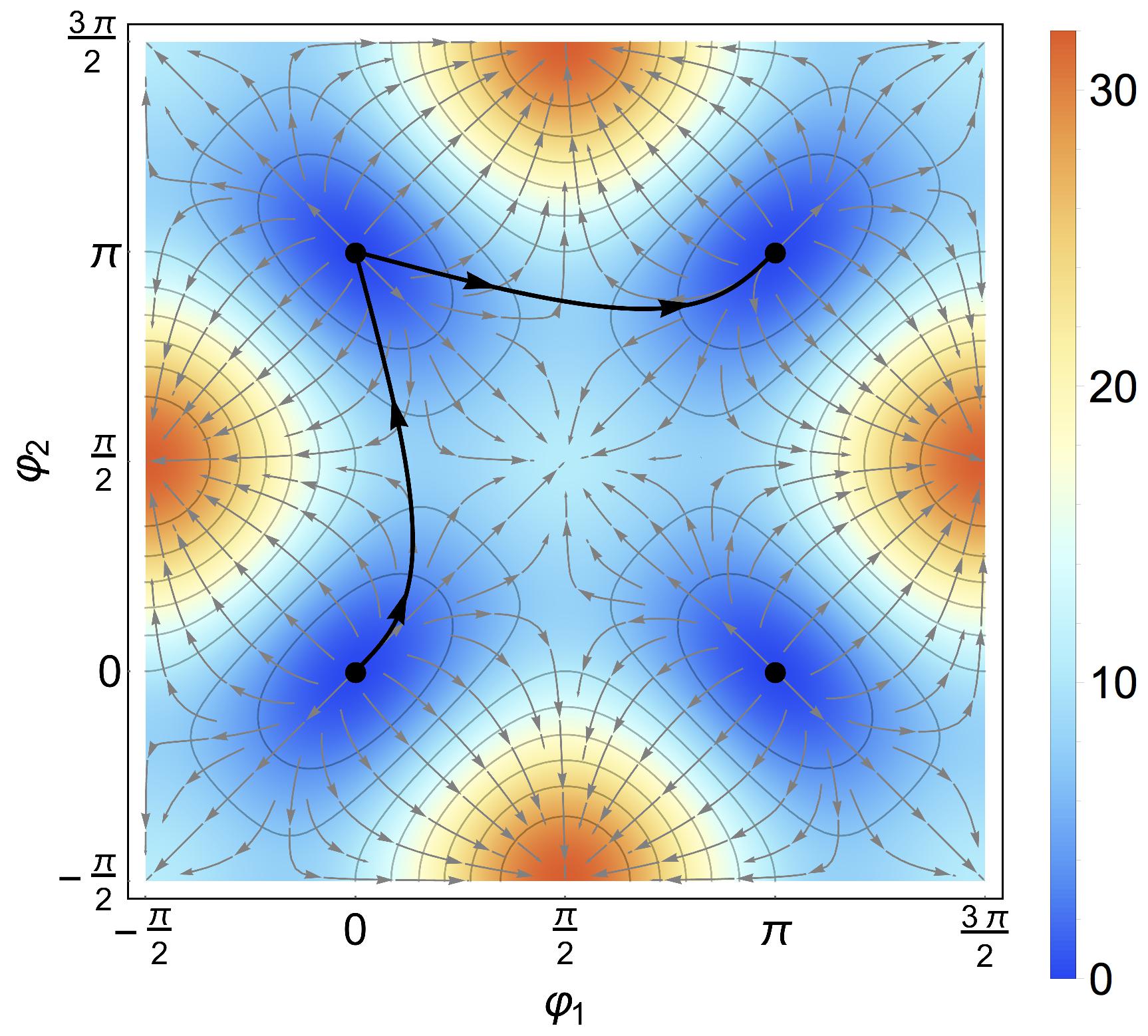}}
\caption{ Potential $V(\varphi_1,\varphi_2)$ for (a), (d) $\lambda=0.8$ and  (b), (e) $\lambda=0.0$ and (c), (f) $\lambda=-1.0$.   A solid black line represents the numerical BPS solution for each case.}
 \label{fig:2potentials}
\end{figure}

\begin{figure}[h!]
 \centering
   \subfigure[]{\includegraphics[width=0.3\textwidth, height =0.25\textwidth]{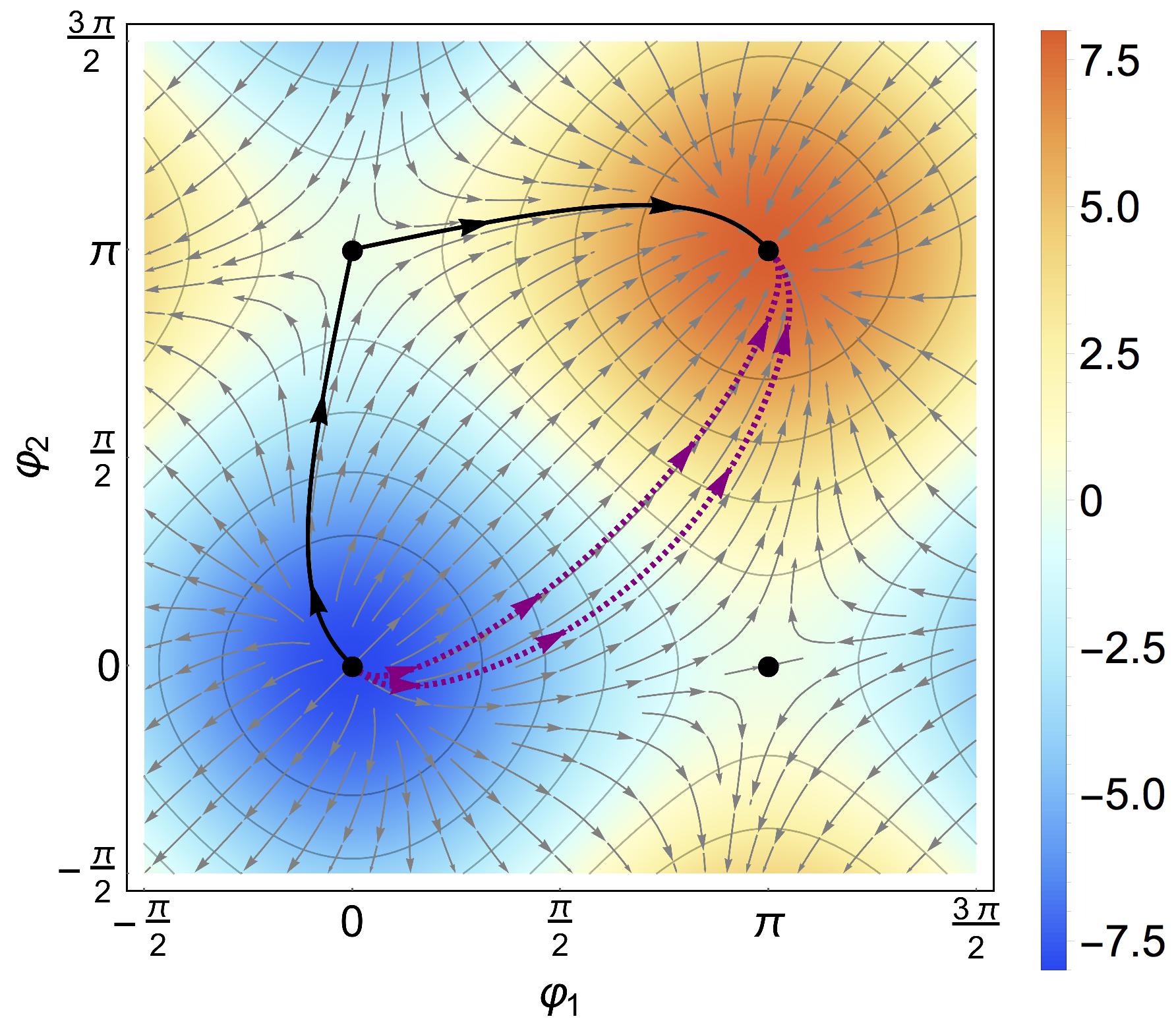}} 
   \hskip0.6cm 
   \subfigure[]{\includegraphics[width=0.3\textwidth, height =0.25\textwidth]{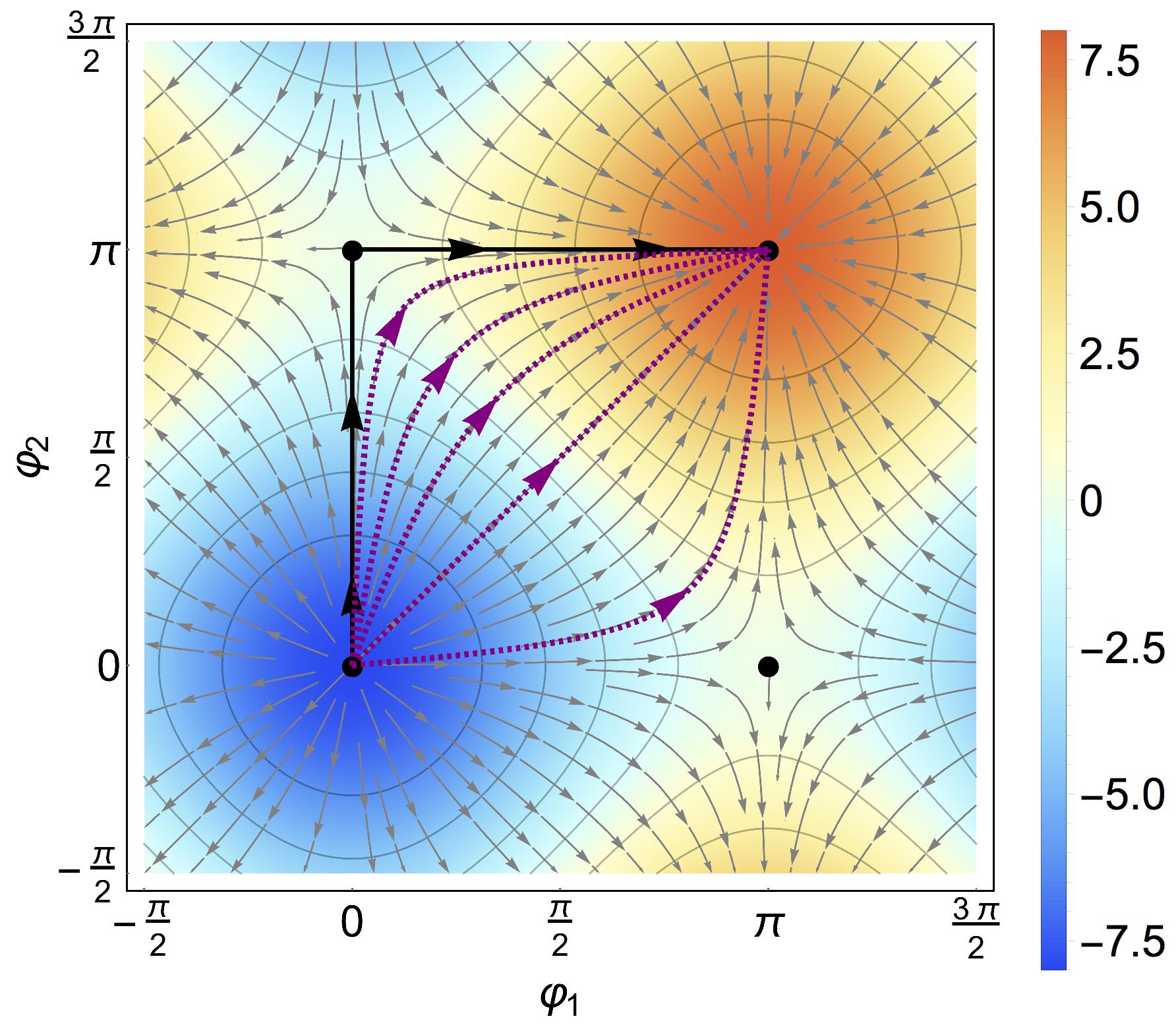}} 
   \hskip0.6cm 
   \subfigure[]{\includegraphics[width=0.3\textwidth, height =0.25\textwidth]{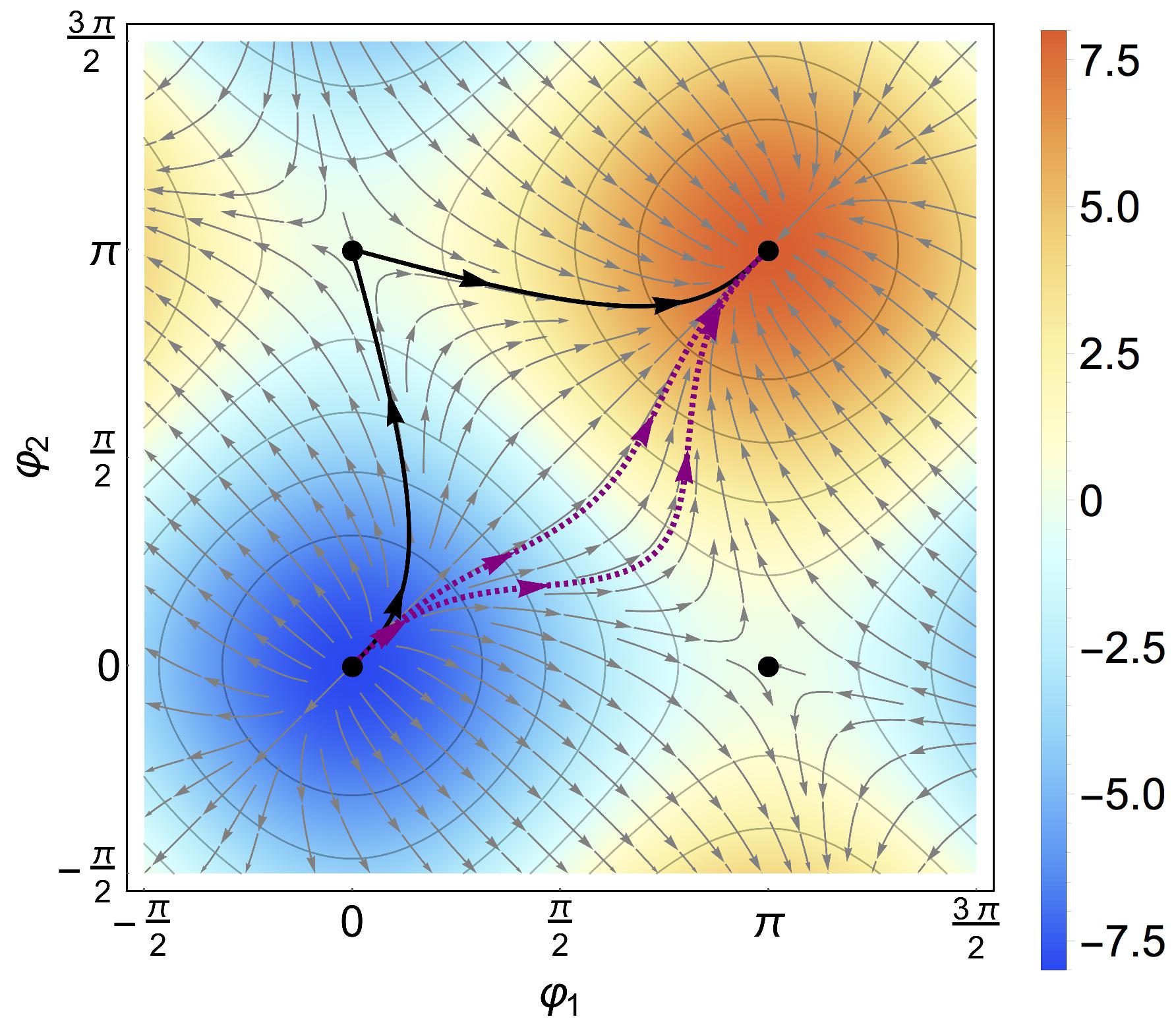}} 
    \hskip-1.5cm\subfigure[]{\includegraphics[width=0.3\textwidth, height =0.25\textwidth]{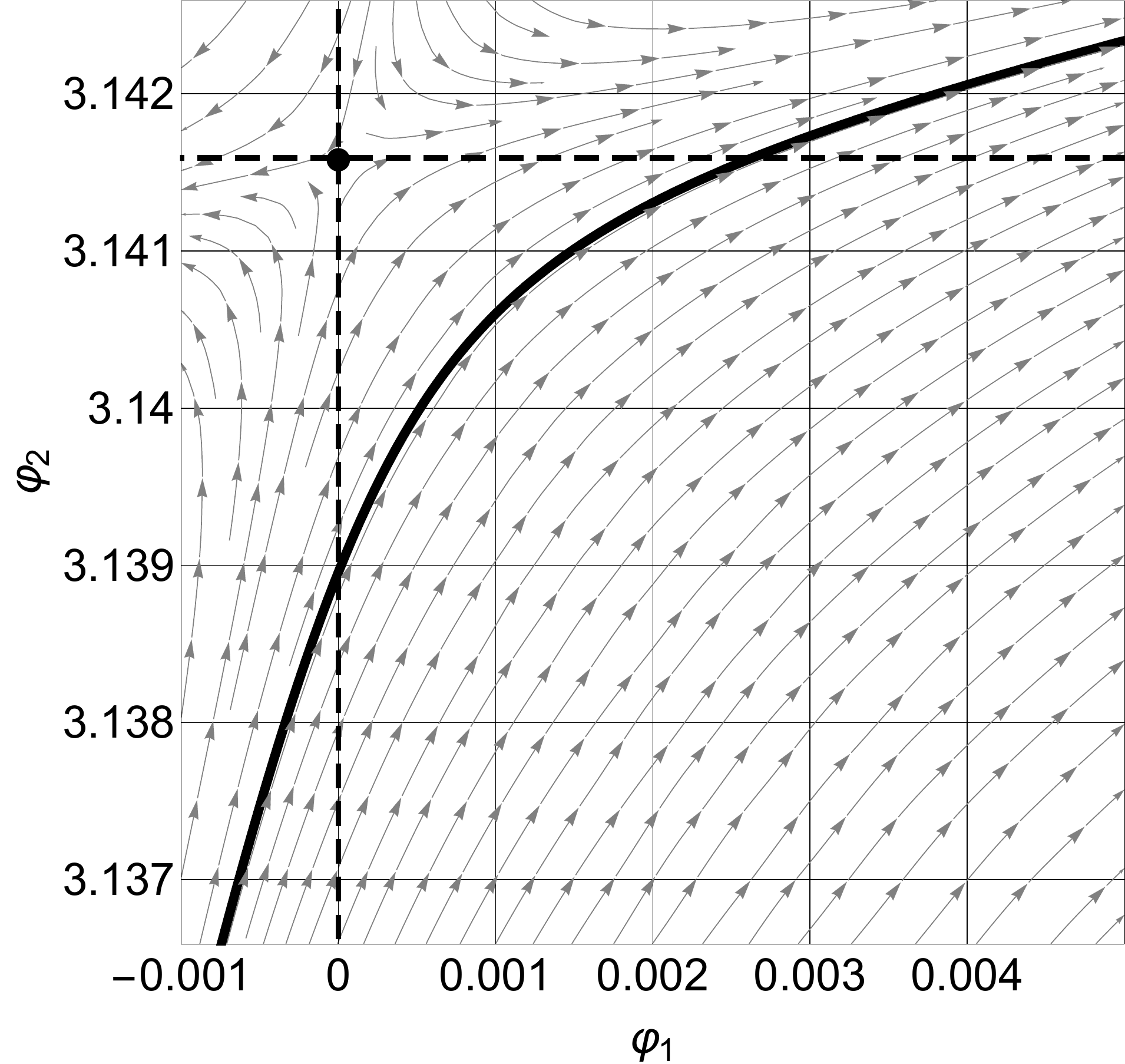}} \hskip0.5cm           
  \subfigure[]{\includegraphics[width=0.3\textwidth, height =0.25\textwidth]{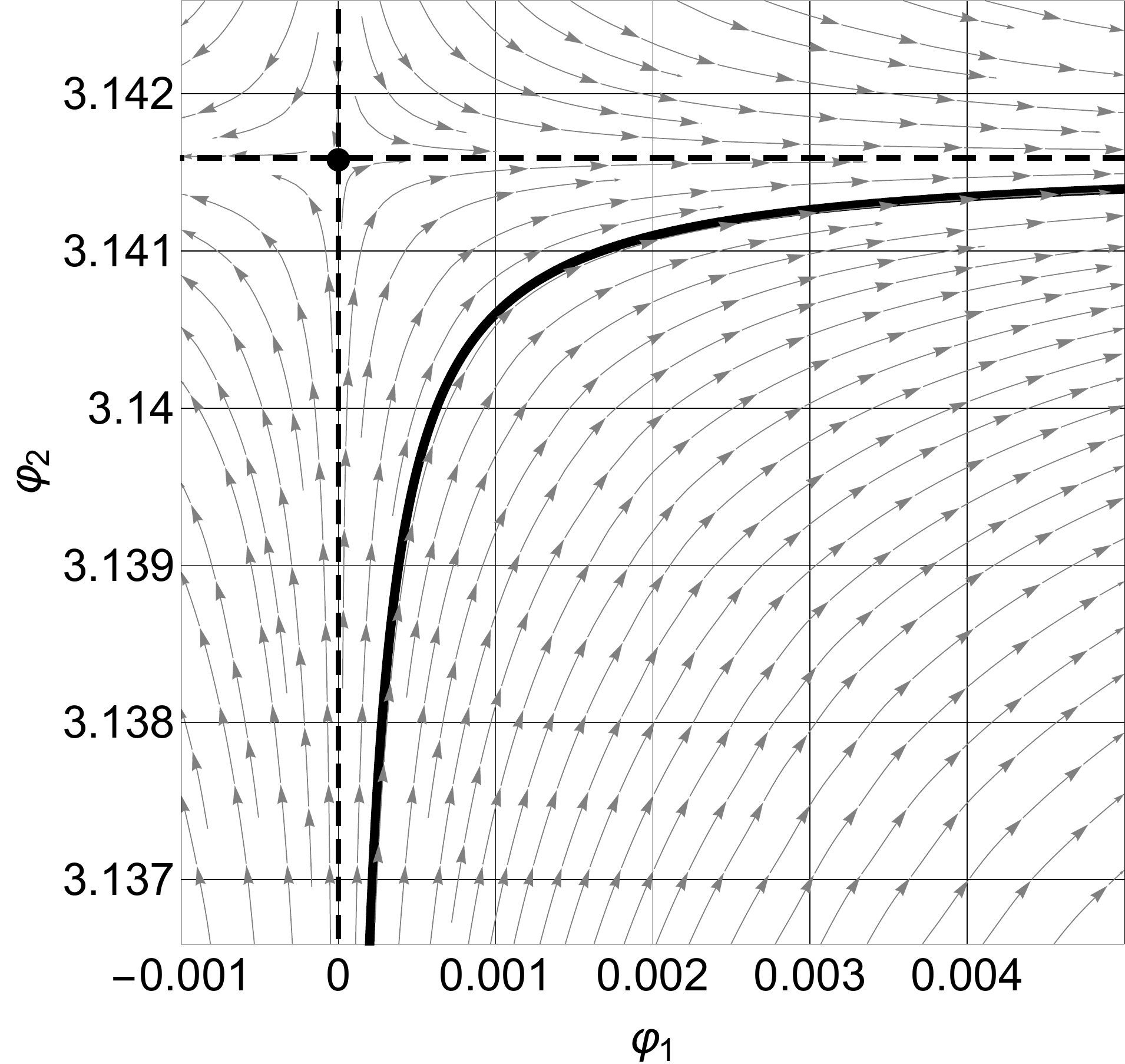}}
  \hskip0.5cm
  \subfigure[]{\includegraphics[width=0.3\textwidth, height =0.25\textwidth]{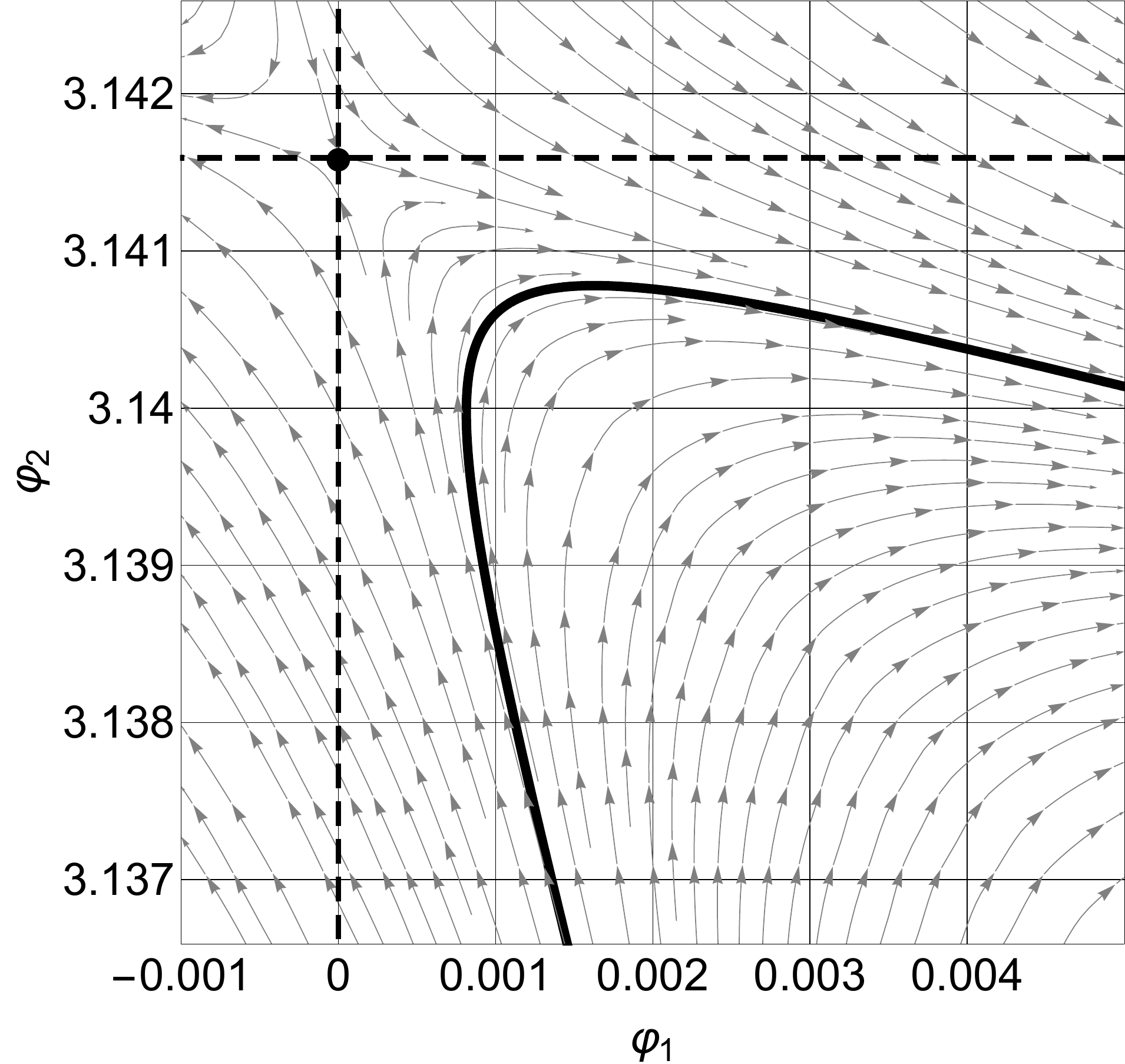}}
\caption{ Pre-potential $U(\varphi_1,\varphi_2)$ and the modified gradient flow $\vec F\equiv+\eta^{-1}\cdot\vec\nabla U$  for (a), (d),  $\lambda=0.8$ and  (b), (e),  $\lambda=0.0$ and (c), (f),  $\lambda=-1.0$. An oriented curve represents the numerical BPS solution for each case.   Dotted curves in (b) represent the analytical BPS solutions for two uncoupled Sine-Gordon models for $x_2=-x_1$ and they correspond with (from right to left) $x_1=\{-0.3,\,0.0,\,0.12,\,0.22,\,0.39\}$. Dotted curves in (a) and (c) stand for numerical solutions of self-dual equations. The solid (numerical) curve has $x_1=1.9$. Figures (d), (e) , (f) show a blow up of the region in  the vicinity of the point $(\vp_1,\vp_2)=(0,\pi)$. }
 \label{fig:3prepotentials}
\end{figure}

The presence of the bumps in our numerical solutions is very consistent with the shape of the potential. In Fig.\ref{fig:2potentials} we present the form of the potential for $\lambda=\{0.8,\, 0.0,\,-1.0\}$.
The vacua (minima) of the potential $V(\vp_1, \vp_2)$ are given by integer multiples of $\pi$
\[
(\vp^{\rm(vac)}_1,\vp^{\rm(vac)}_2)=(n_1\pi, n_2\pi),\qquad n_1,n_2\in{\mathbb Z}.
\]
They correspond to the extrema of the pre-potential $U$. For $(n_1,n_2)=$(even, even) they are minima $U_{\rm min}=-8$; for $ (n_1,n_2)=$(odd, odd) they are maxima $U_{\rm max}=8$ and for $(n_1,n_2)=$(even, odd) or $(n_1,n_2)=$(odd, even) they are saddle points such that the pre-potential takes value $U_{\rm s}=0$. While the value of the  potential is the same for all minima, its value for maxima depends on $\lambda$. Maxima of the potential $V$ for $\lambda=0$ are localized at $(\pm\frac{\pi}{2}+2\pi m_1, \mp\frac{\pi}{2}+2\pi m_2)$, where $m_1,m_2 \in {\mathbb Z}$. When $\lambda$ is different from zero some of these maxima are higher then the others. For $\lambda>0$ higher maxima are those  at $(\frac{\pi}{2}+2\pi m_1, -\frac{\pi}{2}+2\pi m_2)$ and lower ones at $(-\frac{\pi}{2}+2\pi m_1, +\frac{\pi}{2}+2\pi m_2)$ , where $m_1,m_2 \in {\mathbb Z}$. For $\lambda<0$ the meaning of higher and lower maxima  has to be changed round.

In Fig.\ref{fig:2potentials}(d), Fig.\ref{fig:2potentials}(e) and Fig.\ref{fig:2potentials}(f) we plot the BPS solutions in the space of fields on the background provided by potential $V$. The curves interpolate between minima, however,  they do not follow the gradient flow of the potential. Clearly, the gradient flow of the potential $V$ cannot explain their shape. The situation changes if we look at the pre-potential. 
In Fig.\ref{fig:3prepotentials}(a), Fig.\ref{fig:3prepotentials}(b) and Fig.\ref{fig:3prepotentials}(c) we present the same three curves on the background provided by the pre-potential
in which we  have indicated the lines of the modified gradient flow $\vec F = \eta^{-1}\cdot \vec\nabla U$.
The self-duality equations  describe curves in the space of fields.  According to \eqref{BPSeq} a tangent vector $\frac{d\vec\varphi}{dx}$ to any such a curve $\gamma$ representing the BPS solution is given by $\vec F$. Thus all BPS solutions must follow the flow. 
The importance of the flow follows also from the fact that it shows that as $x$ in these
equations increases $U$ grows or decreases monotonically (dependingly on the sign in these equations)
and so, effectively, we have the ``gradient flow'' of $U$.  This follows
from the fact that if the self-duality equations are satisfied we have
\be
\frac{\partial U}{\partial x}\,=\,\frac{\partial U}{\partial \varphi_1}F_1+\frac{\partial U}{\partial \varphi_1}F_2
\,=\,
\frac{4}{4-\lambda^2}\left[ \left(\frac {\partial U}{\partial \varphi_1}\right)^2
\,+\,\lambda \frac {\partial U}{\partial \varphi_1}\frac {\partial U}{\partial \varphi_2}\,+\,
\left(\frac {\partial U}{\partial \varphi_2}\right)^2\right]=2V.
\label{dif}
\ee
For $\vert \lambda \vert<2$ the expression in square brackets in \eqref{dif} is strictly positive and we see that, as 
$x$ changes $U$ changes too.

In contrary to systems with single scalar field, we have here infinitely many BPS solutions (with the same value of energy) that interpolate between the same two vacua. In our example we consider solutions that interpolate between vacua $(0,0)$ and $(\pi,\pi)$. All of them have the same energy $E=16$. This raises the question of how different curves manifest themselves in the set of kink solutions? The answer is obvious for the case $\lambda=0$ where we know the exact form of the solutions \eqref{sgkinks}.  We see that the relative distance between the kinks $x^{(-)}=x_1-x_2$ determines which one of the curves in the space of fields represents the BPS solution. In Fig.\ref{fig:3prepotentials}(b) we have plotted some of such curves corresponding to $x_1=-x_2=\{-0.3,\, 0,\,0.12,\,0.22,\,0.39 \}$.  Note, that the case $\lambda=0$ is rather special as the change of $x^{(-)}$ causes  relative translation between kinks without changing their shape. This is a simple consequence of the fact that for $\lambda=0$ there is no coupling between the fields and so each kink can be shifted in $x$ without affecting the other one.

However, the situation changes when we go to the model with $\lambda\neq 0$. The kinks cannot be shifted independently without changing their shapes. There is only one parameter $x^{(+)}$ that represents the overall translational invariance of the system. This is pretty clear from the particular solution $\vp_1=\vp_2$ which is given by
\be
\vp^{(0)}_1(x)=\vp^{(0)}_2(x)=2\arctan\left[e^{\frac{8}{2-\lambda}\big(x-\frac{1}{2}x^{(+)}\big)}\right].\label{equalneq}
\ee
Note that according to \eqref{equalneq} (and the flow diagram) when kinks are on top of each other the ``bumps'' are absent.
Differently from $\lambda=0$, other BPS solutions cannot be obtained from \eqref{equalneq} by a simple translation like \eqref{transf} because such a transformation would preserve the shapes of kinks. On the other hand, according to the modified flow diagram there are infinitely many BPS solutions which interpolate between the same vacua as the solution \eqref{equalneq} does. They must obviously be related to \eqref{equalneq} by a certain nonlinear mapping which would change positions of kinks and their shapes. In particular such a transformation would be responsible for the creation of ``bumps''.
Some numerical BPS curves for systems with $\lambda=0.8$ and $\lambda=-1.0$ are shown in Fig.\ref{fig:3prepotentials}(a) and Fig.\ref{fig:3prepotentials}(c). They all follow very closely the modified gradient flow.

Looking at Fig.\ref{fig:3prepotentials} we note that all the BPS curves which connect the vacua $(\vp^{\rm(vac)}_1,\vp^{\rm(vac)}_2)=(0,0)$ and $(\vp^{\rm(vac)}_1,\vp^{\rm(vac)}_2)=(\pi,\pi)$ that correspond to the maxima and minima of the pre-potential never reach the vacuum $(\vp^{\rm(vac)}_1,\vp^{\rm(vac)}_2)=(0,\pi)$, which corresponds to the saddle point of the pre-potential. This is pretty clear from figures Fig.\ref{fig:3prepotentials}(d), Fig.\ref{fig:3prepotentials}(e) and Fig.\ref{fig:3prepotentials}(f). Note also, that the numerical BPS curve, which in figures Fig.\ref{fig:2potentials}(b) and Fig.\ref{fig:2potentials}(e) looks like two straight line segments, is indeed a single curve. In particular, for $\lambda=0$ we see that the BPS solution tends to straight segments for $x_1-x_2\rightarrow\infty$ {\it i.e.} for infinitely distant kinks.

\begin{figure}[h!]
 \centering
   \subfigure[]{\includegraphics[width=0.45\textwidth, height =0.3\textwidth]{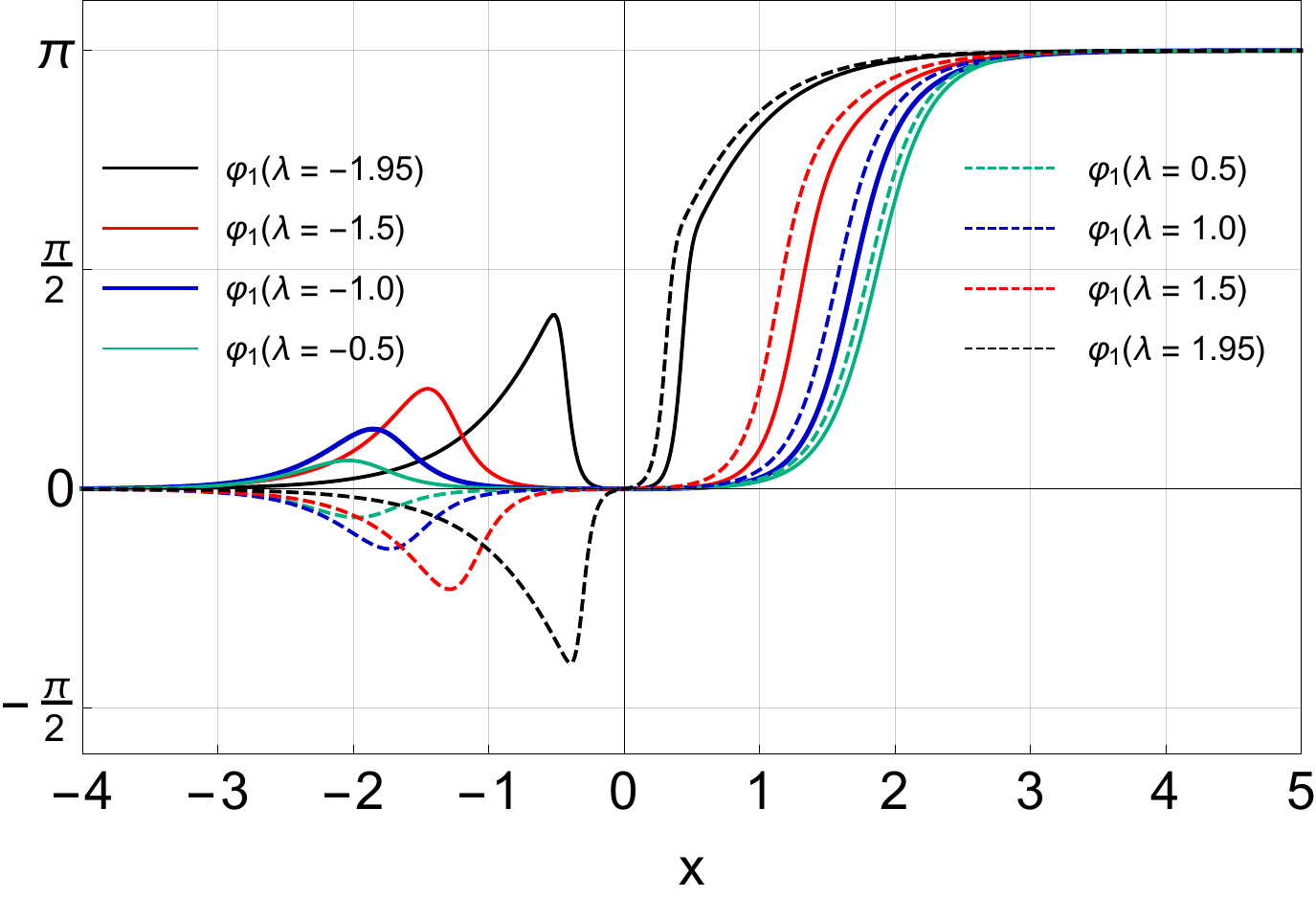}} 
   \hskip0.6cm 
   \subfigure[]{\includegraphics[width=0.45\textwidth, height =0.3\textwidth]{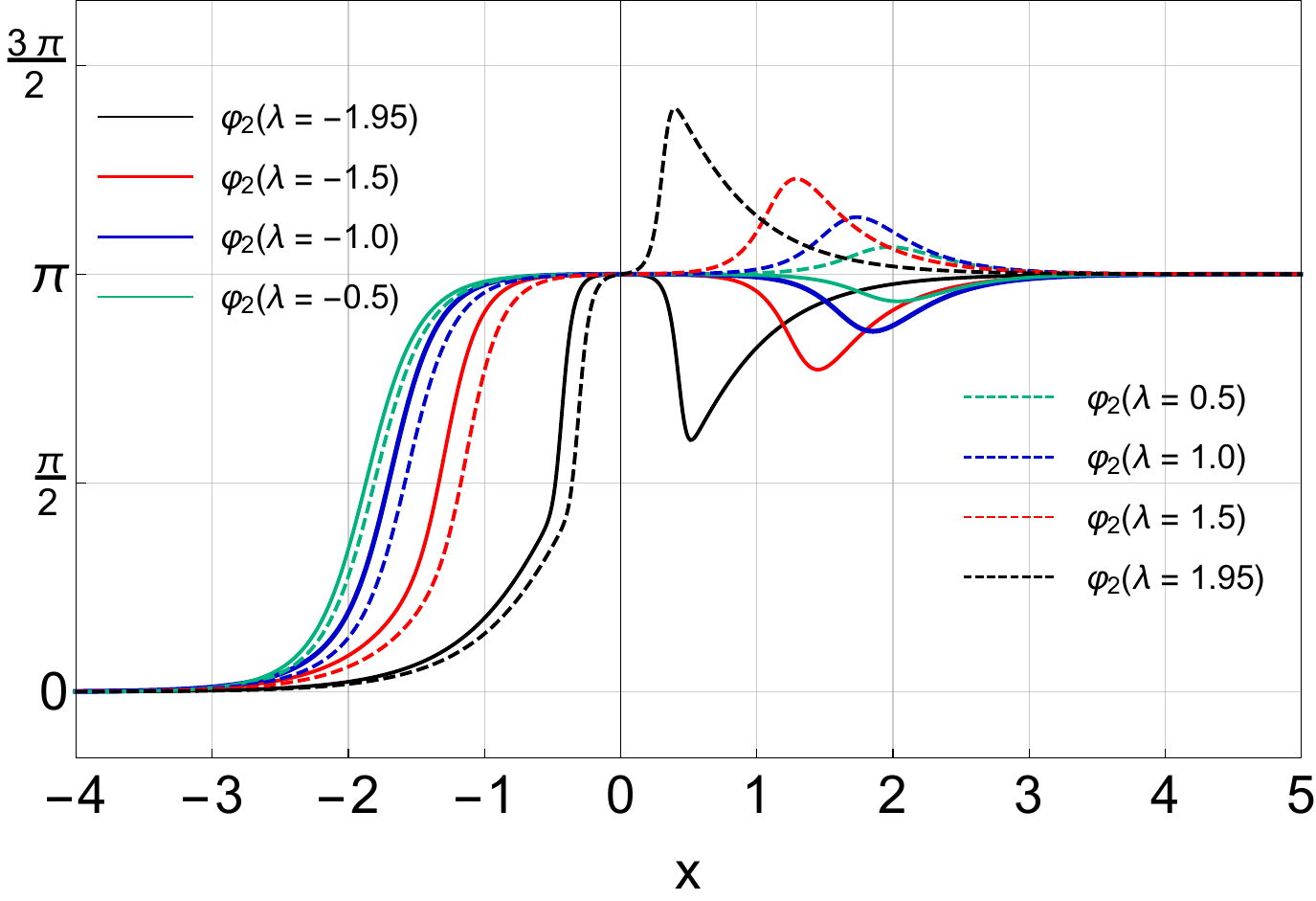}} 
   \subfigure[]{\includegraphics[width=0.5\textwidth, height =0.4\textwidth]{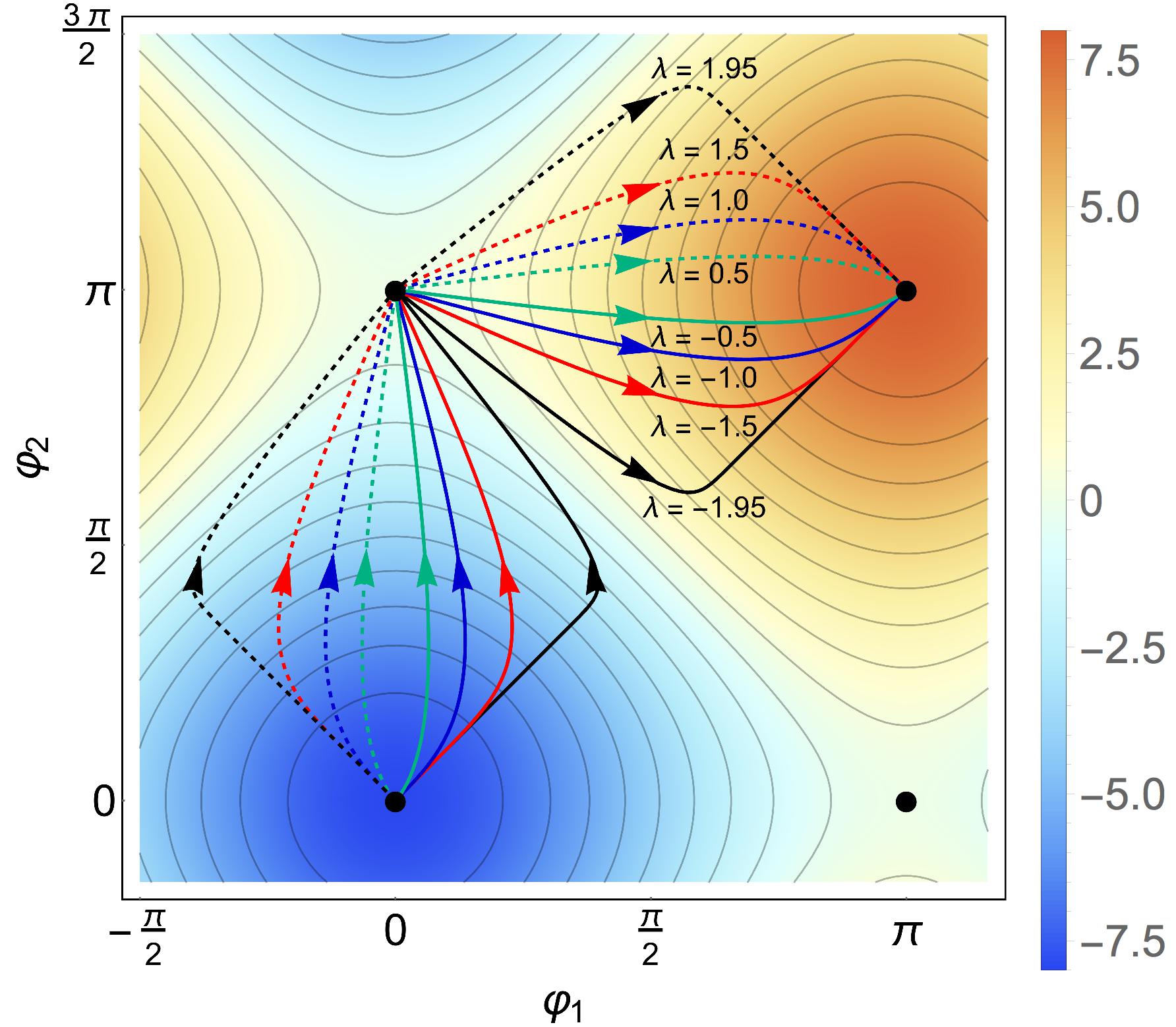}} 
   \hskip0.6cm
   \subfigure[]{\includegraphics[width=0.45\textwidth, height =0.38\textwidth]{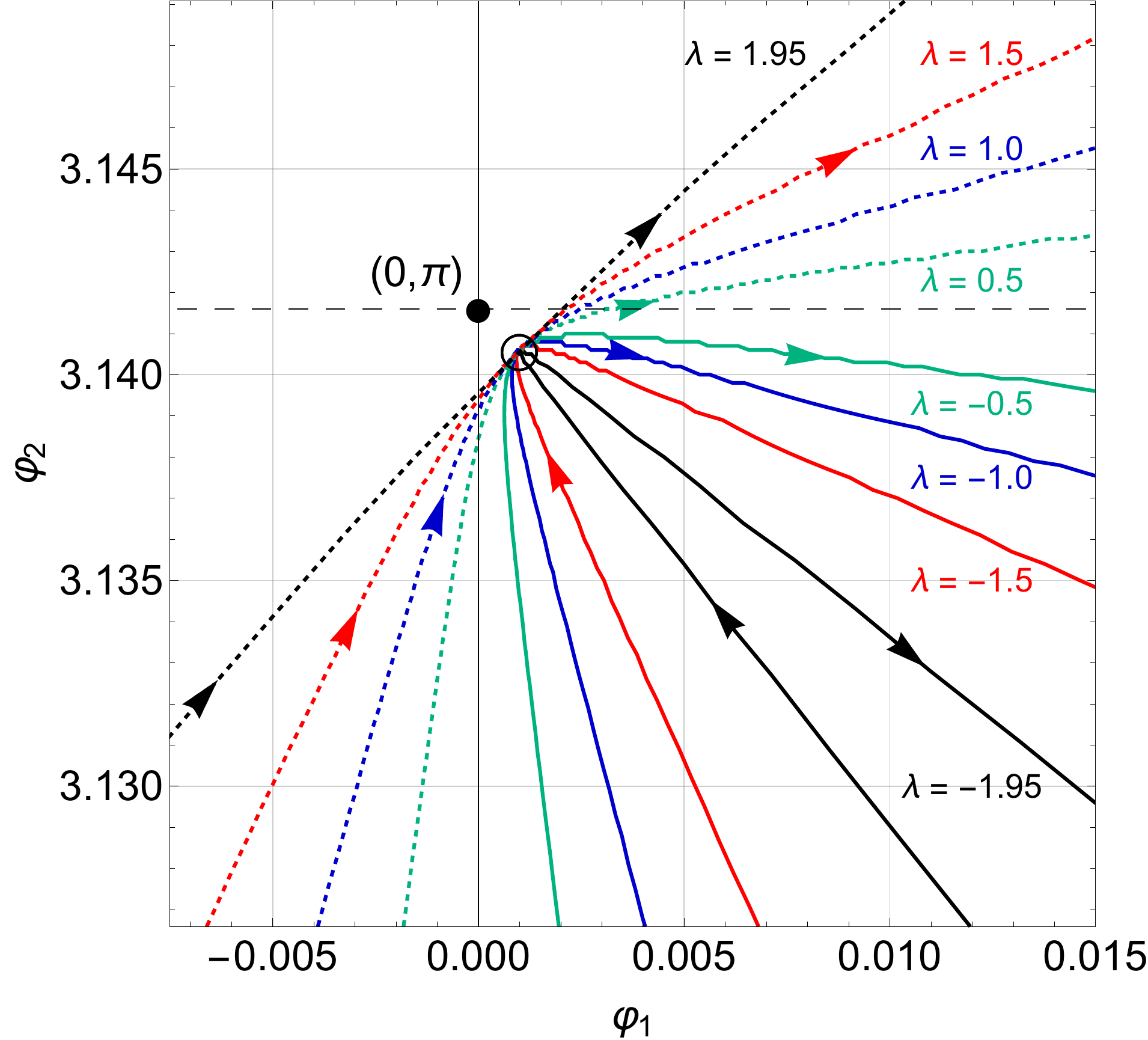}} \hskip0.5cm           
\caption{Dependence of solutions on $\lambda$. Fields (a) $\vp_1$ and (b) $\vp_2$ for solutions that start with the same ``initial'' conditions. (c) Different BPS curves in the space of fields. (d) A blow up of the region in the vicinity of the point $(\vp_1(0),\vp_2(0))=(0.000997, 3.1406)$.}
 \label{fig:extra}
\end{figure}
We have also looked at the energies of our solitonic solutions. These energies have all been 
the same despite the extra bumps in the $\varphi_1$ and $\varphi_2$ fields for nonvanishing $\lambda$.
However, our fields are solutions of the BPS equations and so the values of the energies
are determined by the asymptotic values of the fields and in all our cases these asymptotic values
were 0 for $x\rightarrow -\infty$ and $\pi$ or $-\pi$ for $x\rightarrow +\infty$. Hence, the non-dependence
of energy on $\lambda$ is obvious. In Fig.\ref{fig:extra} we have plotted some solutions that start with the same initial conditions $(\vp_1(0),\vp_2(0))=(0.000997, 3.1406)$. Each solution is obtained in the model with different value of parameter $\lambda$, hence its BPS curve follows a line belonging to a different flow.

Next we have looked at the time dependence properties of our BPS solutions. First we checked that our solutions
were  indeed time independent and Lorentz covariant. To do this we simulated the time dependence of the field
configurations by using a 4th order Runge - Kutta method of solving the equations (\ref{generaleq}).
 We used grids 
of 40000 points with $dx \sim 0.001$ and $dt\sim 0.1dx$. To check Lorentz invariance of our solutions
(for $v\ne0$) we had to find the initial time derivatives of $\varphi_i$. However, for Lorentz invariant
fields we expect the fields being functions of $\frac{x\pm vt}{\sqrt{1-v^2}}$ so for $\frac{\partial \varphi_i}
{\partial t}$ we could use $\frac{\partial \varphi_i}{\partial x}$ multiplied by $\pm v$ and then changing 
$dx$ appropriately. As expected the static fields did not evolve and the time dependent kinks moved as expected.
This confirmed the reliability of our numerical procedures and gave us confidence in our results.
Note that this is not a completely trivial result as close to the additional ``bumps" the time derivatives
terms in each field tried to alter the fields of the bumps in opposite directions but these effects
were compensated by the contributions from the other field. Looking at the plots we did not see any such
problems. In fact, essentially nothing was emitted and so the energies were amazingly well conserved
(to $10^{-7}\,\%$).

\section{Some properties of time dependence of the obtained solutions}


An obvious question then arises at what happens if one starts with a `wrong' BPS solution;
{\it  i.e.} takes a BPS solution corresponding to one value of $\lambda$ and tries to evolve it
with a different $\lambda$.

The answer is simple, the initial field has an extra energy
and it uses this energy to change the profile of its solitonic function - at the same time sending
some waves of energy towards the boundaries {(\it i.e.} also radiating some energy out). The results are very similar when we take initial $\lambda=0$ and the other one very close to it.

\begin{figure}[h!]
\centering
  \subfigure[]{\includegraphics[width=0.35\textwidth,height=0.25\textwidth]{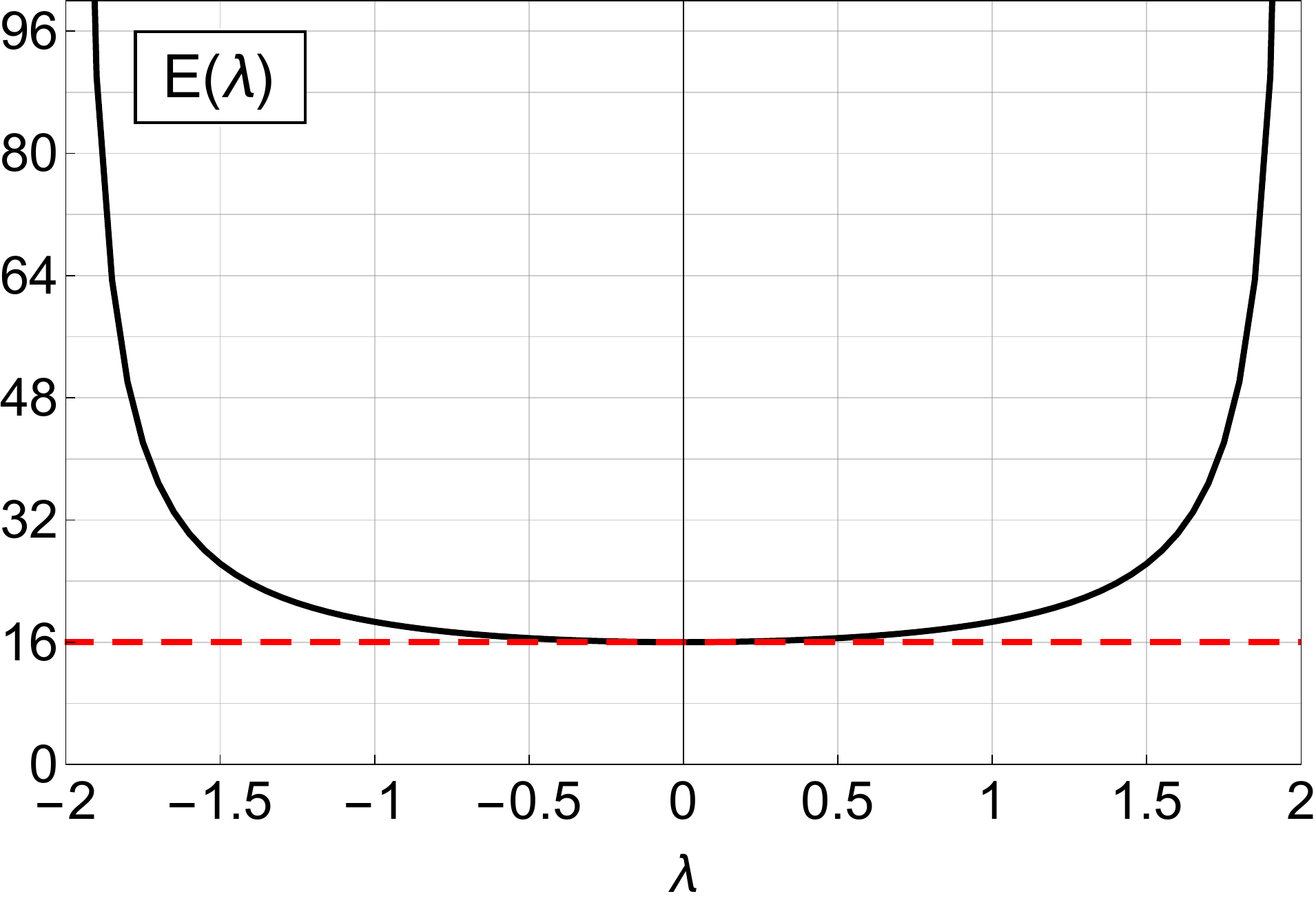}}
  \hskip0.5cm
    \subfigure[]{\includegraphics[width=0.5\textwidth,height=0.3\textwidth]{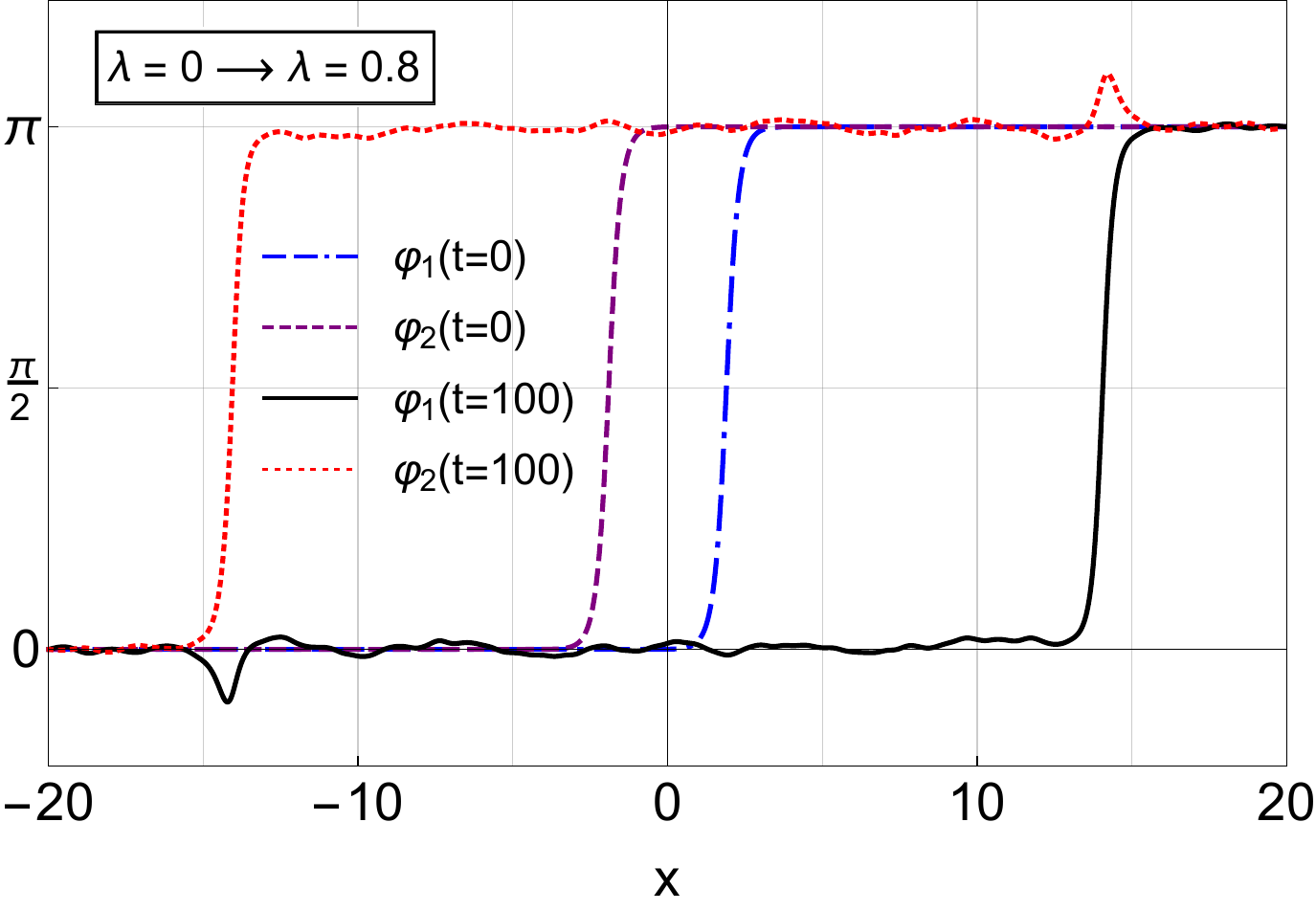}}
    \caption{  (a) Energy of  two Sine-Gordon kinks ($\lambda=0$) in the model with non-zero $\lambda$. (b) Fields $\varphi_1$ and $\varphi_2$ for $\lambda=0.0$ used to carry out time evolution
for the model with $\lambda=0.8$.}
  \label{fig:4absol}
\end{figure}

Let us start with discussion of the case which is the $\lambda=0$ BPS solution given by \eqref{sgkinks}. This solution describes two Sine-Gordon kinks that are not the BPS solutions of the model with nonzero $\lambda$. 
The energy of such a field configuration is given by
\be
E(\lambda)=\frac{8}{4-\lambda^2}\Big[8-\lambda^2-\frac{12\lambda^3}{\sinh(24)}\Big].\label{enemerged}
\ee
Clearly, $E(\lambda)\ge 16$ and $E(\lambda)\rightarrow\infty$ for $\lambda\rightarrow\pm 2$. The minimum $E=16$ is reached for $\lambda=0$. The energy \eqref{enemerged} is shown in Fig.\ref{fig:4absol}(a). In our example $E(0.8)\approx17.523$.

\begin{figure}[h!]
  \centering
   \subfigure[]{\includegraphics[width=0.3\textwidth,height=0.25\textwidth]{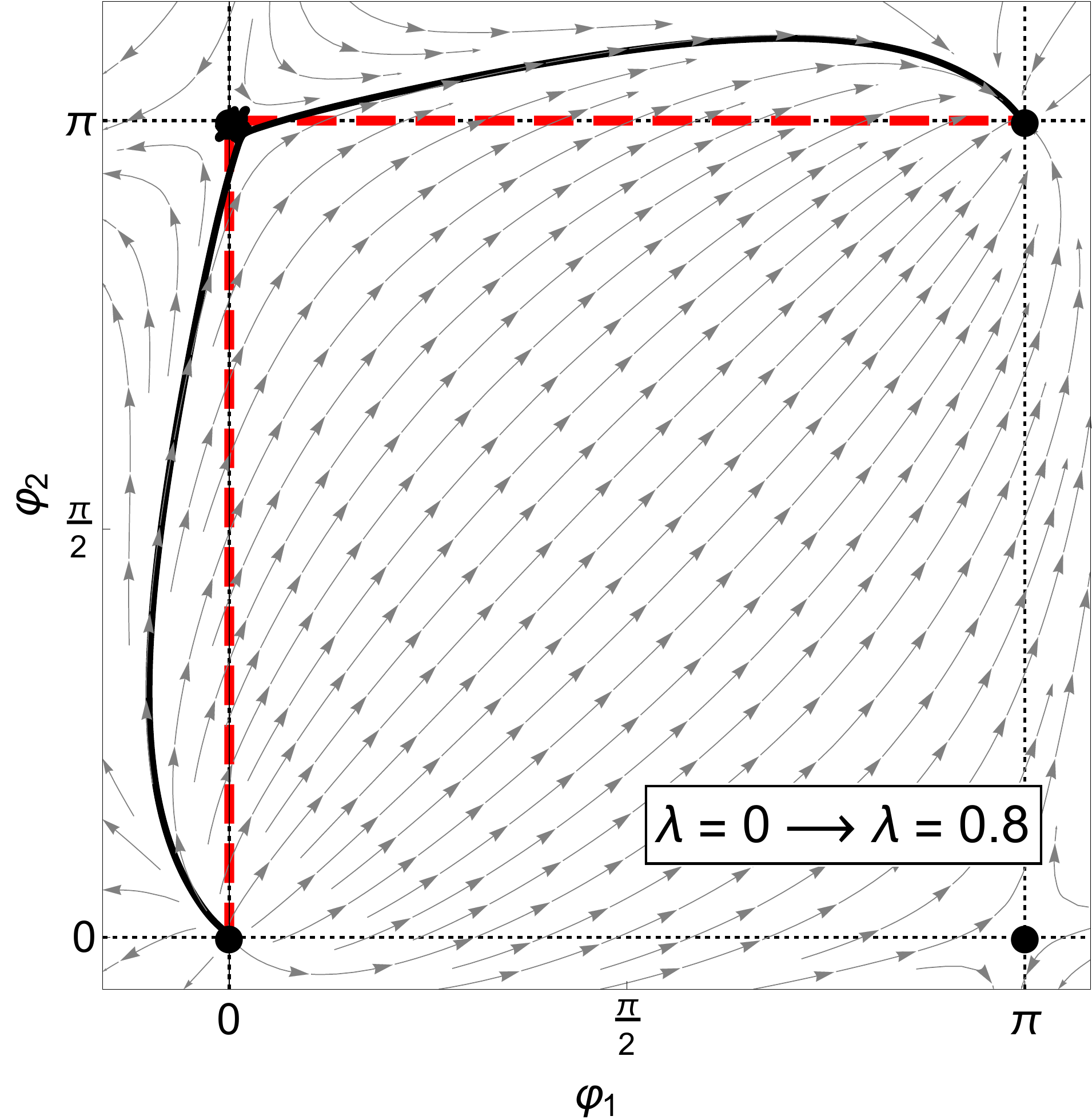}}
   \hskip0.15cm
    \subfigure[]{\includegraphics[width=0.3\textwidth,height=0.25\textwidth]{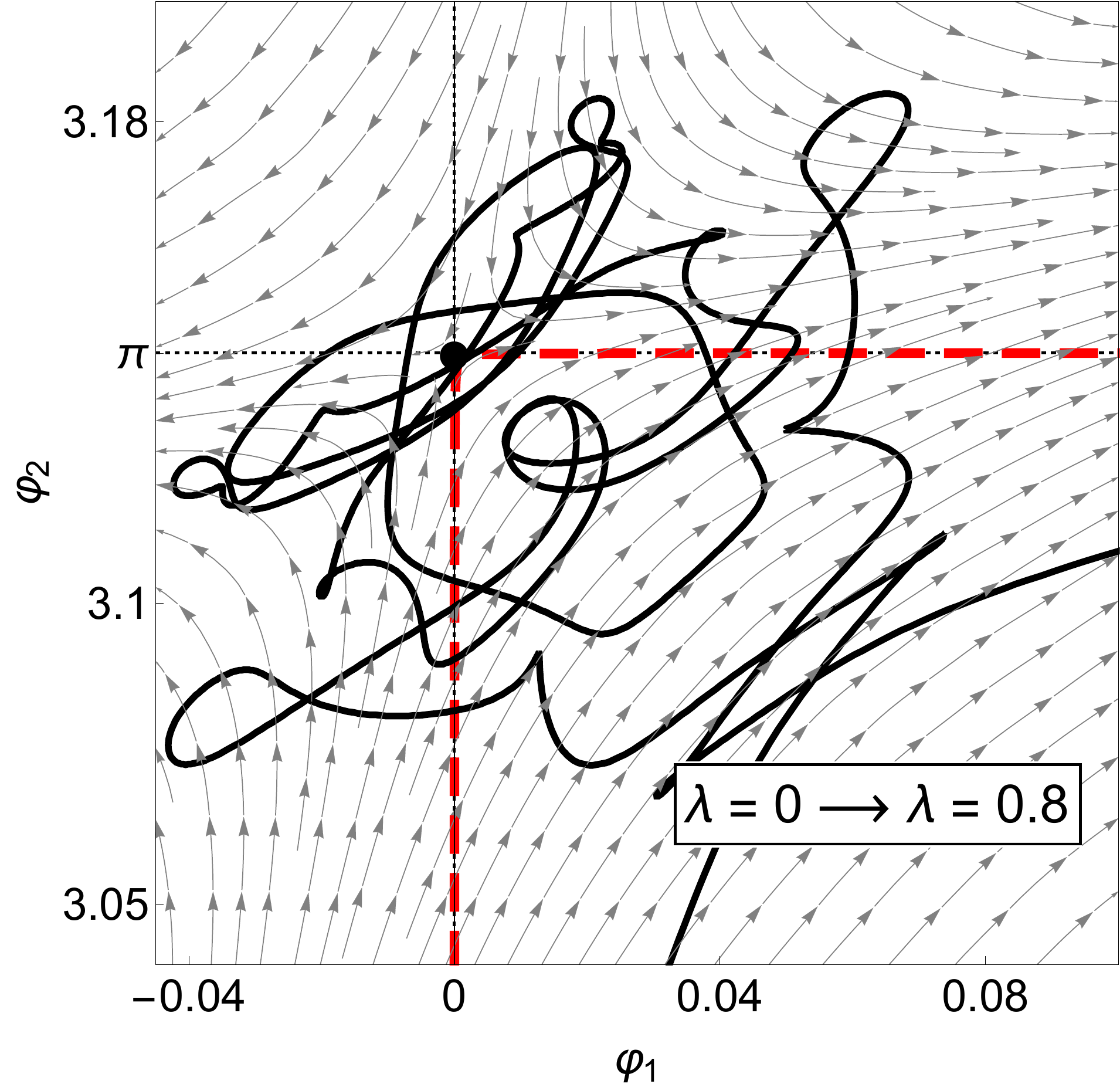}}
    \hskip0.15cm
     \subfigure[]{\includegraphics[width=0.3\textwidth,height=0.25\textwidth]{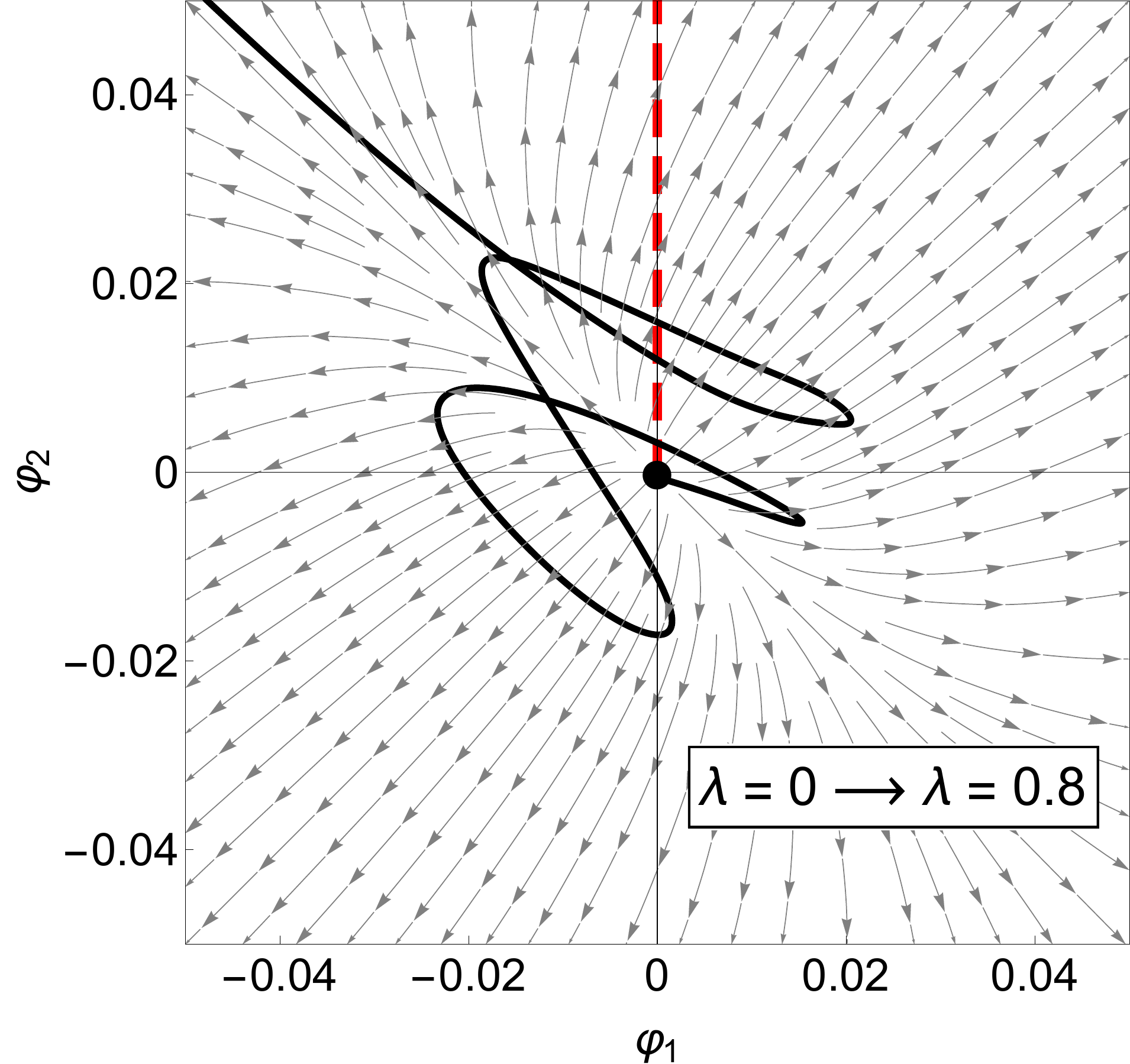}}
   \caption{  Fields  for $\lambda=0$ used to carry out time evolution
in the model with $\lambda=0.8$.  (a) Initial $t=0$ configuration (dashed line) in the space of fields and  the outcome of its evolution in a model with $\lambda=0.8$ at $t=100$ (solid line).  (b) Blow up of a region in vicinity of $(\vp_1,\vp_2)=(0,\pi)$  and (c) another blow up in vicinity of $(\vp_1,\vp_2)=(0,0)$.}
  \label{fig:4sol}
\end{figure}

In figure Fig.\ref{fig:4absol}(b) we show the initial $t=0$ and after $t=100$ profiles of fields $\vp_1$ and $\vp_2$. Looking at the plots we see that 
as the ``bumps'' are being created they move in opposite directions than the kinks. In Fig.\ref{fig:4sol}(a) w plot initial $t=0$ (dashed curve) and final $t=100$ (solid curve) in space of fields $(\vp_1,\vp_2)$. It is pretty clear from this picture that the initial curve (obtained for $\lambda=0$) is deformed so it approaches the lines of the $\nabla_{\eta}U$-flow for $\lambda=0.8$. Thus, the initial field configurations evolve in such a way that the curve in the space of fields gets closer and closer to the "correct" flow.  As one can see from Fig.\ref{fig:4absol}(b) there is some radiation at $t=100$. The presence of this radiation manifests itself in a quite complicated form of the curve in the space $(\vp_1,\vp_2)$ in the vicinity of the vacua (after all, the extra energy is emitted towards them). In figures Fig.\ref{fig:4sol}(b) and (c) we present plots of the blow up of the two regions around $(0,\pi)$ and $(0,0)$. 

As the simulation was performed on a finite segment $x\in[-20,20]$ and we use the ``absorbing'' boundary conditions  the energy of the numerical solution is not conserved. Initially, we had some radiation and the motion of the kinks (and the corresponding ``bumps") towards the ends of the our grid. Once the kinks and the ``bumps" reached the boundary they were absorbed and we had a dramatic decrease of  energy due to this absorption. In this simulation this happened for $t$ close to $t\approx180$.
Of course this is a purely numerical artifact. However, in some cases the evoltion did not lead the the motion of the original kinks (and corresponding ``bumps") so in next few examples we present the plots of the time dependence of initial energy seen in our numerical solutions. They show that the radiation which is generated in the system carries out some quantities of the energy when escaping to spatial infinity.


The extra motion seen in our case is quite interesting so we may wonder what would happen had we started 
other way round {\it i.e.} with the solitons corresponding to $\lambda=0.8$ and evolving them
in the model with $\lambda=0$. In this case the system has to get rid of the energy associated with the extra ``bumps".

 \begin{figure}[h!]
  \centering
 {\includegraphics[width=0.5\textwidth, height=0.3\textwidth]{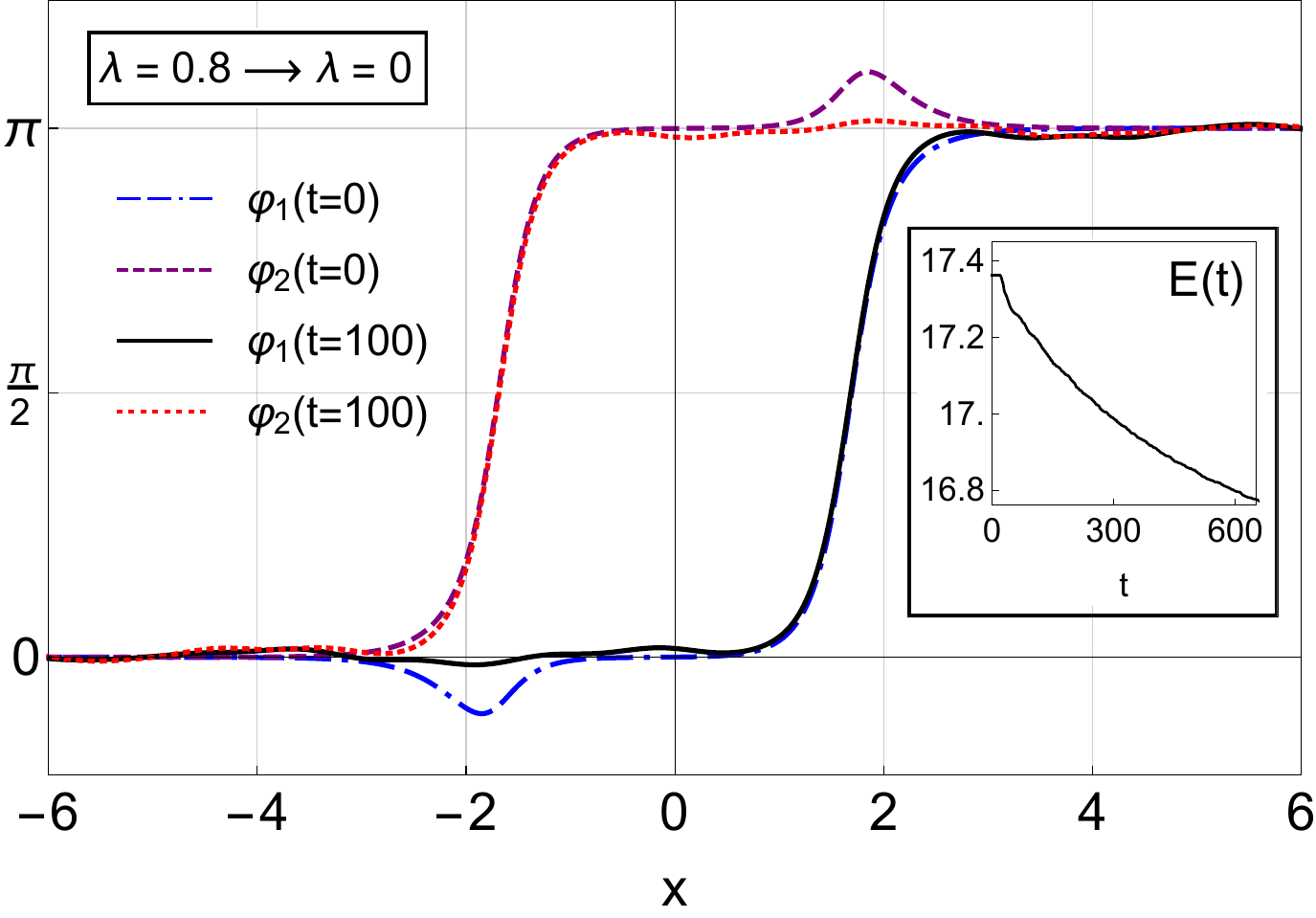}}
   \caption{ Fields $\varphi_1$ and $\varphi_2$ for $\lambda=0.8$ used to carry out time evolution
for the model with $\lambda=0.0$ at $t=0$ and $t=100$. Subfigure shows the energy of the numerical solution on the segment $x\in[-20,20]$.}
  \label{fig:5asol}
\end{figure}

In Fig.\ref{fig:5asol} we present such plots. This time we do not see any final motion. 
 The initial bumps  gradually vanish and some radiation is generated. This radiation hits the absorbing boundaries and the energy of the numerical solution decreases slowly. The subfigure in Fig.\ref{fig:5asol} shows the plot of the time dependence of the energy of our numerical solution.  
 \begin{figure}[h!]
  \centering
   \subfigure[]{\includegraphics[width=0.3\textwidth,height=0.25\textwidth]{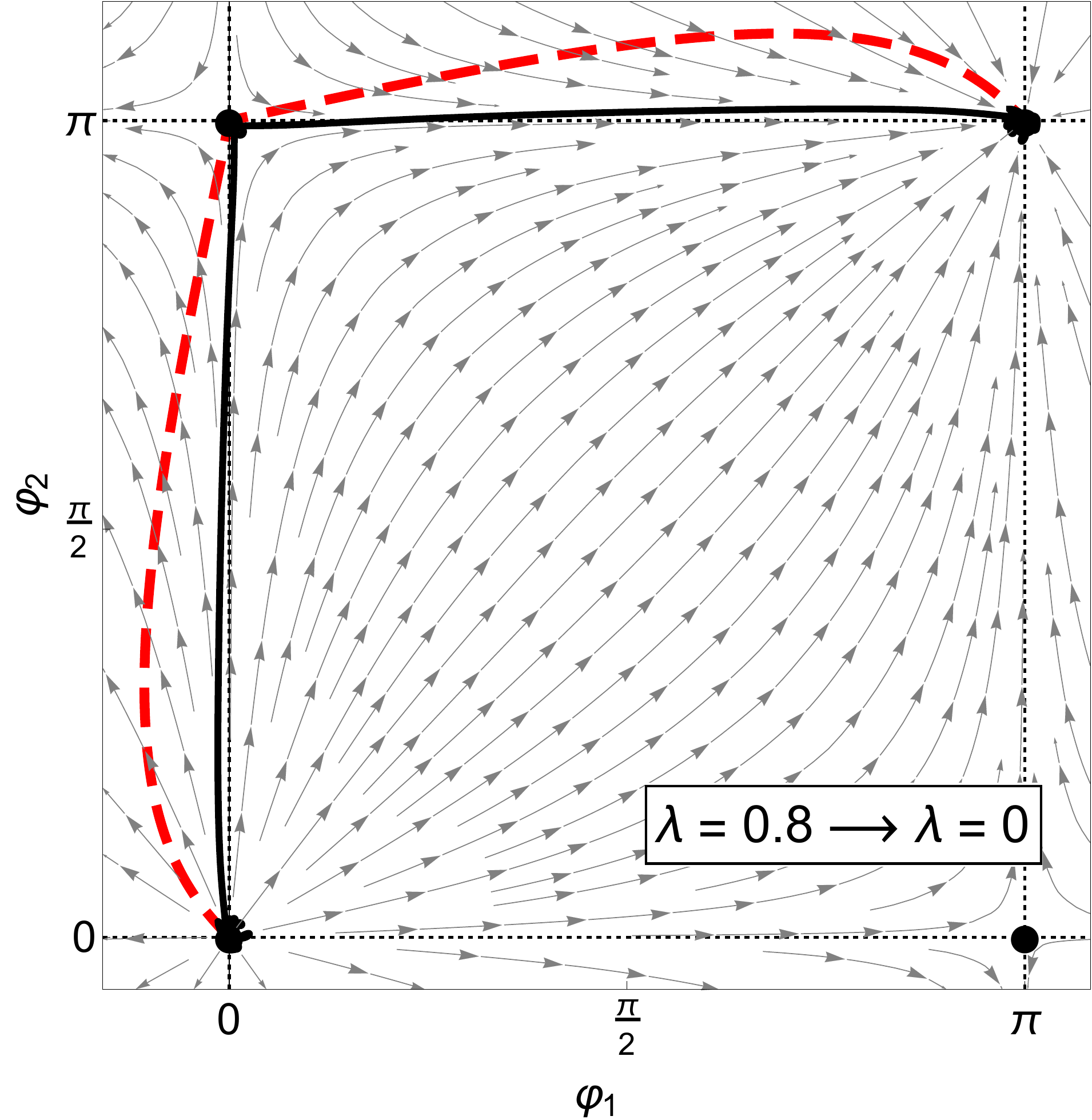}}
    \subfigure[]{\includegraphics[width=0.3\textwidth,height=0.25\textwidth]{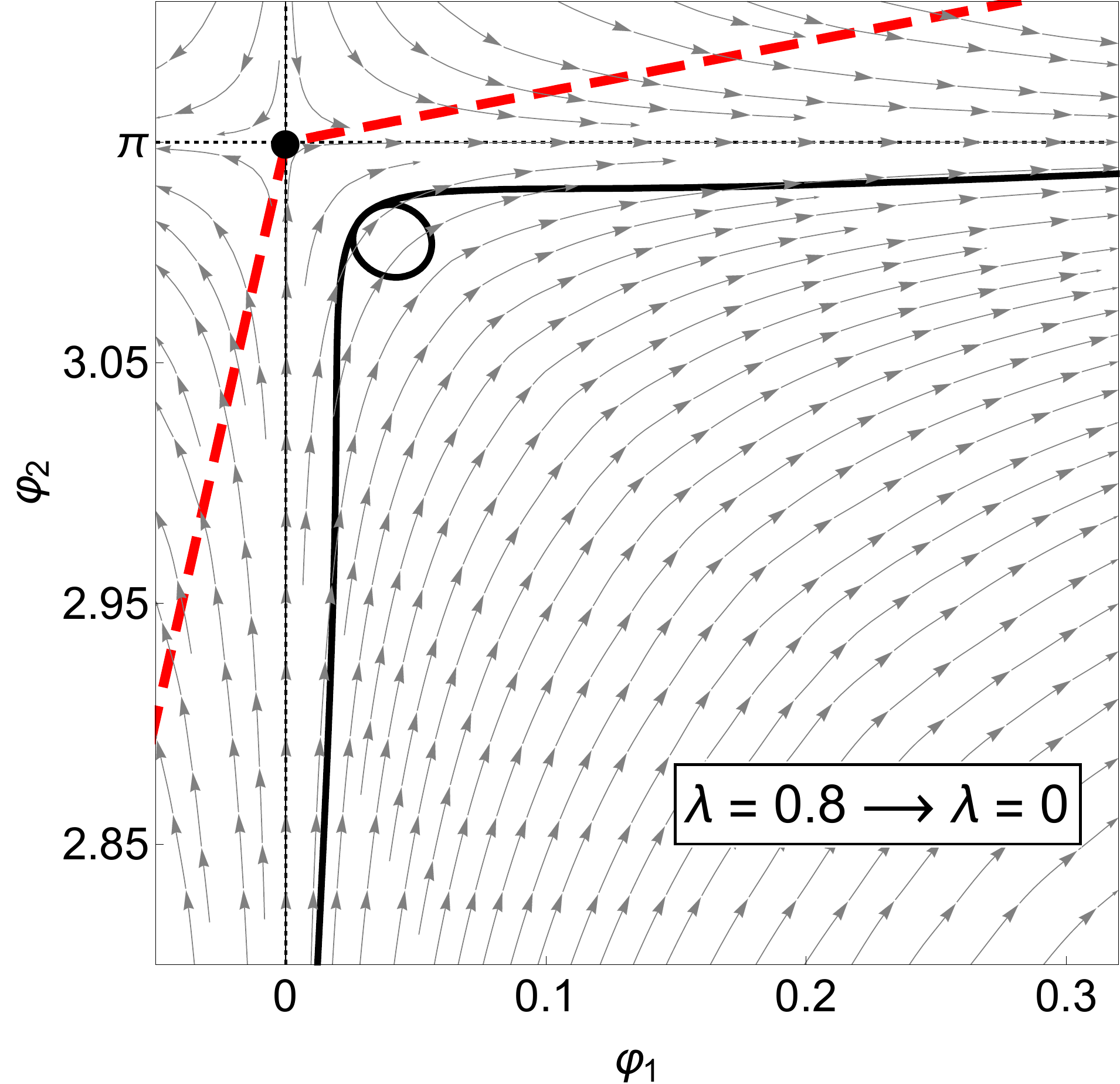}}
     \subfigure[]{\includegraphics[width=0.3\textwidth,height=0.25\textwidth]{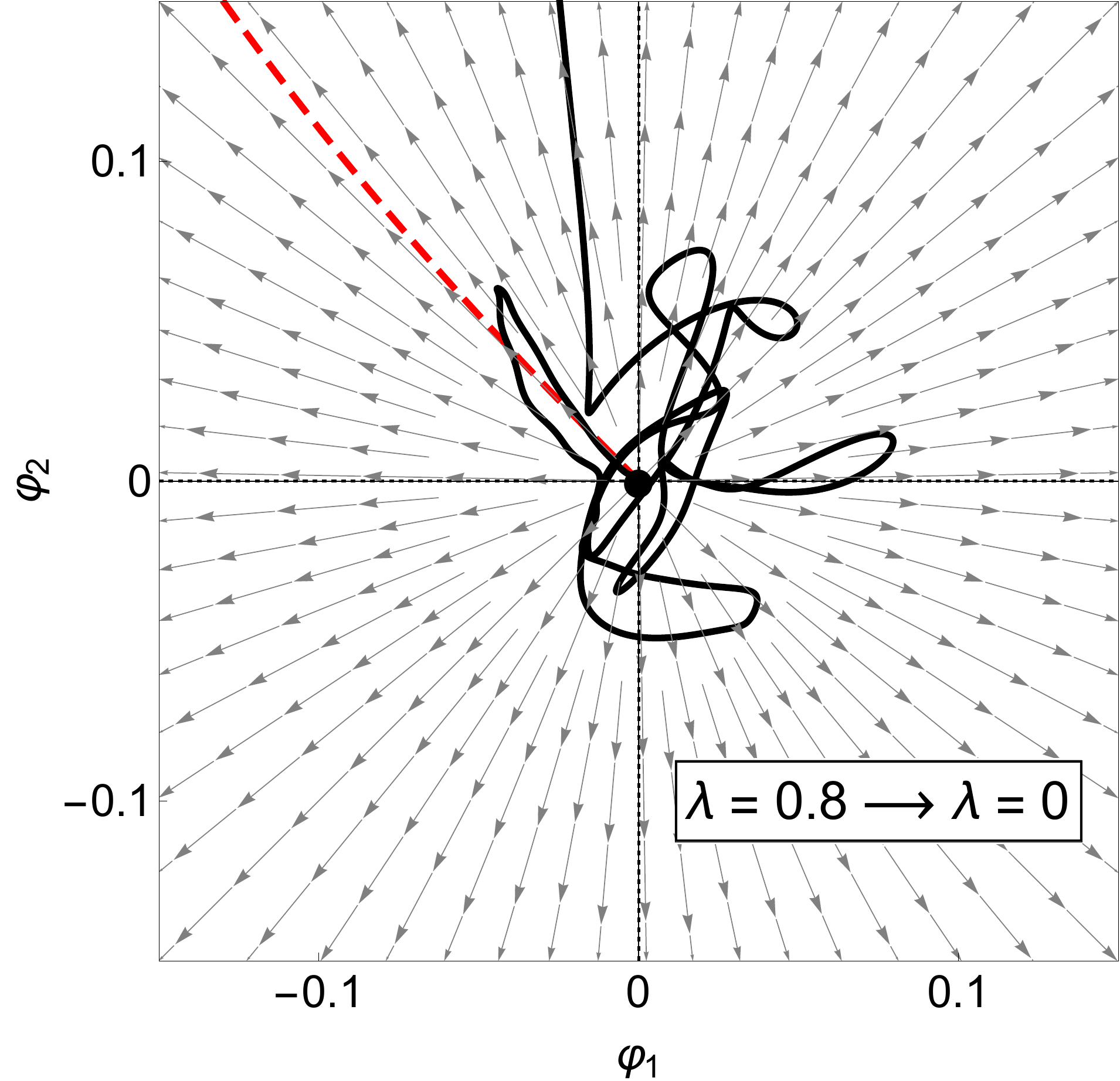}}
   \caption{Fields  for $\lambda=0.8$ used to carry out time evolution
for the model with $\lambda=0$. (a) Initial $t=0$ (dashed curve) and intermediate $t=100$ (solid curve) field configuration with $\lambda=0.8$ emerged in the model with $\lambda=0$ in background provided by a modified gradient flow. A blow up of the region in vicinity of (b) $(\vp_1,\vp_2)=(0,\pi)$  and (c) $(\vp_1,\vp_2)=(0,0)$.}
  \label{fig:5sol}
\end{figure}
In Fig.\ref{fig:5sol}(a) we present the plots of the initial and final curves in the space of fields $(\vp_1,\vp_2)$. Our figure shows very clearly that the system evolves into the "correct" flow {\it i.e} the flow associated with $\lambda=0$. As before we see quite complicated behaviour of the curve in the vicinities of the vacua. We have plotted them in figures Fig.\ref{fig:5sol}(b) and Fig.\ref{fig:5sol}(c).

The difference in these two cases suggests that it would be  interesting to check what happens when we take the initial $\lambda$ and the correct $\lambda$ both different from zero.
 \begin{figure}[h!]
  \centering
  \subfigure[]{\includegraphics[width=0.45\textwidth,height=0.3\textwidth]{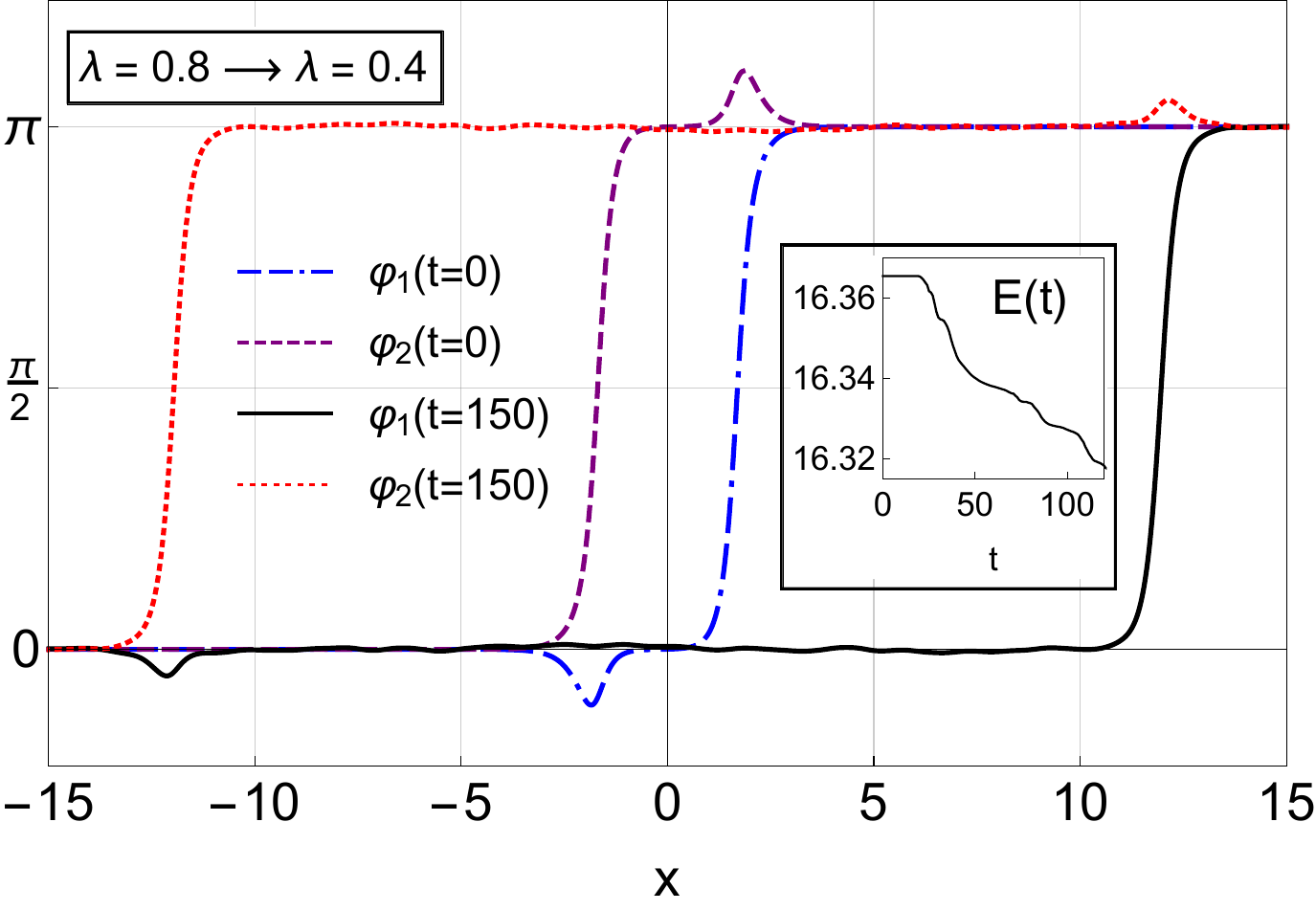}} 
  \hskip 0.5cm               
  \subfigure[]{\includegraphics[width=0.45\textwidth,height=0.3\textwidth]{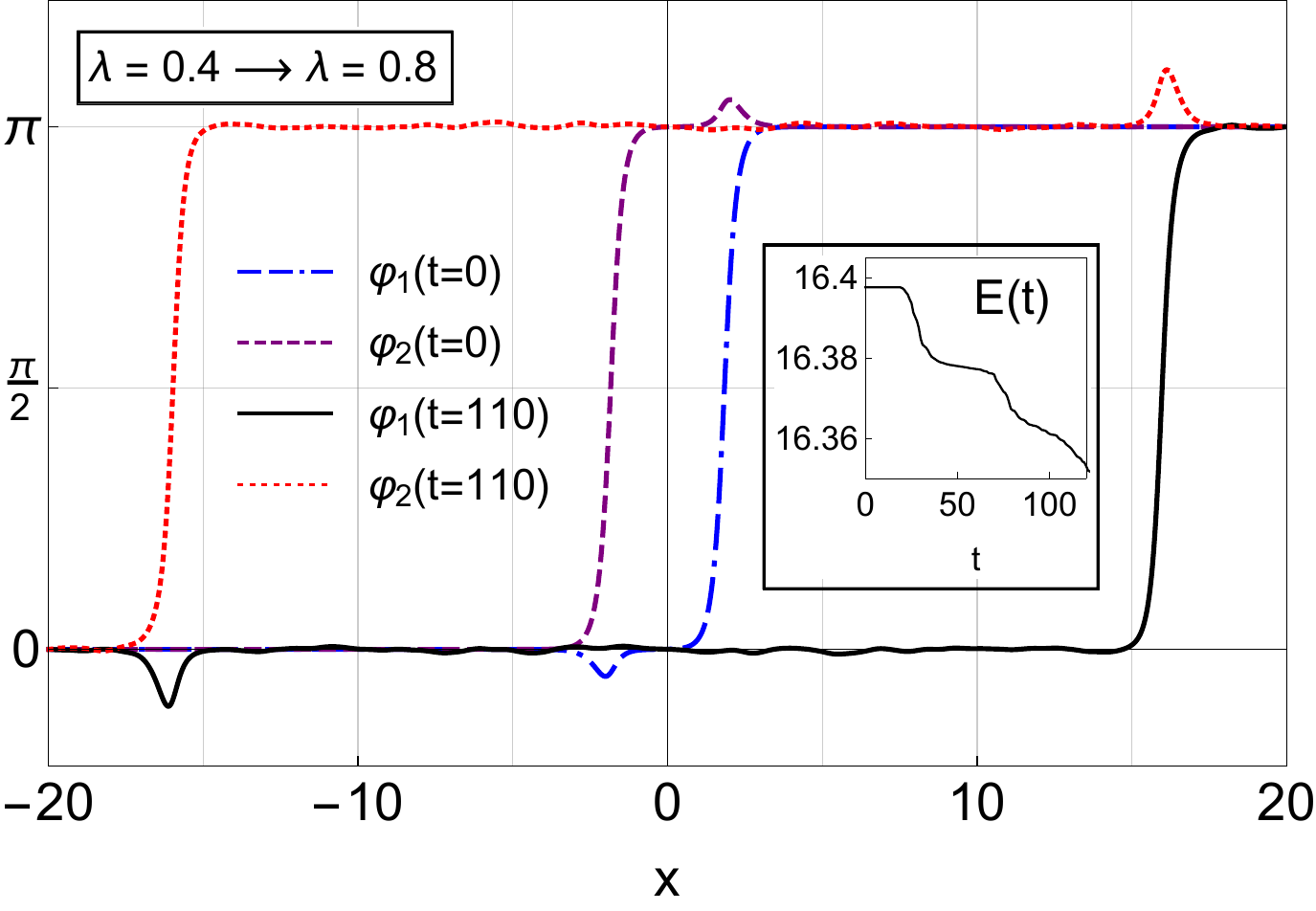}}
   \caption{(a) Fields $\varphi_1$ and $\varphi_2$ initially for $\lambda=0.8$ used to carry out time evolution
for the model with $\lambda=0.4$. (b) The case with interchanged values of $\lambda$.}
  \label{fig:6sol}
\end{figure}

\begin{figure}[h!]
  \centering
\subfigure{\includegraphics[width=0.5\textwidth,height=0.3\textwidth]{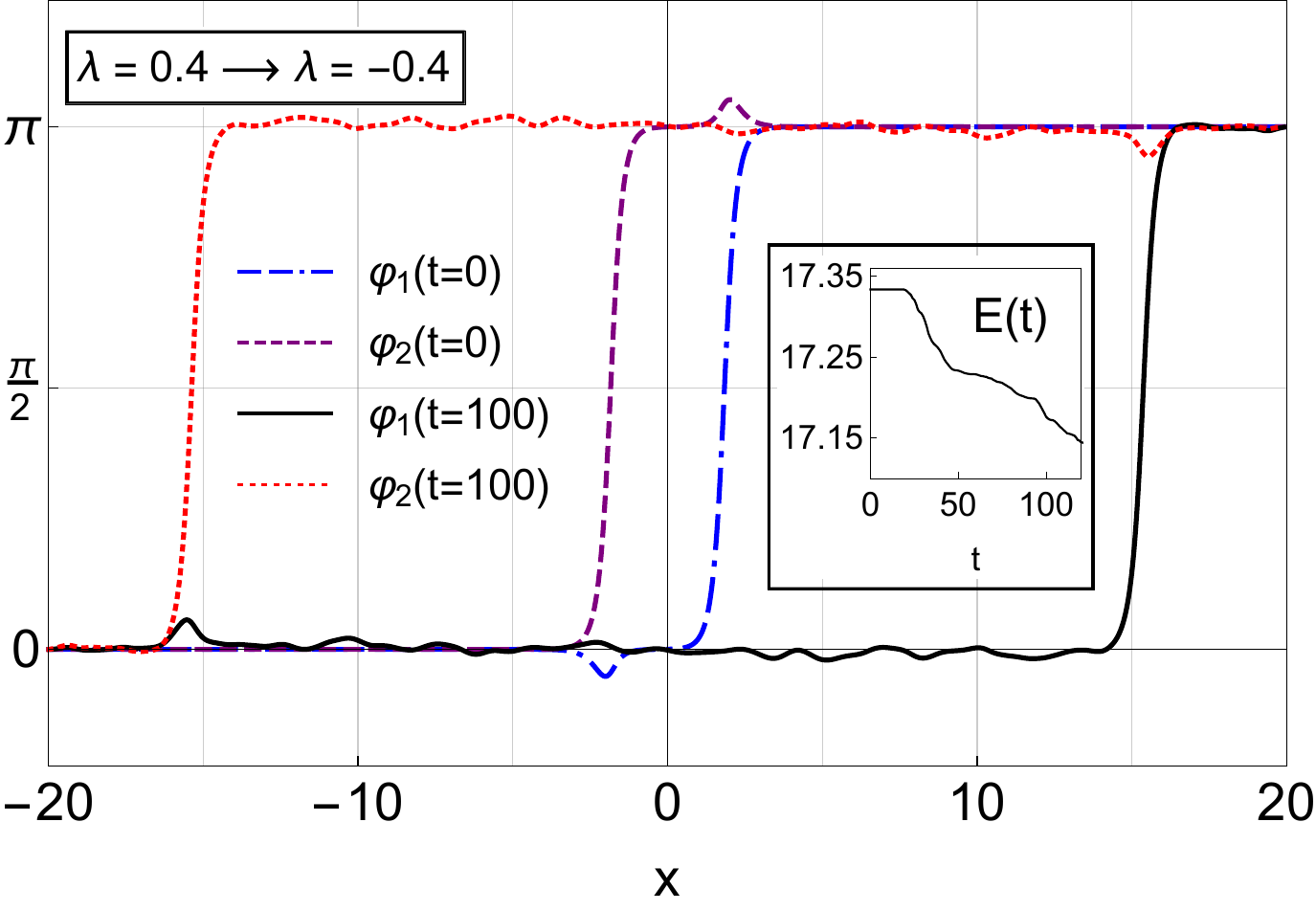}} 
\hskip 0.5cm  
\subfigure{\includegraphics[width=0.3\textwidth,height=0.3\textwidth]{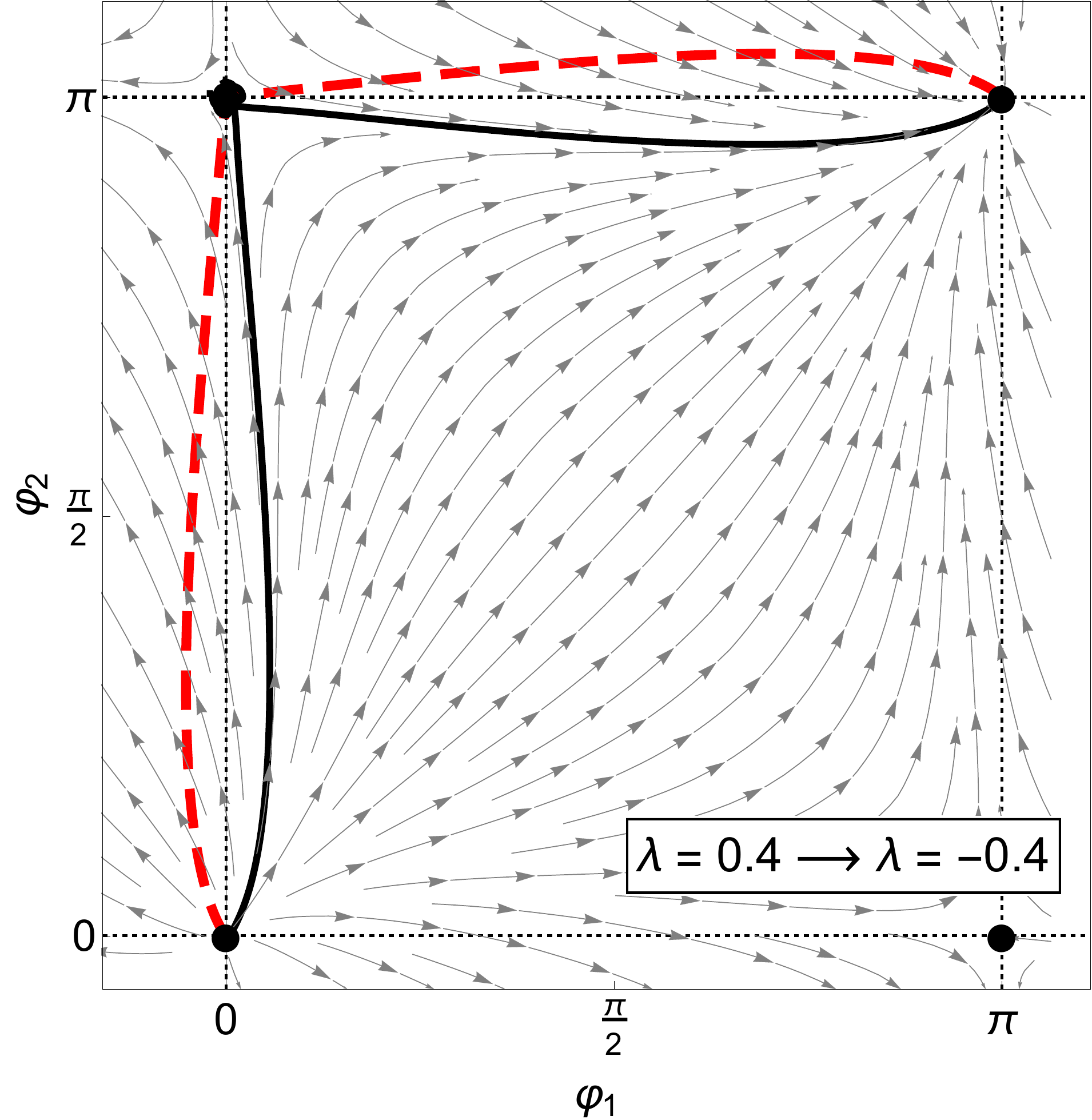}}               
     \caption{(a) Fields $\varphi_1$, $\varphi_2$ initially for $\lambda=0.4$ used to carry out time evolution
of the model with $\lambda=-0.4$. (b) Initial (dashed line) and after $t=100$ (solid line) configuration of fields.}
  \label{fig:9asol}
\end{figure}

In Fig.\ref{fig:6sol}(a) we present plots of the initial configuration corresponding to $\lambda=0.8$ used as the initial condition for the simulation with $\lambda=0.4$. The figures show the initial $\varphi_1$, $\varphi_2$ at $t=0$ and the same fields at $t=150$. We also present the  plot of the energy of the configuration seen in the simulation on $x\in[-20,20]$. In Fig.\ref{fig:6sol}(b) we present similar plots for the simulation started with the configuration for $\lambda=0.4$  in a simulation 
for $\lambda=0.8$ and in Fig.\ref{fig:9asol} for the simulation started with $\lambda=0.4$ configuration but performed  for $\lambda=-0.4$. 
Looking at the plots we see that in each case the initial configuration has a surplus of energy with respect
to the correct one soliton solution of the model. This energy is used on changing the shape of the
field configuration and the rest is emitted and used to move the solitons away from each other.
In all our simulations the motion has always been away from each other.


\section{A more general model}
\setcounter{equation}{0}
Next we have generalised our model by introducing an extra 
parameter $\epsilon=\pm1$ multiplying one of the two terms, say $N$, in \eqref{topo2d}. So we have taken $N(\varphi_2)=4\epsilon  \sin(\varphi_2)$.

The self-duality equations  \eqref{BPSeq} (with upper sign) have now become
\begin{align} 
\partial_x \varphi_1\,&=\,\frac{4}{4- \lambda^2}
\Big(4\sin(\varphi_1)+2\lambda \epsilon \sin(\varphi_2)\Big),\label{aa1}\\
\partial_x \varphi_2\,&=\,
\frac{4}{4-\lambda^2}\Big(2\lambda \sin(\varphi_1)\,+
4\epsilon \sin(\varphi_2)\Big),\label{aa2}
\end{align}
where the rhs of equations \eqref{aa1} and \eqref{aa2} are now given  by \eqref{F1} and \eqref{F2}
and the pre-potential $U$ takes the form $U=-4\big(\cos(\varphi_1)+\epsilon \cos(\varphi_2)\big)$.
The expression for the energy (\ref{ener}) is now 
\be 
E\,=\,-4\Big[\cos(\varphi_1)\,+\,\epsilon \cos(\varphi_2)\Big]_{-\infty}^{\infty}.
\label{energ}
\ee
Note that when $\epsilon=1$ we have the previous model, and when $\epsilon=-1$ we have a model with a kink and an anti-kink solution.
As before we have discussed the case of $\epsilon=1$ let us now look at the case of $\epsilon=-1$.

\subsection{$\epsilon=-1$ case}

When $\epsilon=-1$ and we consider the model with the kink (anti-kink) boundary
conditions {\it i.e.}
\be
\varphi_1(x=-\infty)=\varphi_2(x=\infty)=0;\qquad \varphi_1(x=\infty)=\varphi_2(x=-\infty)=\pi,
\label{energy}
\ee
we find that, again, $E=16$.

To solve the equations (\ref{aa1}) and (\ref{aa2}), like before, we have to choose the values of $\varphi_1$ and $\varphi_2$ at some $x$, say $x_0=0$,
and then numerically solve these equations, from this value of $x$, using positive value
of $dx$, getting the expressions for larger values of $x$. Then we repeat the procedure from the same original value
of $x$ using negative $dx$. The final trajectory is obtained by ``sawing'' together both sets of results. 
This is the same procedure we used before except that this time it is slightly harder to predict the final shape of the curves. When we used this procedure before we knew that if we took the initial values of $\varphi_1$
and $\varphi_2$ positive and $\lambda$ was also  positive we would end up with two kinks, but this time
this is less clear. This is due to the fact that the self-dual equations are just numerical equations, and when one
solves them the system does not know/care about the topology. One has to get the initial values of $\varphi_i$ 
correctly.

We have performed many simulations of such static systems for various values of $\lambda$ (ranging from -1.99 to 1.99).
They all were amazingly static (all numerical artifacts were so small that we saw no overall energy change).
Of course, when $\vert\lambda\vert$ is larger the interaction between the fields is stronger so we run some simulations
for very long times and so no motion and no energy change.

\begin{figure}[h!]
  \centering
  {\includegraphics[width=0.6\textwidth,height=0.4\textwidth, angle =0]{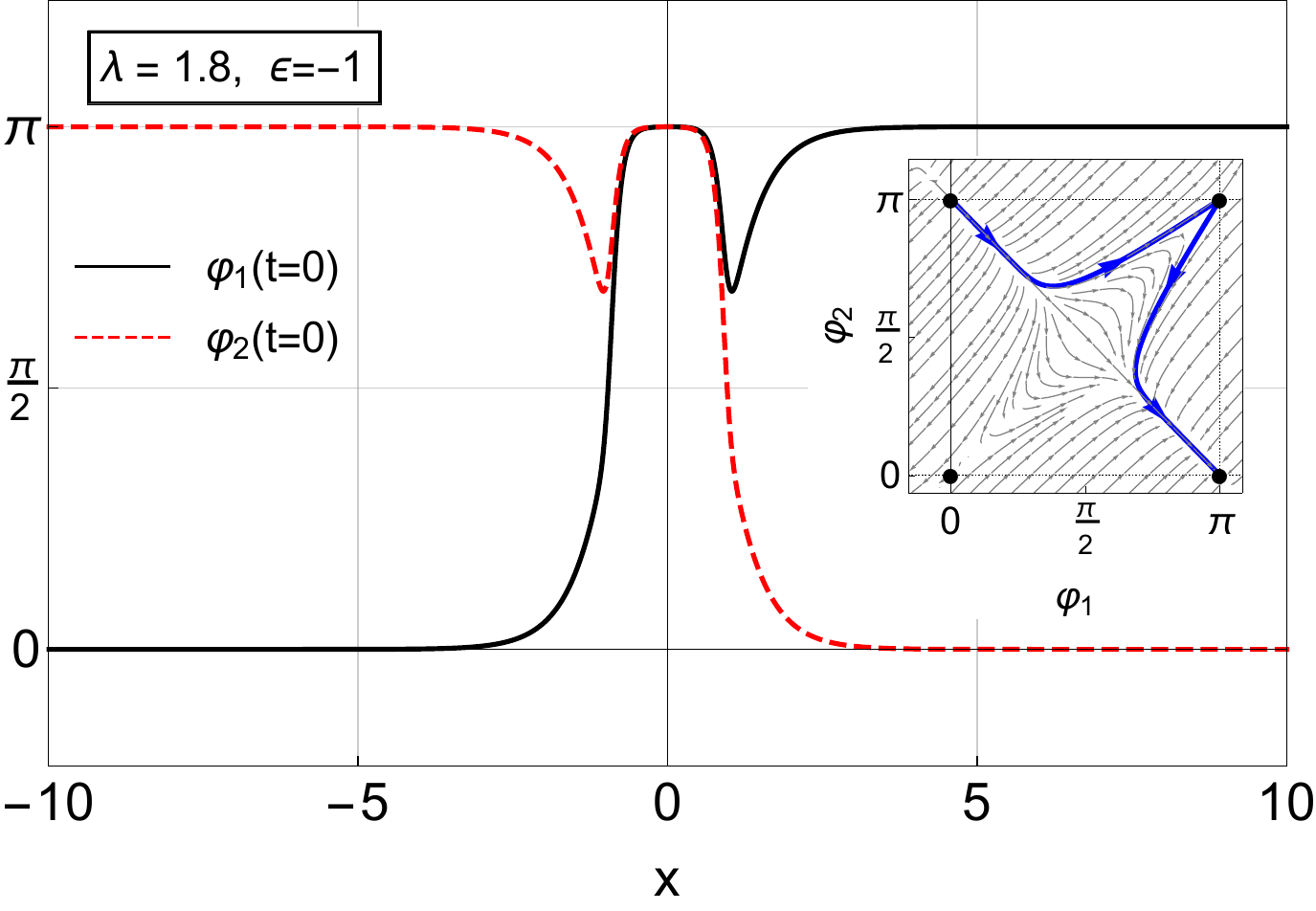}}                
  \caption{ The static  configuration of  fields $\varphi_1$ and $\varphi_2$ for $\lambda=1.8$ and $\epsilon=-1$. The subfigure shows the BPS curve in the space of fields.}
  \label{fig:Xa}
\end{figure}

In Fig.\ref{fig:Xa} we plot fields $\varphi_1$ and $\varphi_2$ for a simulation with $\lambda=1.8$. We evolve this initial configuration using time dependent Euler-Lagrange equations.

As the initial configuration was given by the static BPS solution the forms of these fields have never changed. The fields vary a lot around $\frac{\pi}{2}$ which happens to be
a possible definition of the positions $x_a(t)$ of kink and antikinks, where $a=1,2$. They satisfy $\varphi_a(x_a)=\frac{\pi}{2}$. We did not observe any motion of the kinks, {\it i.e.} $x_a(t)=const$, in this simulation. We skip the plot of the time dependence of energy corresponding with this simulation as the values of the energy have not changed even by $10^{-7}\,\%$!
This is very interesting, as one might expect, that although the fields should be static,
the small numerical errors could make them evolve in time.
However, the errors somehow cancel and the field configurations are extremely stable.

Next we decided to evolve the fields {\it i.e.} by giving them  small {\it initial velocities}. As we stressed before
the model is relativistically invariant and so giving both kink and anti-kink the same velocity does not change anything
and the structures evolve without radiating. And as the solitonic structures are well localised, and they 
really interact with each only when they are close together the only interesting tests would involve sending the structures towards each other.

As before, when we made the solitons move - we did this by defining the initial time derivatives of the fields 
by exploiting the Lorentz covariance of the model; {\it i.e.} by taking the time derivative 
of each field proportional to the spatial derivative of this field, with the constant of the proportionality given by the velocity. In our simulations, we have always taken the two structures move with the same
speed towards each other - so that the interaction would take place more or less in the middle of our grid.

We have performed many such simulations, for various values of the velocity and for various values of $\lambda$.
To reduce the potential error of such a procedure we restricted our attention to small velocities and tried
to localise the initial solitons at reasonable distances from each other. This was quite complicated 
as we first had to solve the self-duality equations and only then boost the fields. The results 
 of our simulations were basically very similar but they exhibited many interesting properties
which we have studied in detail and which we describe below.

\subsection{Some results}

In most cases the solitons passed through each other very smoothly - emitting very little radiation.
This was particularly true for small values of $\lambda$ ({\it i.e.} when the solitons interacted with each
very little). However, for larger values of $\vert\lambda\vert$ we saw reflections.

\begin{figure}[h!]
  \centering
  \subfigure[]{\includegraphics[width=0.45\textwidth,height=0.33\textwidth, angle =0]{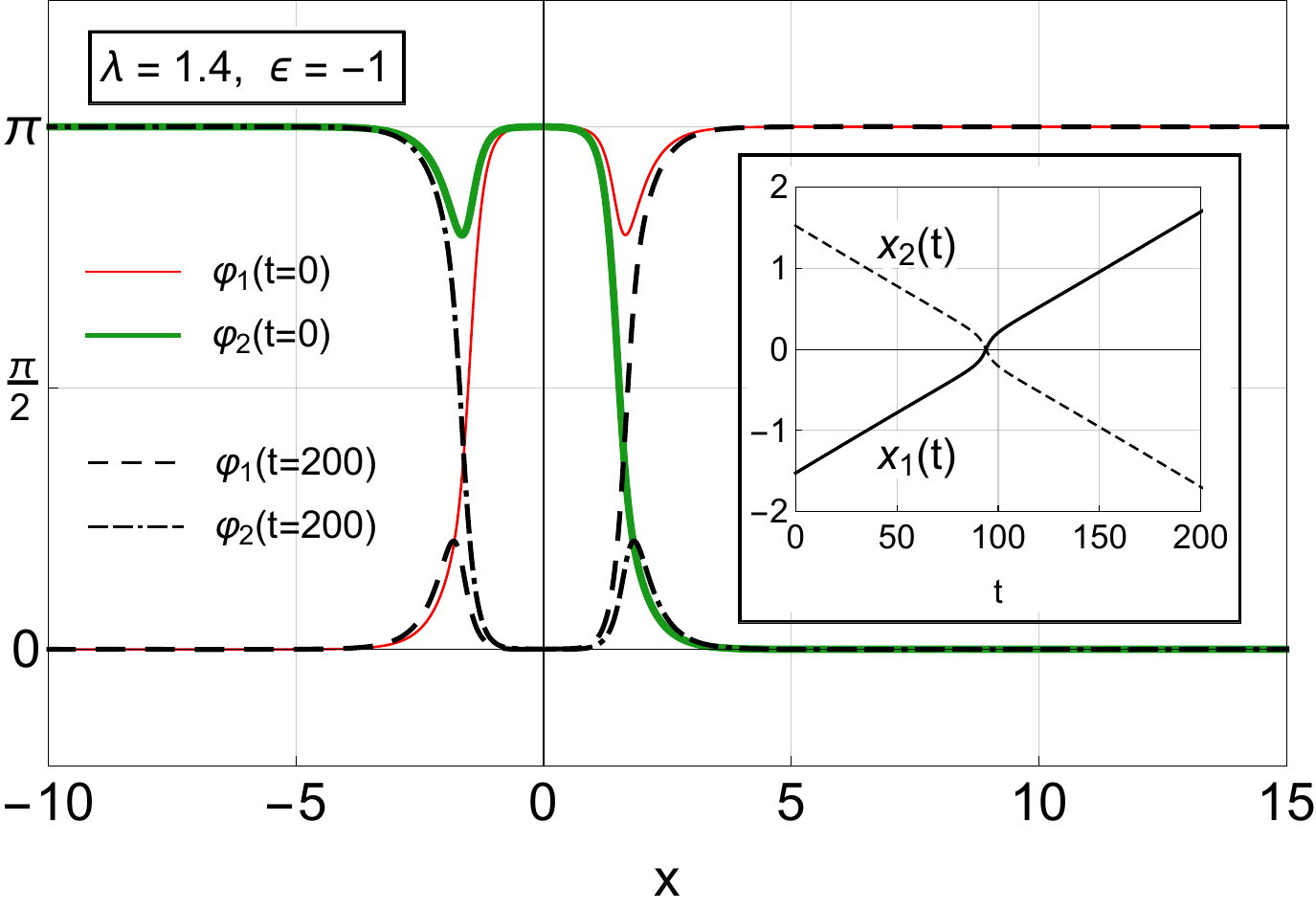}} 
  \hskip0.5cm               
  \subfigure[]{\includegraphics[width=0.45\textwidth,height=0.33\textwidth, angle =0]{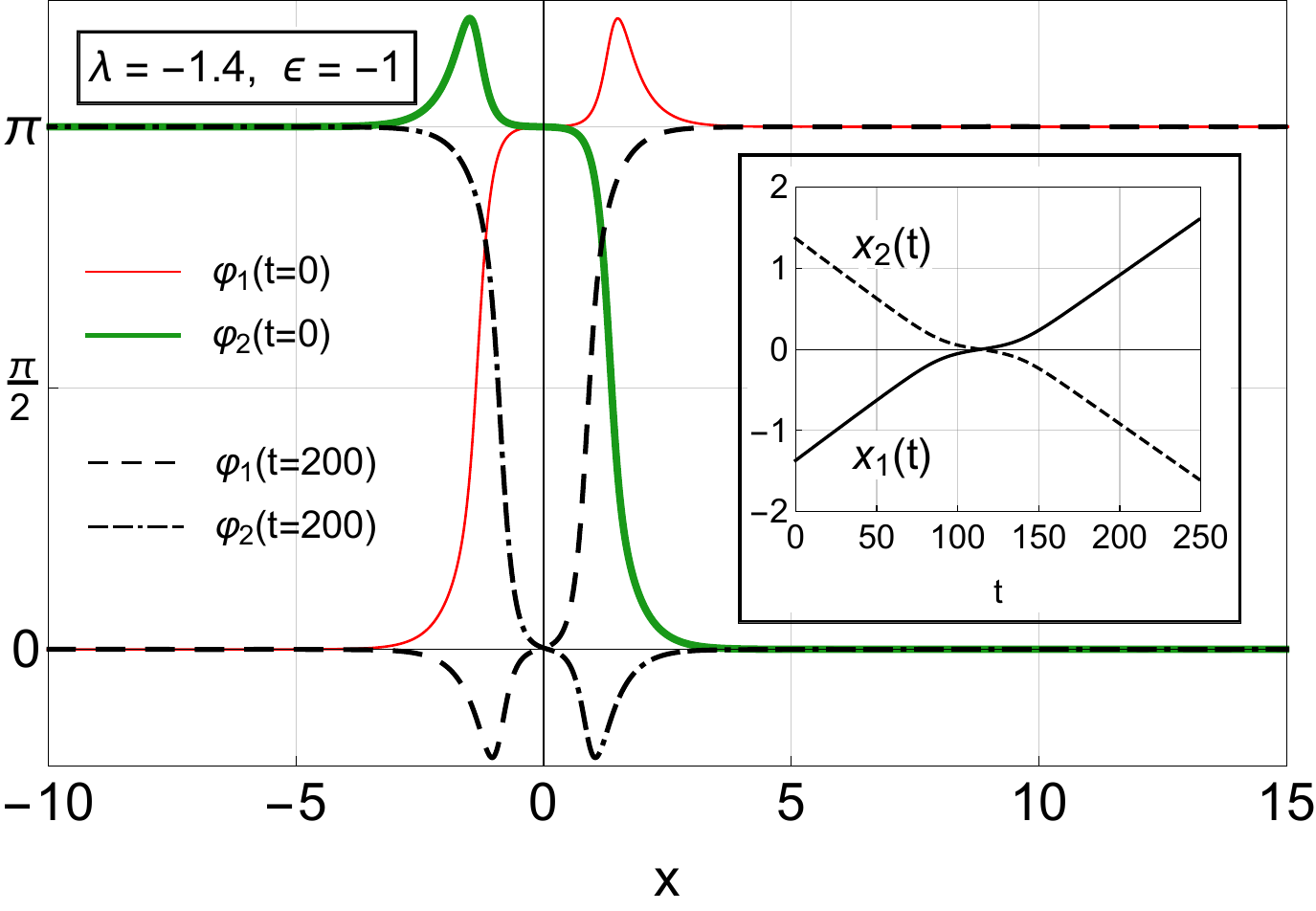}}
  \caption{Time evolution of the fields and the kink trajectories $x_a(t)$ for $a=1,2$ where  $\varphi_a(x_a)=\frac{\pi}{2}$.  (a)  $\lambda=1.4$ case, (b) $\lambda=-1.4$ case.}
  \label{fig:Xc}
\end{figure}

Let us discuss first the cases of small $\vert\lambda\vert$.
In figure Fig.\ref{fig:Xc}(a) and (b) we present plots of the initial, $t=0$, and after the scattering, {\it i.e.} at $t=200$, fields $\vp_1$ and $\vp_2$ for $\lambda=1.4$ and $\lambda=-1.4$. In insertions we present also the plots of the trajectories $x_1(t)$ and $x_2(t)$ of the kink and antikink described by these solutions.

Looking  at the structures described by the fields and their scattering we note a few things:
\begin{itemize}
\item Not much overall change in the linearity of the trajectories nor in the shape of the fields themselves.
\item Basic difference - ``bumps'' on soliton fields pointing in different directions
and small overall deformations of the trajectories. For $\lambda>0$ we see a positive shift along the trajectory and 
perhaps a slight slowing down of the soliton, while for $\lambda<0$ the shift is negative and also the slight
slowing down of the solitons.
\end{itemize}

This resembles very closely the familiar phase shift of solitons during their scatterings \cite{Ablowitz}.
So, the obvious question is: how does this depend on $\lambda$?  Clearly, for vanishing $\lambda$ the effects
disappear, as the fields become independent, so we have looked at many, and particular larger, values of $\vert \lambda\vert$.

\begin{figure}[h!]
  \centering
 {\includegraphics[width=0.6\textwidth,height=0.35\textwidth]{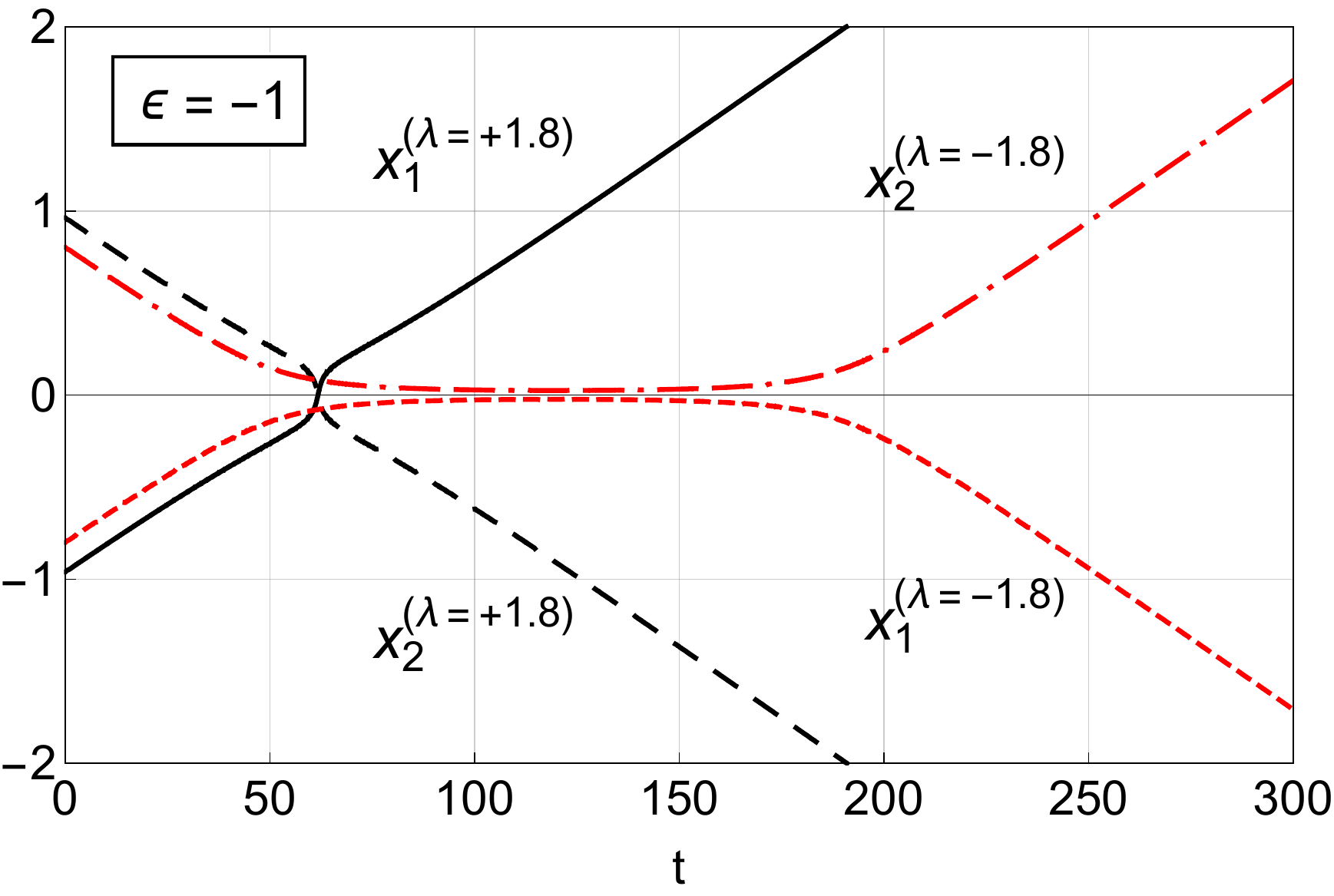}}
  \caption{Trajectories  $x_a(t)$ where $a=1,2$ and $\varphi_a(x_a)=\frac{\pi}{2}$  in the model with $\lambda=\pm1.8$ and $\epsilon=-1$.}
  \label{fig:Xf}
\end{figure}

\begin{figure}[h!]
  \centering
  \subfigure[]{\includegraphics[width=0.45\textwidth,height=0.3\textwidth]{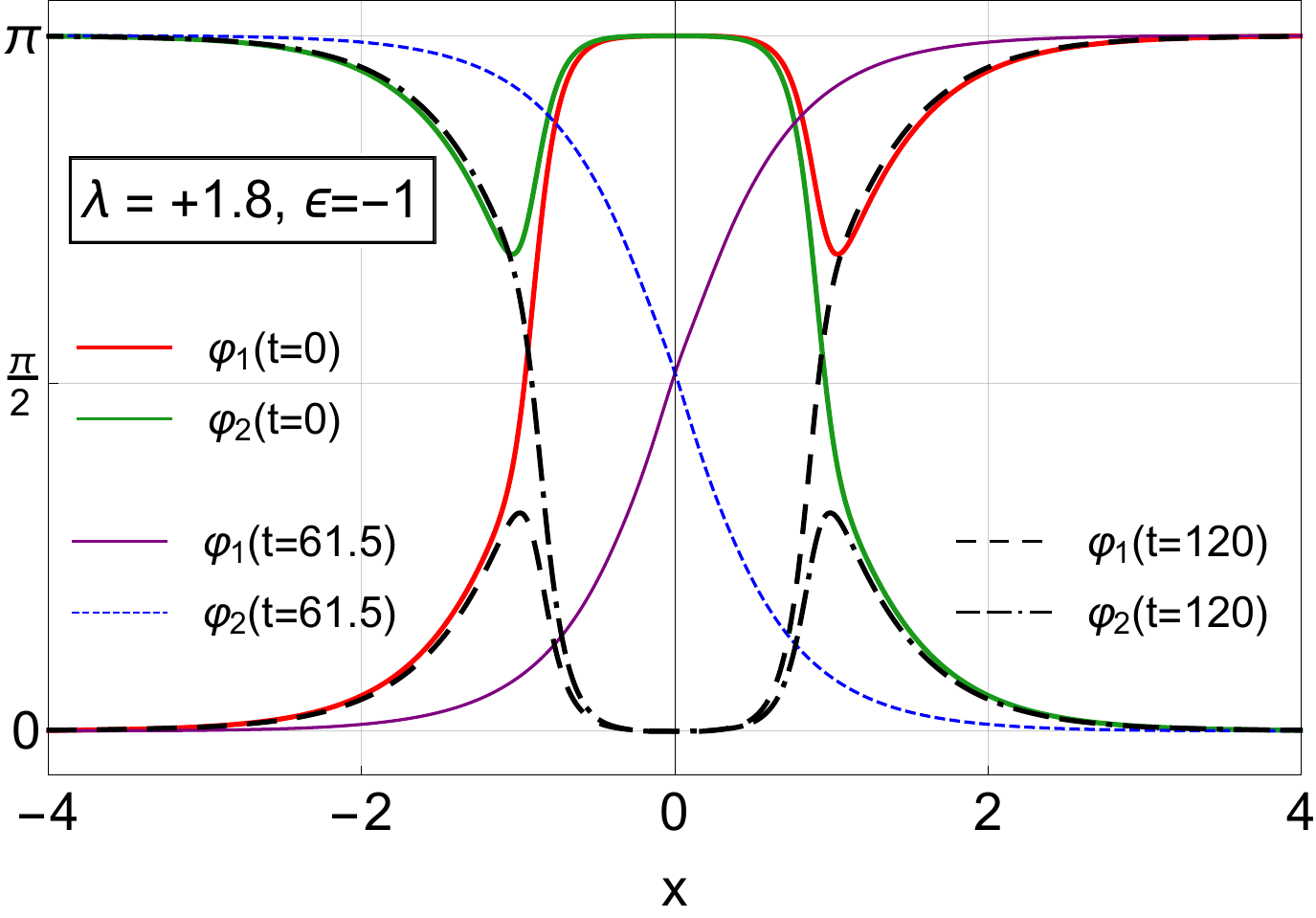}}\hskip0.9cm                
  \subfigure[]{\includegraphics[width=0.45\textwidth,height=0.3\textwidth]{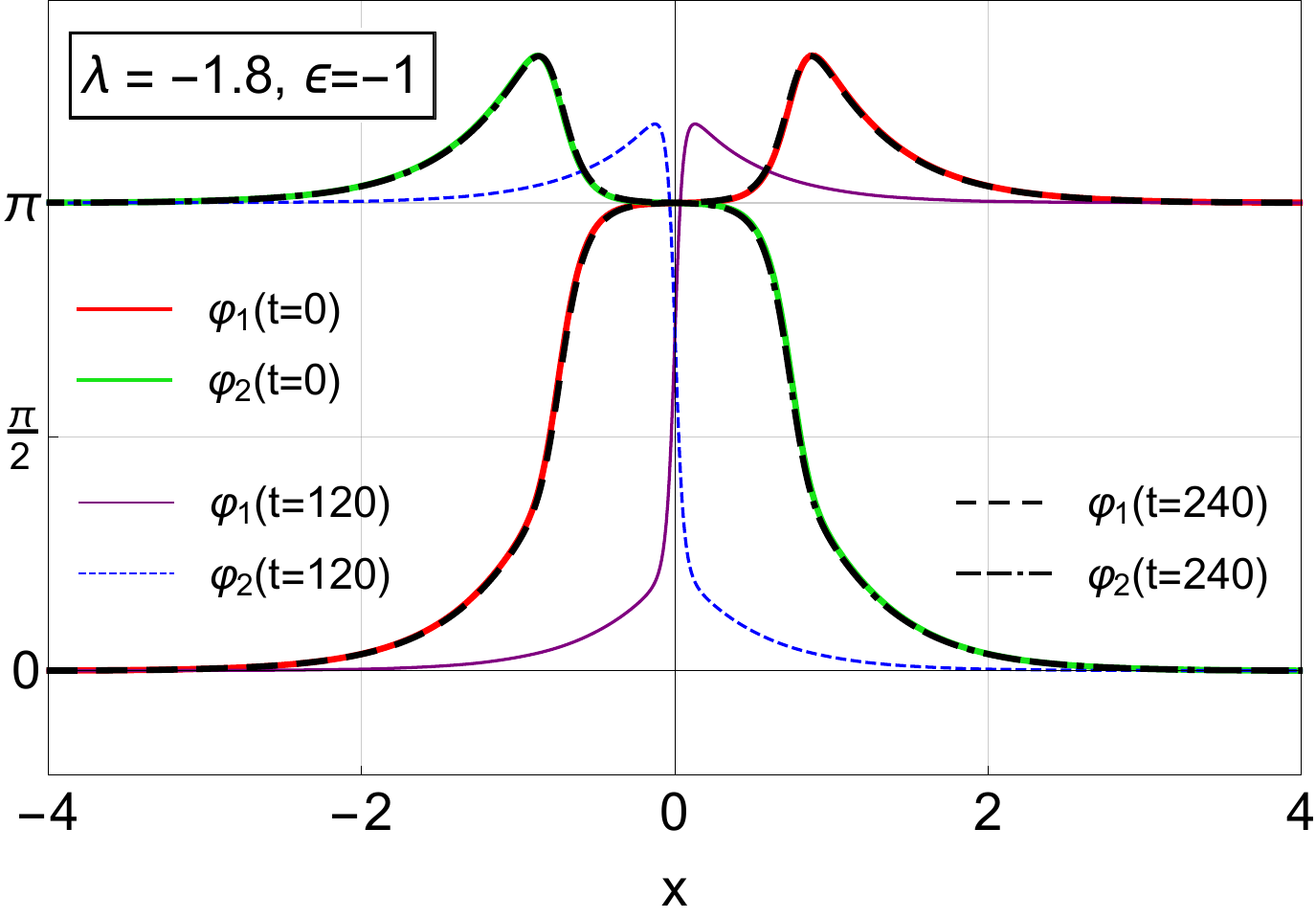}}
  \caption{The case $\epsilon=-1$. The fields $\varphi_1$ and $\varphi_2$ for (a) $\lambda=1.8$ (transmission) and (b) $\lambda=-1.8$ (reflection).}
  \label{fig:Xg}
\end{figure}

We carried out many simulations for increasing values of $\vert \lambda\vert$. For smaller ones there was no significant difference until we came close to $\vert\lambda\vert=1.8$ and there everything was different. For positive values of $\lambda$ {\it i.e.} $\lambda=1.8$, the results were not that different from those for $\lambda=1.4$, shown in Fig.\ref{fig:Xc}(a) but for $\lambda=-1.8$ we had a reflection! In Fig.\ref{fig:Xf} we present the plots of some of the observed trajectories. 
The plots of fields at different instants of time are shown in Fig.\ref{fig:Xg}(a) and (b). We note that for $\lambda=-1.8$ the fields at $t=0$ and $t=240$ are quite similar. 

\begin{figure}[h!]
  \centering
  \subfigure[]{\includegraphics[width=0.32\textwidth,height=0.32\textwidth, angle =0]{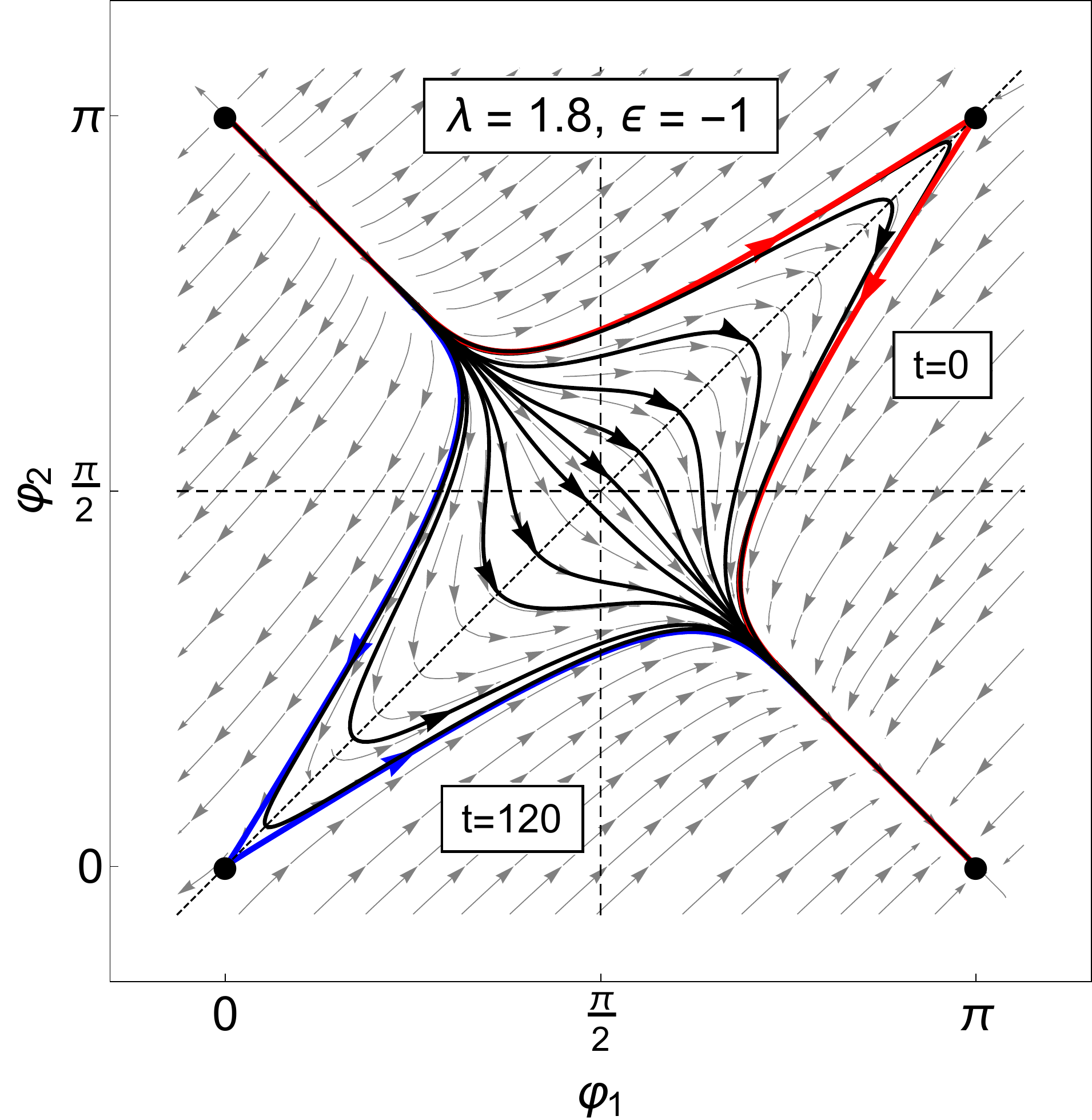}}                
  \subfigure[]{\includegraphics[width=0.32\textwidth,height=0.32\textwidth, angle =0]{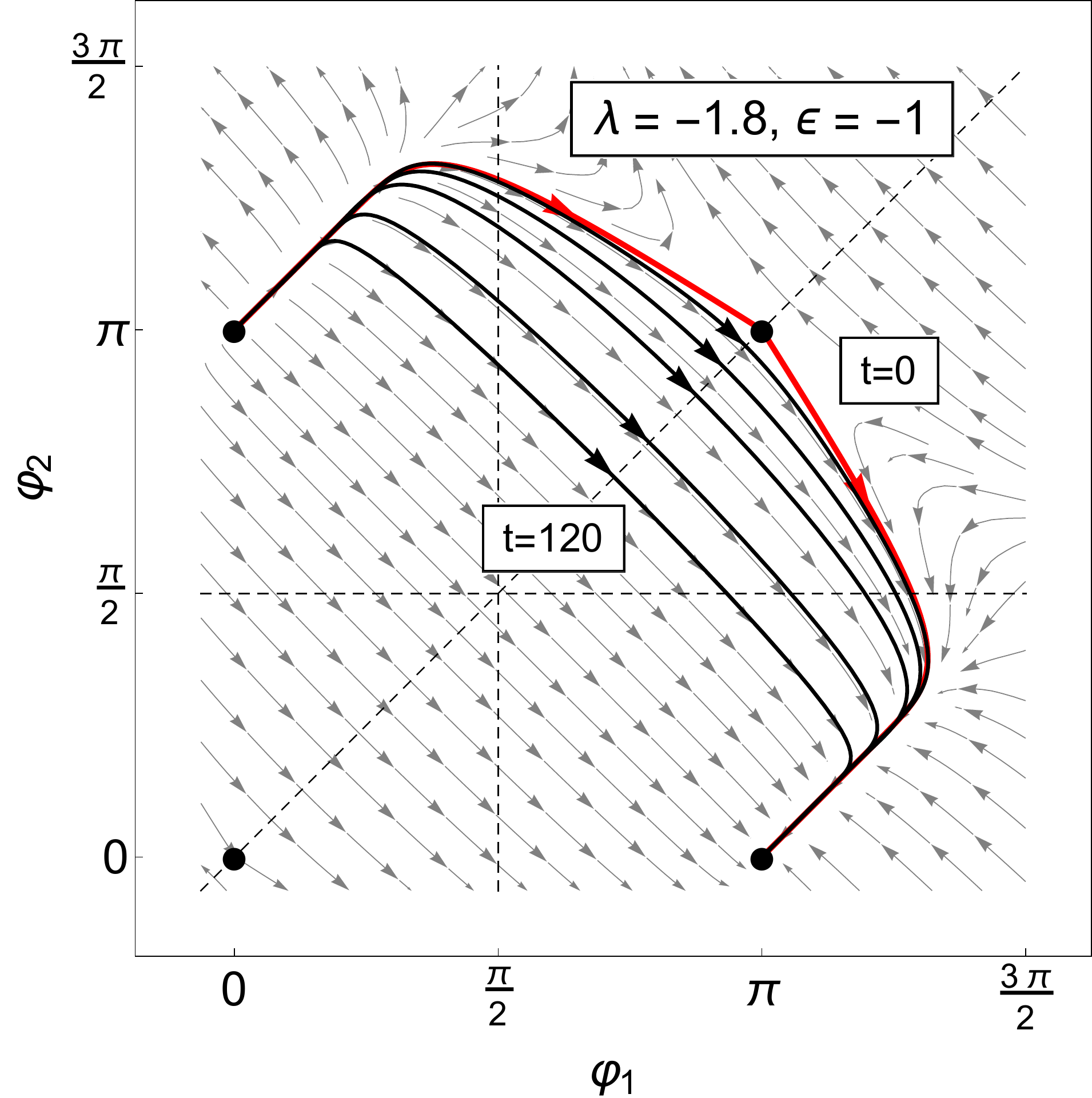}}
   \subfigure[]{\includegraphics[width=0.32\textwidth,height=0.32\textwidth, angle =0]{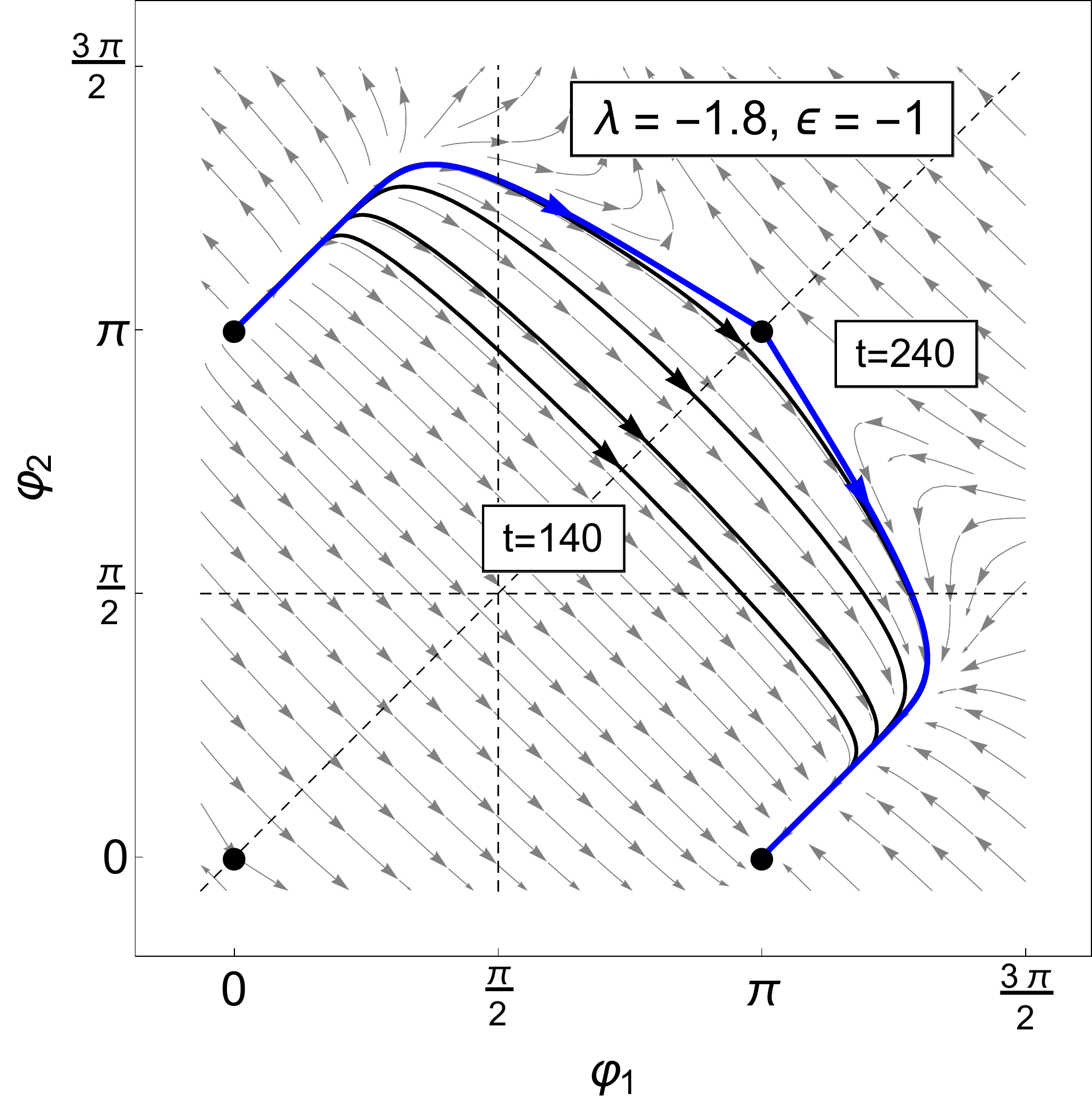}}
   \caption{The time evolution of the fields for $\lambda=\pm1.8$ and $\epsilon=-1$. (a)  Transmission; the curves   correspond to instants of time  $t=0, 40, 50, 58, 60, 61, 61.5, 62, 63, 64, 70, 80, 120$. (b) Evolution before the reflection at instants $t=0, 40, 50, 60,80,120$   and (c) after the reflection at $t=140,160, 180, 200, 240$.}
 \label{fig:Xh}
\end{figure}

Next we have looked at the time evolution of the fields for these two cases ($\lambda=1.8$ and $\lambda=-1.8$) in the space of functions $\varphi_1$ and $\varphi_2$. The plots of the fields are shown in Fig.\ref{fig:Xh}.
We see very clearly the difference in their behaviour.  Fig.\ref{fig:Xh}(a) shows curves that represent evolution of the fields $\varphi_1$ and $\varphi_2$ for $\lambda=1.8$. The initial, ({\it i.e.} $t=0$) curve  is clearly very different from the final one (at $t=120$). The $\varphi_1$-kink - $\varphi_2$-antikink reflection is shown in  Fig.\ref{fig:Xh}(b) and (c) where (b) shows the curves, at various times, before the reflection and (c) shows them  after the reflection. Although  the numerically obtained curves at specific values of time are not the BPS curves they amazingly well follow the modified gradient flow of the pre-potential $\eta^{-1}\cdot\vec \nabla U$ which gives the curves of the solutions of self-dual equations for different values of the initial condition.

Given these interesting results for $\lambda=-1.8$ we have decided to look in more detail at the dependence of the evolution on the values of $\lambda$ and tried to determine the  value of  $\lambda$ at which we have this change of the behaviour. First we have found that for all values of $\lambda>0$ we have a transmission so we concentrated our attention on $\lambda<0$. As we said before; 
for small values of negative $\lambda$ we also had the transmissions. So we have tried to determine
the value of $\lambda$ at which the change from the transmission to the reflection takes place and also determine
how this takes placed.   We have found that this transition is really quite complicated.

\begin{figure}[h!]
  \centering
 \subfigure[]{\includegraphics[width=0.45\textwidth,height=0.3\textwidth]{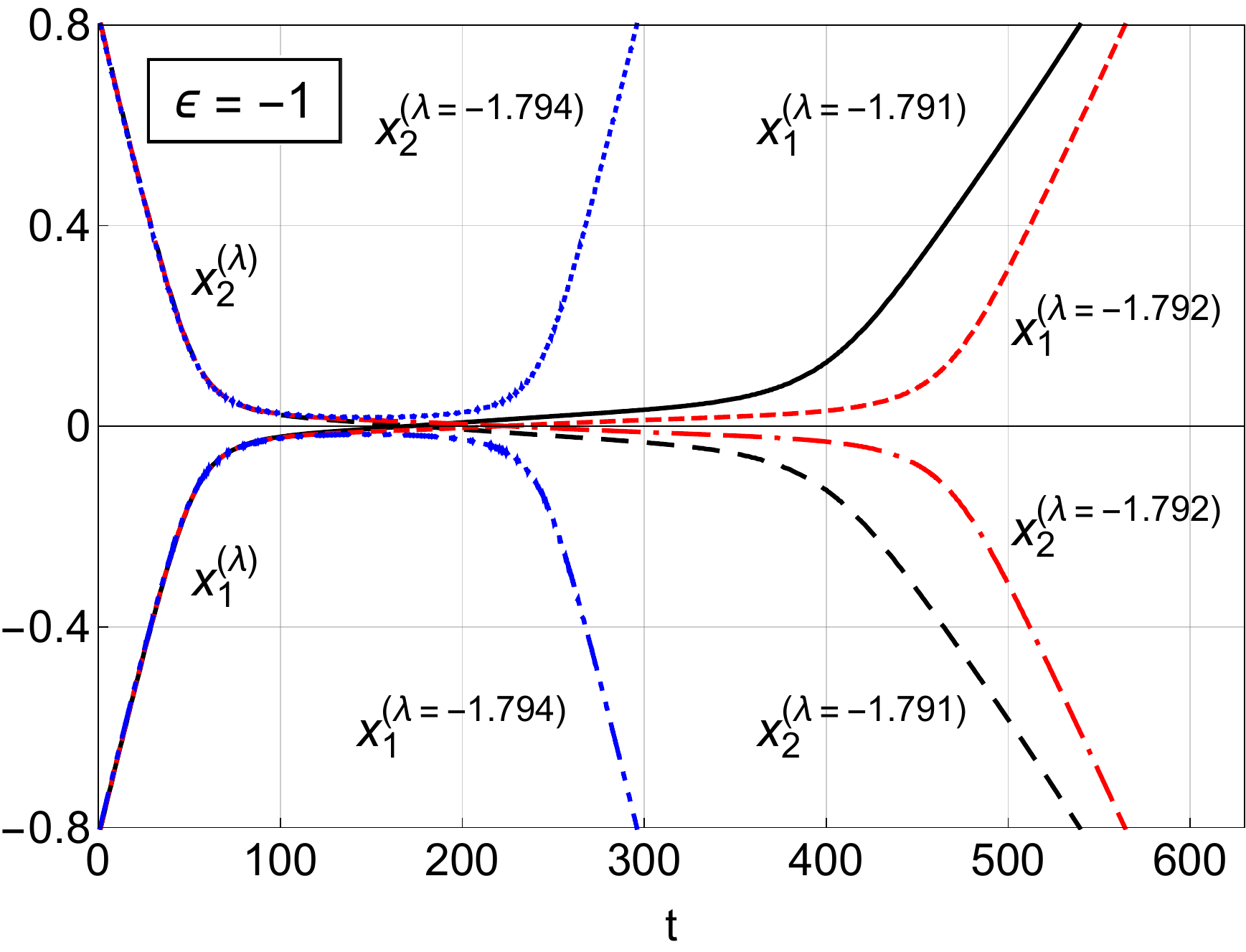}}\hskip0.9cm                
  \subfigure[]{\includegraphics[width=0.45\textwidth,height=0.3\textwidth]{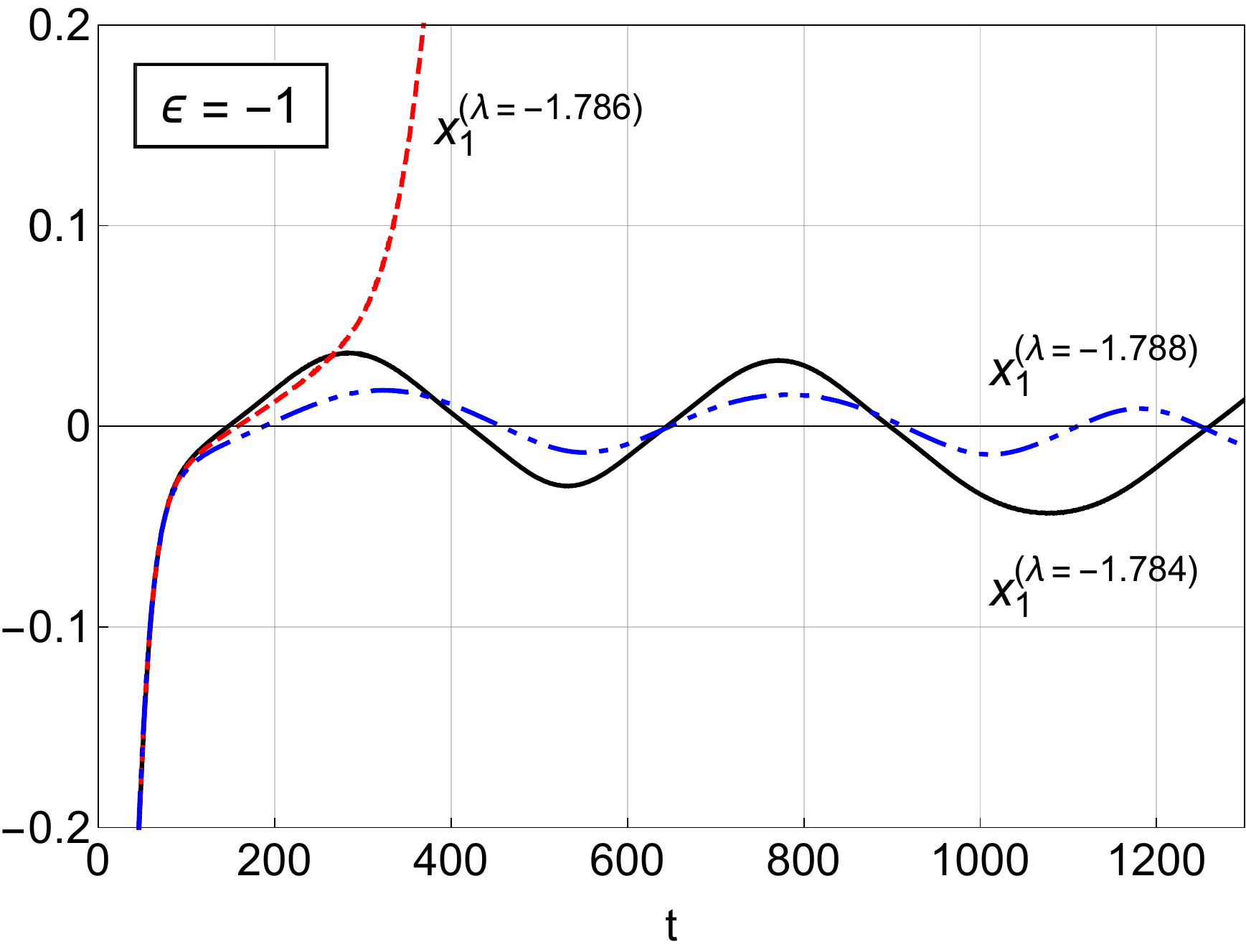}}
  \caption{ (a) Kink ($x_1(t)$) and anti-kink ($x_2(t)$) trajectories  of  solutions with  $\lambda=-1.791$ (transmission), $\lambda=1.792$ (transmission) and $\lambda=-1.794$ (reflection). (b) Trajectories $x_1(t)$ near to the transmission/reflection for $\lambda=-1.784$, $\lambda=-1.786$ and $\lambda=-1.788$.}
  \label{fig:Xk}
\end{figure}

In the end we have found that this change takes place around $\lambda=-1.793$ as for  $\lambda=-1.792$ we still had a transmission and for $\lambda=-1.784$ we had a reflection.
In Fig.\ref{fig:Xk}(a) we present the plots of the kink and anti-kink  trajectories of $\varphi_1$ and $\varphi_2$ seen in both cases.

However, looking at some values below $\lambda=-1.793$, 
like $\lambda=-1.784$ or even $\lambda=-1.788$, we have found the system a bit uncertain as to what to do
(although for $\lambda=-1.786$, a value in-between, we saw a transmission).
In Fig.\ref{fig:Xk}(b) we plot the corresponding trajectories of only $\varphi_1$ kinks (to make the picture somewhat clearer)  as seen in three simulations ($\lambda=-1.784$, $-1.786$ and $-1.788$).

An obvious question then arises - how does this change from the transmission to the reflection take place? This is quite complicated but one can show that, in part, this is associated with the gradual flattening of the soliton trajectory during the scattering.  This is clear from Fig.\ref{fig:Xk}(a) when comparing trajectories seen in the simulations for $\lambda=-1.791$ and $\lambda=-1.792$, {\it i.e.} just before the reflection.  We clearly see a slightly larger distortion of the trajectory for $\lambda=-1.792$.

\subsection{$\epsilon=+1$ case}
\begin{figure}[h!]
  \centering               
 {\includegraphics[width=0.6\textwidth,height=0.35\textwidth]{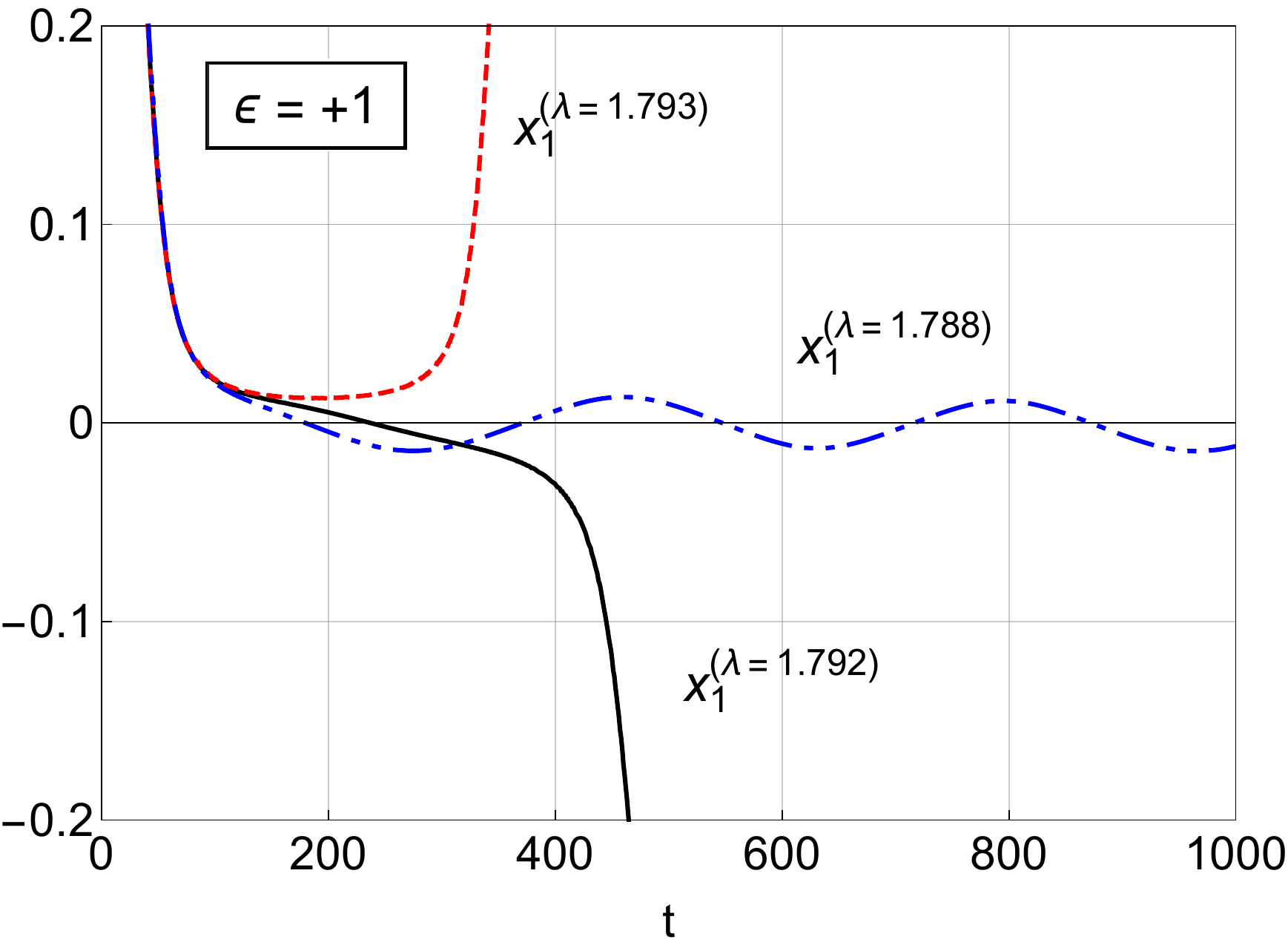}}
  \caption{The kink trajectories $x_1(t)$ seen for $\lambda=1.788$, $\lambda=1.792$ and $\lambda=1.793$ in the model with $\epsilon=1$. }
  \label{fig:Xn}
\end{figure}

Given the results we had obtained for $\epsilon=-1$ we have decided to return to the $\epsilon=1$ case. Looking at our 
self-duality equations (\ref{aa1}) and (\ref{aa2}) we note that, in fact, if we take the case $\epsilon=-1$ and define $\phi_2=\pi-\varphi_2$ and furthermore, change the sign of $\lambda$ we get the equations for $\epsilon=+1$.
We see that the two cases are very similar and related and so we can expect, using the previous functions $\varphi_1$ and $\varphi_2$, also to be able to find a transition between the transmission and the reflection olf the kinks in this case too.
We have carried out many such simulations, starting from slightly different initial values of $\varphi_1$ and $\varphi_2$ and, as expected, have found essentially the same results as before (but this time for the values of $\lambda$ with an opposite sign). Thus we saw a transmission for all the negative values of $\lambda$ and for positive $\lambda$ up to 1.792. From $\lambda=1.793$ onwards we saw reflections.
In Fig.\ref{fig:Xn} we present three trajectories of the $x_1(t)$ kinks, {\it i.e.} of the field $\varphi_1$, for $\lambda=1.788$, $\lambda=1.792$ and for $\lambda=1.793$.

It is amazing that, given that we have taken the initial values of the fields for the two values of $\epsilon$ completely independently, the results are very similar (with 
the above mentioned changes of the sign of $\lambda$). They show that the general nature of our results (the values of $\lambda$ at which the transition takes place) is close to the vanishing of the numerators of equations (\ref{aa1}) and (\ref{aa2}). But, in addition, it also shows that the results are very little dependent on the initial values of the fields or on the inavoidable numerical errors.

\subsection{Adiabatic motion with fields along the lines of flow}

In most of our numerical calculations we have seen that the relative  motion of the kinks (or kinks and anti-kinks) was such that, at each time, the fields were aligned to the lines of the pre-potential flow.
This may appear somewhat surprising but it can be partly explained by the adiabatic nature of our simulations.

Our simulations were started by taking the initial values of
$\frac{d\phi_i}{dt}$   proportional to    $\frac{d\phi_i}{dx}$. This was due to the fact that considered a small velocity of each kink in our simulations and we relied on the fact that the initial solitons were far apart and well localised. Moreover, as the model is Lorentz covariant, the constant of proportionality was   $\frac{v}{\sqrt{1-v^2}}$.

However, from the self-duality equations which gave us the expressions for the fields at the initial time we have that
\begin{align}
F_1 &= \partial_x \phi_1 = A[2\partial_{\phi_1}U+\lambda\partial_{\phi_2}U]\\
F_2 &= \partial_x \phi_2 = A[2\partial_{\phi_2}U+\lambda\partial_{\phi_1}U]
\end{align}
where $A$ is a constant.
But as $\eta^{-1}$ is proportional to 
\be
\eta^{-1}_{ab}\,=\, \frac{4}{4-\lambda^2}\left(\begin{array}{cc}
1& \frac{\lambda}{2}\\
\frac{\lambda}{2}& 1 \end{array}\right)
\lab{inveta}
\ee
we see that the flow $\vec F\,=\,\eta^{-1}\vec \nabla U$ gives us 
\be
\frac{\partial \phi_1}{\partial \phi_2} = \frac{F_1}{F_2} \,=\, \frac{ 2\partial_{\phi_1}U+\lambda\partial_{\phi_2}U}{2\partial_{\phi_2}U+\lambda\partial_{\phi_1}U}, 
\ee
which is exactly the expression above. So we see that our procedure, sends the fields from one set of flow lines to another.

Note that is very similar to what happens in the usual geodesic approximation where the motion of the structures is approximated by the motion in the potential valley with
 the parameters of the static solutions evolving the solution along the valley. 
Here this happens implicitly and the approximation works very well as our velocities have always been very small.



\section{Conclusions}

In this paper we have looked in detail at some properties of the simplest BPS models in (1+1) dimensions involving more that one field; namely, the interacting BPS generalisation of 
of two Sine-Gordon fields. This generalisation has emerged 
out of our previous studies (\cite{e} and \cite{AFHWZ}) of BPS systems involving more than one field. The two fields are 
 coupled together with strength $\lambda$. Of course, when this strength vanishes we have two non-interacting fields and nothing special happens, {\it i.e.} each field possesses a kink or antikink solution.
However for $\lambda \ne0$ the fields interact and the presence of one of them affects the other one. These interactions change the shape
of the kink fields in an interesting way (in the form of a small kink-antikink configuration added at the place where the kink of the other field is located).

In this paper we have analysed many properties of such models paying particular attention to the dependence
of these properties on the coupling between these two Sine-Gordon fields. In our work we have found it very convenient 
to think of our static BPS fields as curves in the space of fields (as discussed in detail in cite{e}) and then compare the more general fields to the generalised gradient flow of the appropriate pre-potential $U$.

First we looked at the solutions of the self-duality equations themselves and evolved them using these solutions
as the initial conditions for the full time dependent equations. Not surprisingly, they did not evolve. But this was true to an incredible degree of accuracy. This demonstrated to us that such solutions are really very stable,
and all the perturbations introduced by the numerical simulations did not changed this.

Then we altered the initial conditions - by giving the kink in one field, and the kink or antikink in the other
one small velocity towards each other. As such initial fields were not solutions of the self-duality
equations (which require the fields to be static) they did evolve. We looked at small velocities and we were surprised to see that, that despite their evolutions, the fields at all times resembled the solutions
of the self-duality equations, {\it i.e.} the fields evolved through such solutions. So they tried to align themselves 
with the lines of the generalised flow.  For small values 
of $\vert \lambda\vert$ one could perhaps describe their time evolutions by appropriate collective
coordinates. However, the fields also showed the well known phase shift of the solitons, see {\it e.g.} \cite{Ablowitz}, along
their trajectories. This phase shift increases with the interaction ({\it i.e.} depends on $\vert\lambda\vert$)
but for larger values we observed a reflection. These reflections occured for the values of $\lambda$ 
when the solitons, undergoing their phase shift came closer together. At some values the solitons 
have got trapped (they were oscillating staying close together gradually settling towards a static solution
described by two solitons together). Of course, adding solitons some velocity, increased the energy of the system 
(by extremely small amounts) and during these oscillations this extra energy was slowly emitted.

We have presented some explanations of our observations but we want to go further and consider  also multisoliton configurations of each field and their interactions.
This work is currently in progress and we hope to report some
concrete results in the near future.

However, at this stage we can state that our work has shown that systems involving more interacting fields
are quite complicated even for small values of the coupling constant $\lambda$ connecting
the fields together.  Recently, some papers have also appeared presenting various studies of multi-field models in (1+1) dimensions. 
A good recent paper of such a class is \cite{shnir} which carried out similar work for two coupled $\lambda \phi^4$ in (2+1) model.

Both classes of results suggest that, when the coupling constants are small; {\it i.e.} when $\lambda\ne0$ in our case, the system is non-integrable - but for small values of $\lambda$ the models are not that different
from being integrable. Hence, its properties partially support the ideas of quasi-integrability  \cite{us}. At the same time the property
of the field theory of `being BPS' does not appear to be extremely important. Of course, this is all in (1+1) dimensions.
This may be very different for the field theories in higher dimensions - and, in particular, in models in which multisoliton solutions
can be determined by self-duality ({\it i.e.}) like monopoles in (3+1) dimensions or baby skyrmions in (2+1) dimensions.
We plan to look in more detail at such theories also in our future work.

\section*{Acknowledgements}
 LAF and WJZ want to thank the Royal Society for 
its grant which supported their research and they want
to thank FAPESP and Durham University for their joint 'Collaboration grant' that supported their reciprocal visits to Sao Carlos and Durham which have made the research described in this paper possible.
WJZ work was supported in part also by the Leverhulme Trust Emeritus Fellowship.

\end{document}